\documentclass[aps,preprint]{revtex4-1}

\usepackage{mathtools}
\usepackage{amsmath,bm}
\usepackage{amssymb}
\usepackage{subfigure}
\usepackage{setspace}
\usepackage{dcolumn}
\usepackage{color}
\usepackage{graphicx}

\usepackage[hidelinks]{hyperref}

\begin{document}

\title{Dynamics and topology of non-Hermitian elastic lattices with non-local feedback control interactions}

\author{Matheus I. N. Rosa}
\affiliation{ School of Mechanical Engineering, Georgia Institute of Technology, Atlanta GA 30332}
\author{Massimo Ruzzene}
\affiliation{ Department of Mechanical Engineering, University of Colorado Boulder, Boulder CO 80309}

\date{\today}

\begin{abstract}
We investigate non-Hermitian elastic lattices characterized by non-local feedback control interactions. In one-dimensional lattices, we show that the proportional control interactions produce complex dispersion relations characterized by gain and loss in opposite propagation directions. Depending on the non-local nature of the control interactions, the resulting non-reciprocity occurs in multiple frequency bands characterized by opposite non-reciprocal behavior. The dispersion topology is also investigated with focus on winding numbers and non-Hermitian skin effect, which manifests itself through bulk modes localized at the boundaries of finite lattices. In two-dimensional lattices, non-reciprocity is associated with directional dependent wave amplification. Moreover, the non-Hermitian skin effect manifests as modes localized at the boundaries of finite lattice strips, whose combined effect in two directions leads to the presence of bulk modes localized at the corners of finite two-dimensional lattices. Our results describe fundamental properties of non-Hermitian elastic lattices, and open new possibilities for the design of metamaterials with novel functionalities related to selective wave filtering, amplification and localization. The results also suggest that feedback interactions may be a useful strategy to investigate topological phases of non-Hermitian systems.
\end{abstract}


\maketitle

\section{Introduction}\label{Introduction}
 
Metamaterials and phononic crystals are periodic structures designed to manipulate acoustic and elastic waves~\cite{lu2009phononic,hussein2014dynamics}. Potential applications include vibration attenuation~\cite{huang2009wave}, noise reduction~\cite{yang2010acoustic}, wave focusing~\cite{lin2009gradient}, cloaking~\cite{cummer2007one}, and the design of seismic barriers~\cite{miniaci2016large}. Recent breakthroughs in topological insulators in solid state physics~\cite{hasan2010colloquium} and photonics~\cite{lu2014topological} have motivated the search for topology-based functionalities in mechanical and acoustic metamaterials. This has culminated in the consolidation of topological mechanics~\cite{huber2016topological} and acoustics~\cite{yang2015topological} as active research fields~\cite{ma2019topological}. Topological states have been successfully observed in several platforms~\cite{fleury2016floquet,mousavi2015topologically,susstrunk2015observation,wang2015topological,nash2015topological,pal2017edge, miniaci2018experimental,liu2018tunable,chaunsali2018subwavelength}, and have been pursued to achieve robust, diffraction-free wave motion. Additional functionalities have been explored in the context of topological pumping~\cite{rosa2019edge,grinberg2019robust,chen2019mechanical,riva2019edge,brouzos2019non}, quasi-periodicity~\cite{apigo2019observation,ni2019observation,Pal_2019}, and non-reciprocal wave propagation in active~\cite{fleury2014sound,trainiti2016non,wang2018observation,chen2019nonreciprocal, marconi2019experimental} or passive non-linear~\cite{coulais2017static,bunyan2018acoustic,darabi2019broadband,mojahed2019tunable} systems. These works and the references therein illustrate a wealth of strategies for the manipulation of elastic and acoustic waves, and suggest intriguing possibilities for technological applications in acoustic devices, sensing, energy harvesting, among others.

Considerable efforts have been recently devoted towards the exploration of non-Hermiticity in various physical platforms such as in optical~\cite{makris2008beam,zhao2019non}, optomechanical~\cite{xu2016topological}, 
acoustic~\cite{fleury2015invisible}, and mechanical~\cite{ghatak2019observation,brandenbourger2019non} systems. Non-Hermitian systems are non-conservative systems where loss and/or gain are inherently present from interactions with the environment. In this context, the realization that parity-time (PT) symmetric non-Hermitian Hamiltonians may exhibit purely real spectra~\cite{bender1998real} has sparkled renewed interest in non-Hermitian physics~\cite{longhi2018parity,el2018non}. Indeed, a large portion of recent studies has focused on PT symmetry and the role of exceptional points~\cite{miri2019exceptional}, whose intriguing properties lead to unconventional phenomena such as unidirectional invisibility~\cite{lin2011unidirectional,fleury2015invisible}, single-mode lasers~\cite{feng2014single} and enhanced sensitivity to perturbations~\cite{hodaei2017enhanced,chen2017exceptional}. 
Initial interest~\cite{gong2018topological,shen2018topological,ghatak2019new,torres2019perspective} revolved around exceptional points exhibiting unique topological features with no counterparts in Hermitian systems, such as Weyl exceptional rings~\cite{xu2017weyl}, bulk Fermi arcs and half-integer topological charges~\cite{zhou2018observation}. Further observations of a seemingly breakdown of the bulk-bundary correspondence principle~\cite{lee2016anomalous,xiong2018does} has led to proposals for a general classification of the topological phases of non-Hermitian systems~\cite{gong2018topological,shen2018topological,kawabata2019symmetry}. A particular point of interest is the observation of the \textit{non-Hermitian skin effect}~\cite{yao2018edge,alvarez2018non,lee2019anatomy,longhi2019probing}, whereby all eigenstates of one-dimensional (1D) systems are localized at a boundary, in sharp contrast with the extend Bloch modes of Hermitian counterparts. This intriguing feature of non-Hermitian lattices has recently been experimentally demonstrated using topoelectrical circuits~\cite{hofmann2019reciprocal} and quantum walks of single photons~\cite{xiao2019observation}. Further theoretical investigations have also shown higher order skin modes localized at corners and edges of 2D and 3D non-Hermitian lattices~\cite{lee2019hybrid}. 

While most studies have so far focused on non-Hermitian optical and condensed matter systems, a few works have explored non-Hermiticity in elastic and acoustic media, most of which focus on PT phase transitions and exceptional points~\cite{zhu2014p,fleury2015invisible,christensen2016parity,liu2018unidirectional,zhang2019non,
lopez2019multiple,wu2019asymmetric,hou2018tunable}. More recently, feedback control has been pursued to establish non-reciprocal interactions in a mechanical metamaterial that emulates the non-Hermitian Su-Schrieffer-Heeger (SSH) model~\cite{ghatak2019observation}. Such setting was used to experimentally demonstrate the existence of zero-frequency edge states in the non-Hermitian topological phase, and also to realize unidirectional wave amplification~\cite{brandenbourger2019non}. Motivated by these notable contributions, we here investigate a family of 1D and 2D elastic lattices with non-local, proportional feedback interactions and explore a series of unconventional phenomena stemming from their non-Hermiticity. Starting from a wave propagation perspective, we demonstrate that the frequency bands of 1D lattices are entirely non-reciprocal, due to the presence of gain and loss in opposite propagation directions. Such behavior is tunable based on the non-locality of the feedback interactions, which can be exploited to establish multiple frequency bands with interchanging non-reciprocal behavior. We also show that the bulk eigenmodes of finite lattices are localized at a boundary according to the non-Hermitian skin effect, and that their localization edge is well predicted by the winding number of the complex dispersion bands, which is aligned with recent findings on quantum lattices~\cite{gong2018topological}. Our analysis is then extended to 2D lattices where non-reciprocity manifests itself as a preferential direction for wave amplification, which is defined by the control interactions. We show that the non-local control in 2D lattices establishes multiple non-reciprocal frequency/wavenumber bands with different preferential directions of amplification. Finally, we investigate skin modes in finite lattice strips and show that their combined effect in two directions leads to bulk modes localized at the corners of finite 2D lattices. Our work provides fundamental perspectives on a new class of non-Hermitian elastic lattices with feedback interactions and contributes to recent efforts in exploring non-Hermiticiy for the design of metamaterials with novel functionalities~\cite{ghatak2019observation,brandenbourger2019non}. 

This paper is organized as follows: following this introduction, the analysis of wave propagation and topological properties of 1D lattices with feedback interactions is presented. Next, results are extended to 2D lattices where directional wave amplification and bulk corner modes are demonstrated. Finally, we summarize the main results of the work and outline future research directions.

\section{One-dimensional elastic lattices with feedback interactions}\label{sec2}

\begin{figure}[b!]
	\centering
			\includegraphics[height=0.2\textwidth]{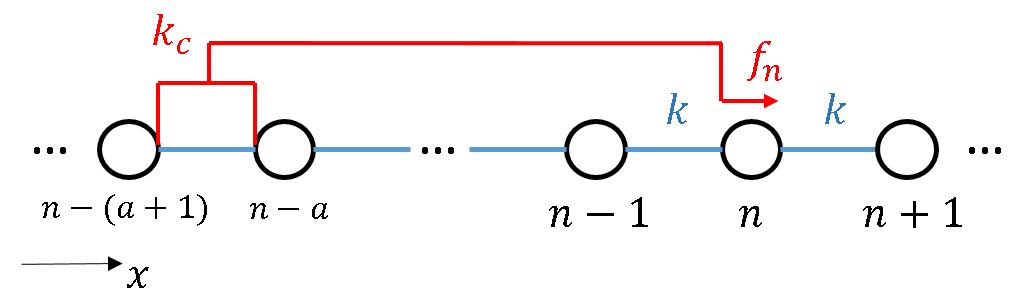}\label{Fig1a}
	\caption{One-dimensional lattice of equal masses $m$ connected by springs of stiffness $k$ with feedback control interactions. A force $f_n=k_c(u_{n-a}-u_{n-(a+1)})$ is applied to each mass along the lattice, corresponding to a reaction proportional to the strain of a spring $a$ units behind.}
	\label{Fig1}
\end{figure}

We consider 1D elastic lattices of equal masses $m$, separated by a unit distance, and connected by springs of equal stiffness $k$ (Fig.~\ref{Fig1}). Control interactions are introduced by considering an additional force, applied to the $n$-th mass, that reacts proportionally to the elongation of a spring at location $n-a$ ($a\in\mathcal{I}$). This force is expressed as $f_n=k_c(u_{n-a}-u_{n-(a+1)})$, where $k_c$ denotes the proportional control gain, and $u_n$ is the displacement of mass $n$ along the $x$ axis. In the absence of external forces, the equation governing the harmonic motion of mass $n$ is given by
\begin{equation}\label{goveq1d}
(2k-\omega^2 m) u_n -k(u_{n+1}+u_{n-1}) - k_c(u_{n-a}-u_{n-(a+1)})=0.
\end{equation}

The considered lattices are non-Hermitian since their dynamic stiffness matrix $\mathbf{D}=\mathbf{K}-\omega^2 \mathbf{M}$ is real but not symmetric, \textit{i.e.} $\mathbf{D}^T \neq \mathbf{D}$. These lattices are non-conservative systems where gain and loss are introduced by the feedback interactions, leading to intriguing properties discussed throughout this paper. Although active components would be required for a practical implementation, these systems can be mathematically treated in a linear and autonomous form (as in Eqn.~\eqref{goveq1d}), which motivates the investigations presented herein in terms of non-reciprocity and of topological properties of the bulk bands and their relation to the Non-Hermitian skin effect~\cite{yao2018edge,alvarez2018non,lee2019anatomy,longhi2019probing}. 

\subsection{Dispersion relations and non-reciprocity}

Wave propagation is investigated by imposing a Bloch-wave solution of the form $u_n=Ue^{i(\omega t - \mu n)}$, where $\omega$ and $\mu$ respectively denote angular frequency and non-dimensional wavenumber. Substitution in Eqn.~\eqref{goveq1d} yields the dispersion relation
\begin{equation}\label{disp1d}
\Omega^2 = 2(1-\cos\mu)-\gamma_c (1- e^{i\mu}) e^{i\mu a}
\end{equation}
where $\Omega=\omega/\omega_0$ is a normalized frequency, with $\omega_0=\sqrt{k/m}$, and $\gamma_c=k_c/k$. The feedback interaction makes the right-hand side of Eqn.~\eqref{disp1d} generally complex, which results in complex frequencies $\Omega=\Omega_r + i\Omega_i$ that come in pairs $\{ \Omega,-\Omega \}$. Without loss of generality we focus on the solution $\Omega$ with positive real part ($\Omega_r>0$), which corresponds to a wave $u_n=Ue^{i(\Omega_r \tau-\mu n)}e^{-\Omega_i \tau}$, ($\tau =t\omega_0$), that travels along the positive (negative) $x$ direction when $\mu$ is positive (negative), and that is exponentially attenuated (amplified) in time when $\Omega_i$ is positive (negative). 

\begin{figure}[b!]
	\centering
		\subfigure[]{
			\includegraphics[height=0.35\textwidth]{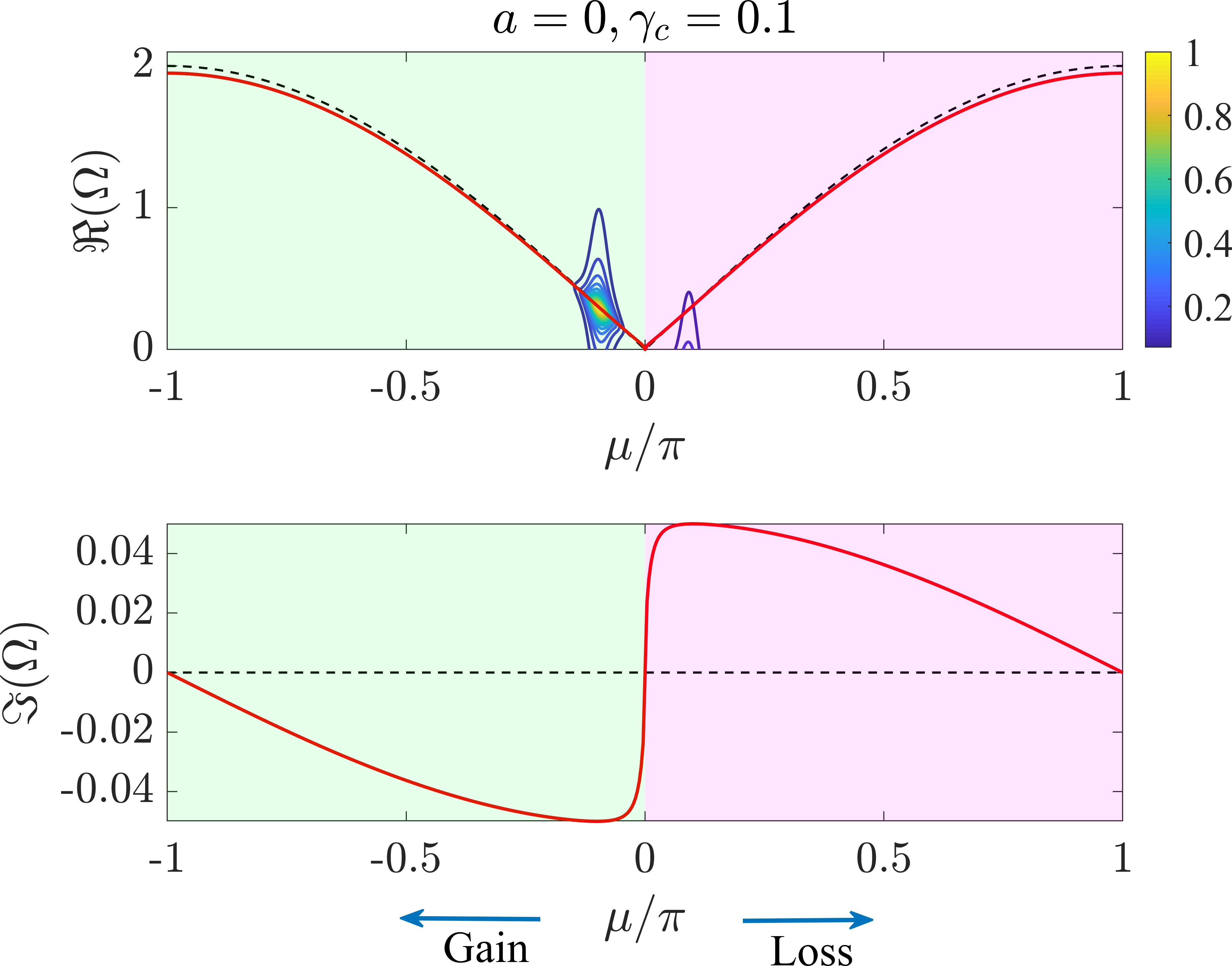}\label{Fig2a}}
		\subfigure[]{
			\includegraphics[height=0.35\textwidth]{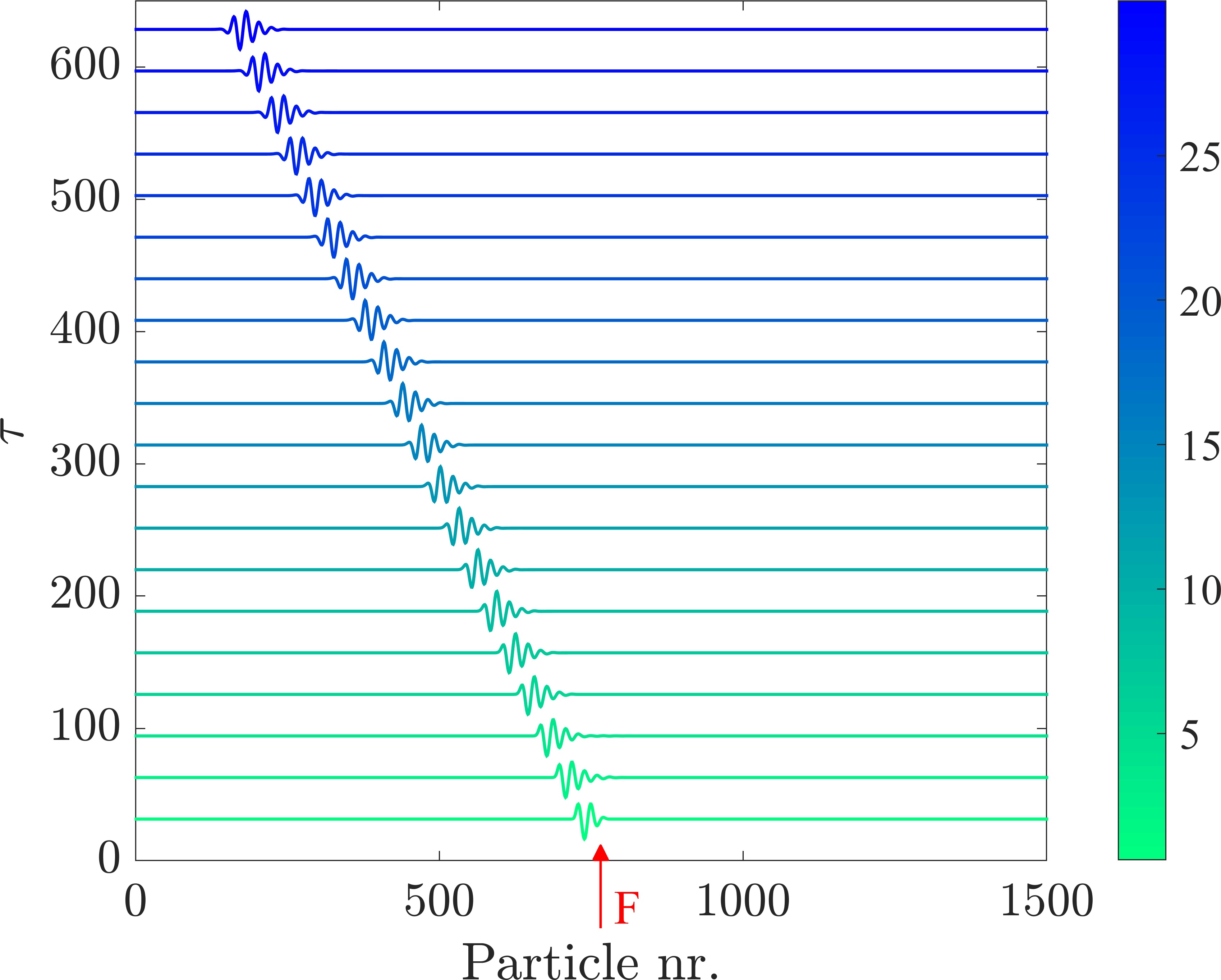}\label{Fig2b}}
		\subfigure[]{
			\includegraphics[height=0.35\textwidth]{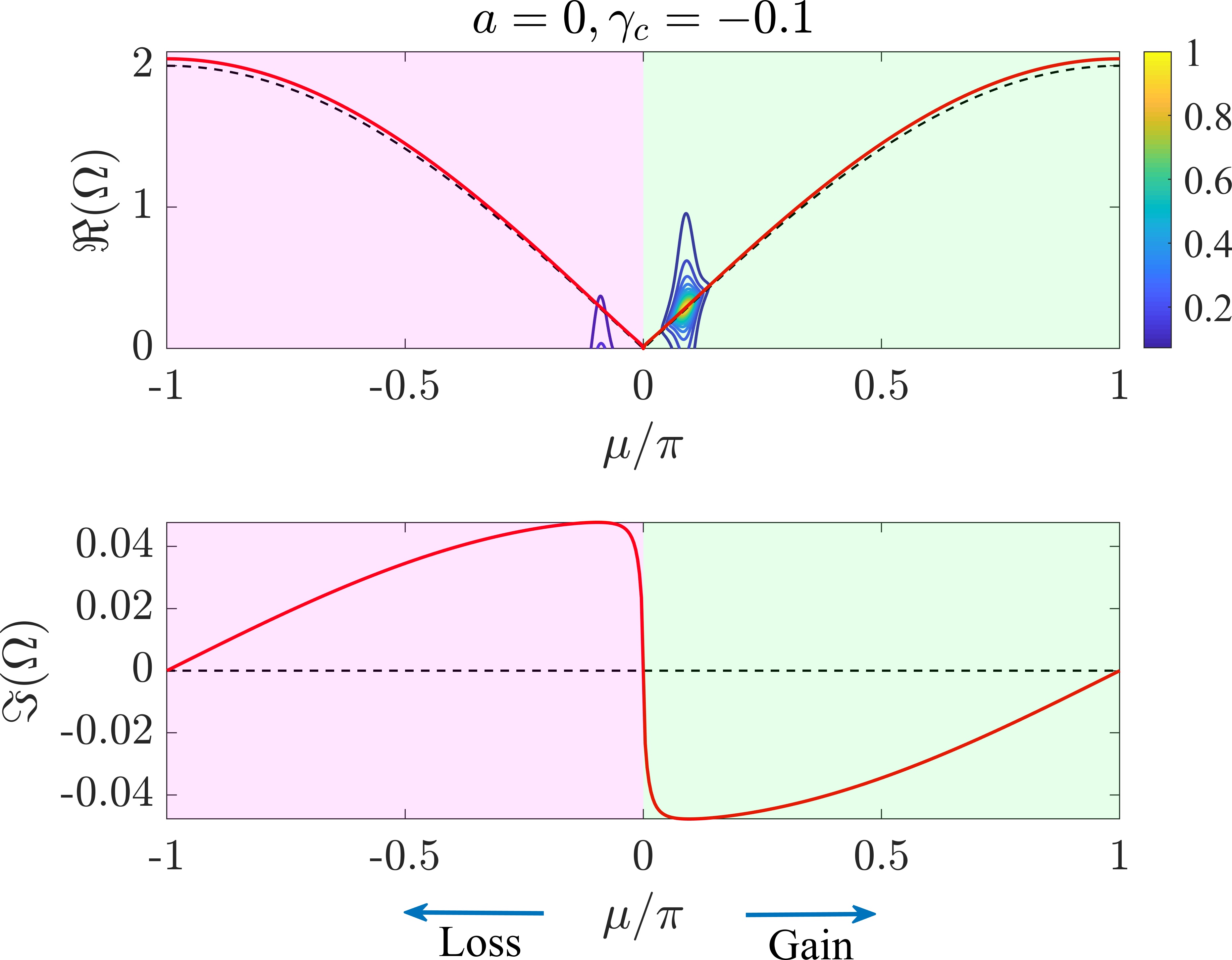}\label{Fig2c}}
		\subfigure[]{
			\includegraphics[height=0.35\textwidth]{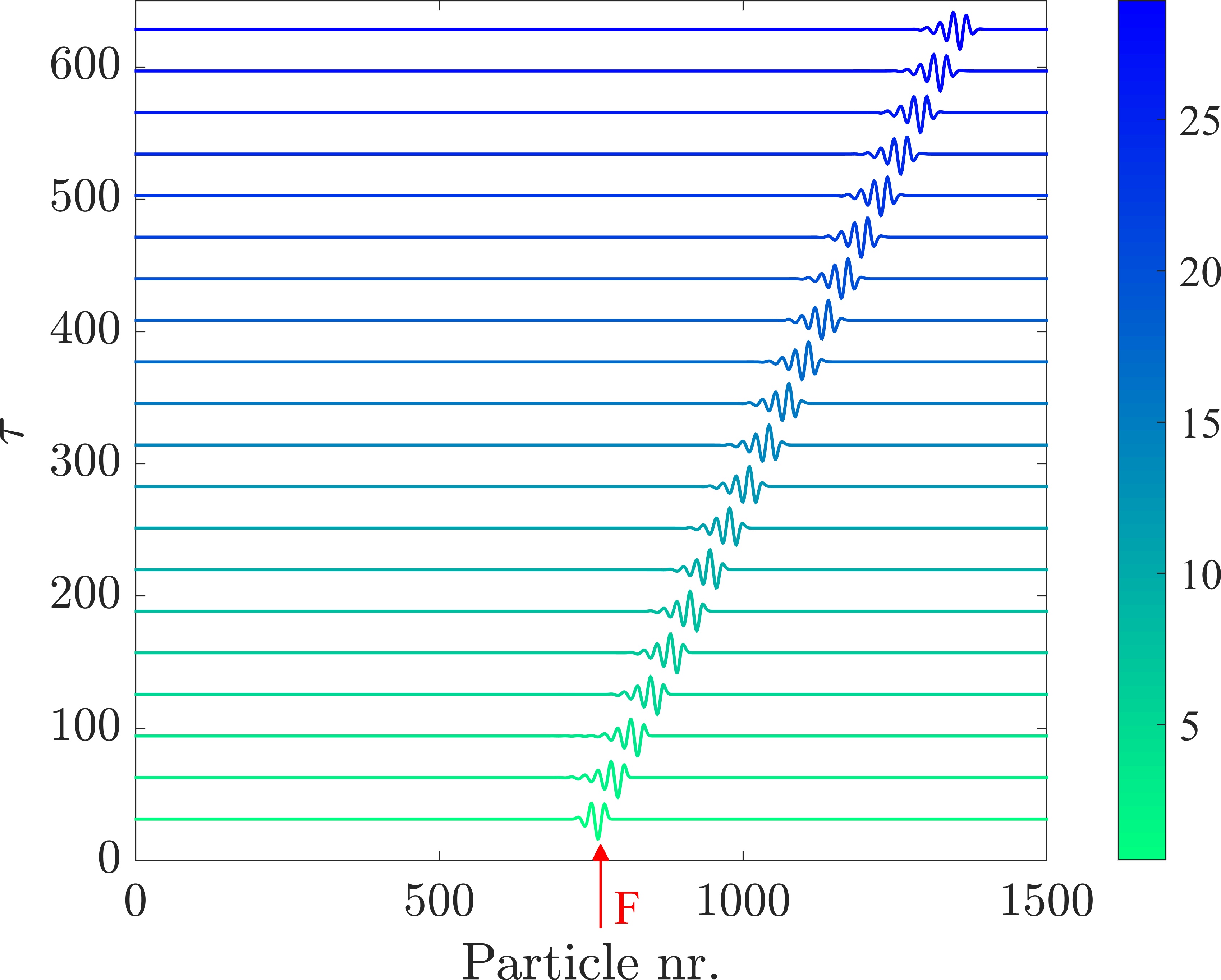}\label{Fig2d}}
		\subfigure[]{
			\includegraphics[width=0.45\textwidth]{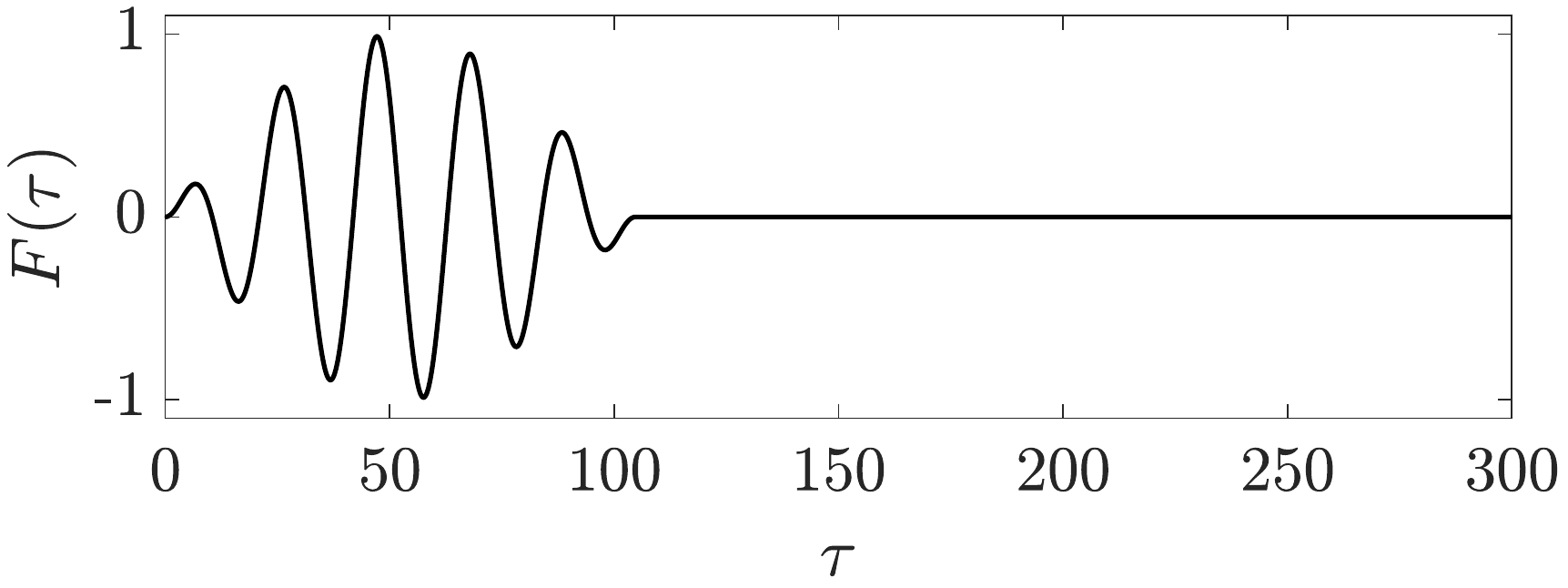}\label{Fig2e}}
		\subfigure[]{
			\includegraphics[width=0.45\textwidth]{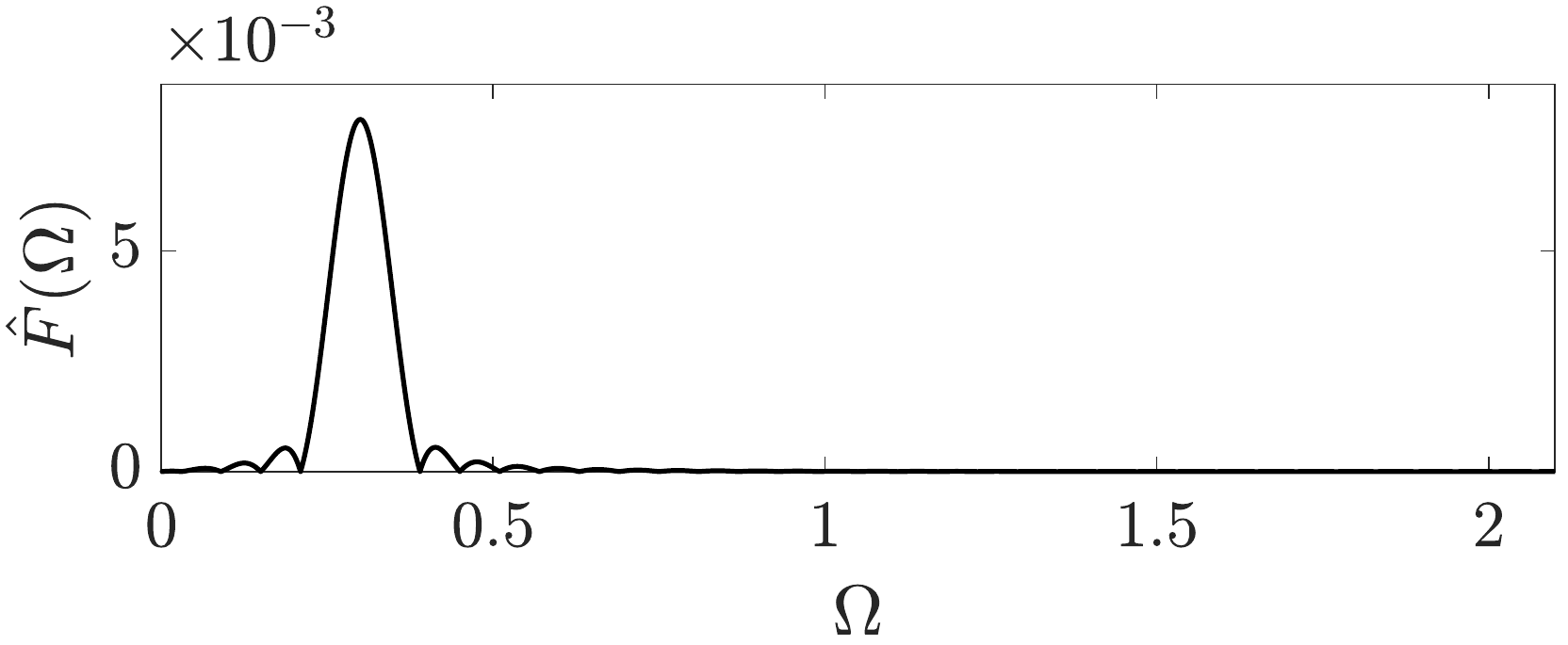}\label{Fig2f}}
	\caption{Non-reciprocal amplification and attenuation of waves in lattices with local feedback interactions ($a=0$). The dispersion $\Omega(\mu)$ for $\gamma_c=0.1$ and $\gamma_c=-0.1$ are respectively shown in (a) and (c) (solid red lines), superimposed to the dispersion of the passive lattice with $\gamma_c=0$ (dashed black lines). Attenuation and amplification zones are identified by shaded pink and green areas revealing non-reciprocal behavior: the lattice with $\gamma_c=0.1$ amplifies waves traveling to the left and attenuates waves traveling to the right, while $\gamma_c=-0.1$ results in a opposite behavior. Transient simulation results reported as waterfall plots in (b) and (d) illustrate the non-reciprocal behavior, which is further confirmed by their dispersion estimated through FT operations (contours in (a,c)). The force applied to mass $n=750$ is displayed in the time and in the frequency domains in (e) and (f), respectively.}
	\label{Fig2}
\end{figure}

We first investigate the case of local control ($a=0$), \emph{i.e.} with the feedback force proportional to the elongation of the left adjacent spring. Figure~\ref{Fig2a} displays the dispersion for $\gamma_c=0.1$ (solid red lines), superimposed to the dispersion $\Omega=\sqrt{2(1-\cos\mu)}$ of a lattice with no feedback interactions $\gamma_c=0$ (dashed black lines). A remarkable feature of the dispersion lies in its imaginary component: positive wavenumbers are associated with loss due to positive $\Omega_i$ values (represented by shaded pink areas), while negative wavenumbers are associated with gain due to negative $\Omega_i$ values (shaded green areas). Therefore, the lattice with $\gamma_c=0.1$ amplifies waves traveling to the left and attenuates waves traveling to the right, while an opposite behavior is observed for $\gamma_c=-0.1$ (Fig.~\ref{Fig2c}). The non-reciprocity associated with gain and loss is confirmed by time domain simulations, where a 5-cycle sine burst of center frequency $\Omega=0.3$ (Figs.\ref{Fig2}(e,f)) is applied to the center mass of a chain of $N=1500$ masses. The resulting transient responses evaluated by numerical integration are displayed in Figs.~\ref{Fig2}(b,d) in the form of waterfall plots. For visualization purposes, the displacement along the lattice for each time instant is normalized by the instantaneous $L_\infty$ norm (along $x$), which is employed in the associated log-scale colormap. A wave packet is amplified as it propagates to the left for $\gamma_c=0.1$, and to the right for $\gamma_c=-0.1$. The frequency/wavenumber content of the wave packets is evaluated by computing the two-dimensional Fourier Transform (FT) to recover the displacement in reciprocal space $\hat{u}(\omega,\mu)$. The results displayed as contours plots are superimposed to the theoretical dispersion curves in Figs.~\ref{Fig2}(a,c), to confirm the expected non-reciprocal behavior highlighted by the concentration of the spectral content of the transients in the gain (green) portions of the reciprocal space.

\begin{figure}[t!]
	\centering
		\subfigure[]{
			\includegraphics[height=0.35\textwidth]{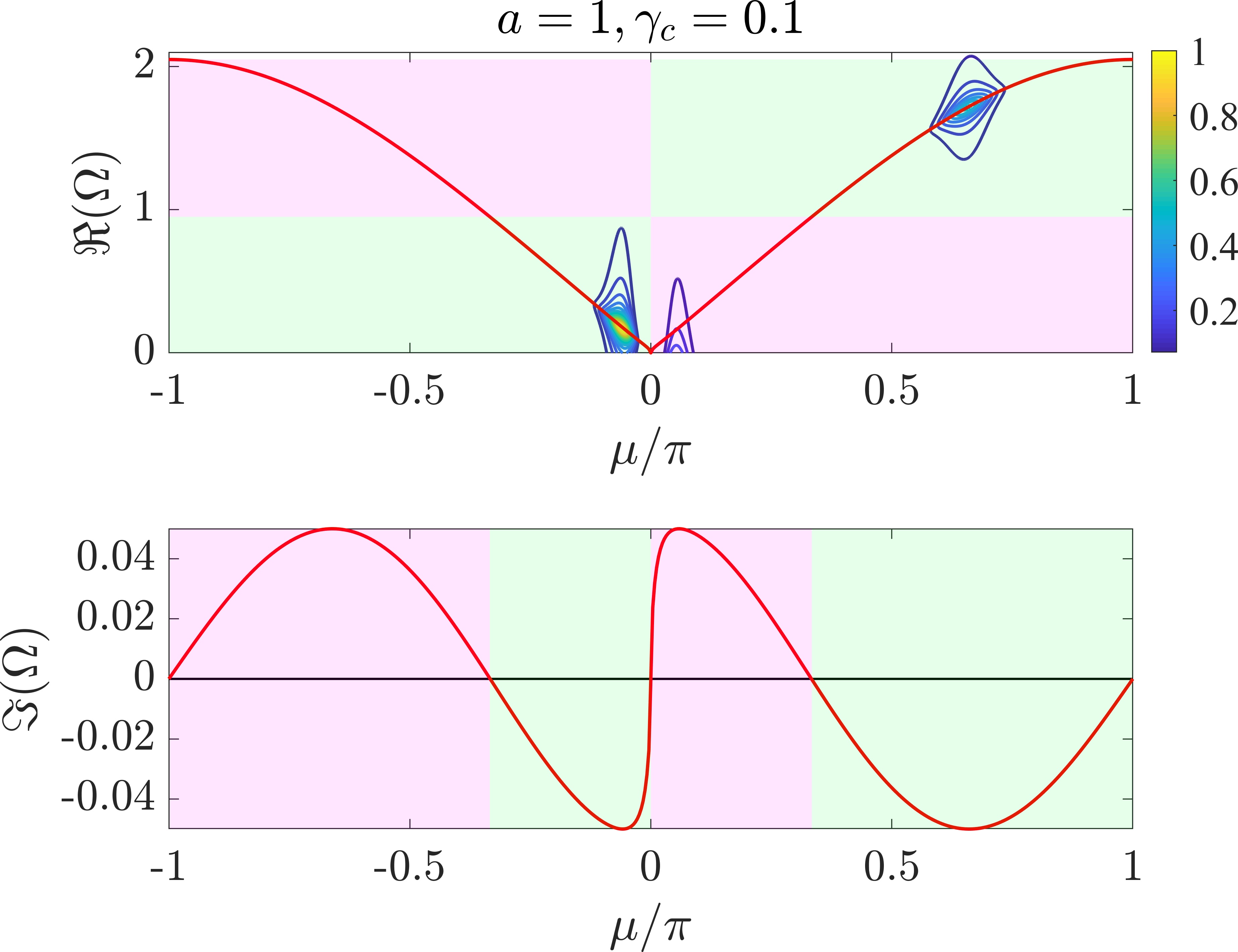}\label{Fig3a}}
		\subfigure[]{
			\includegraphics[height=0.35\textwidth]{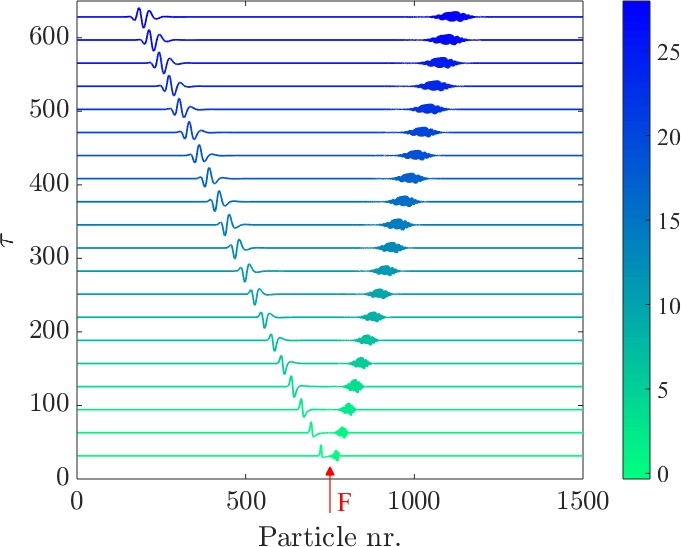}\label{Fig3b}}
			\subfigure[]{
			\includegraphics[height=0.35\textwidth]{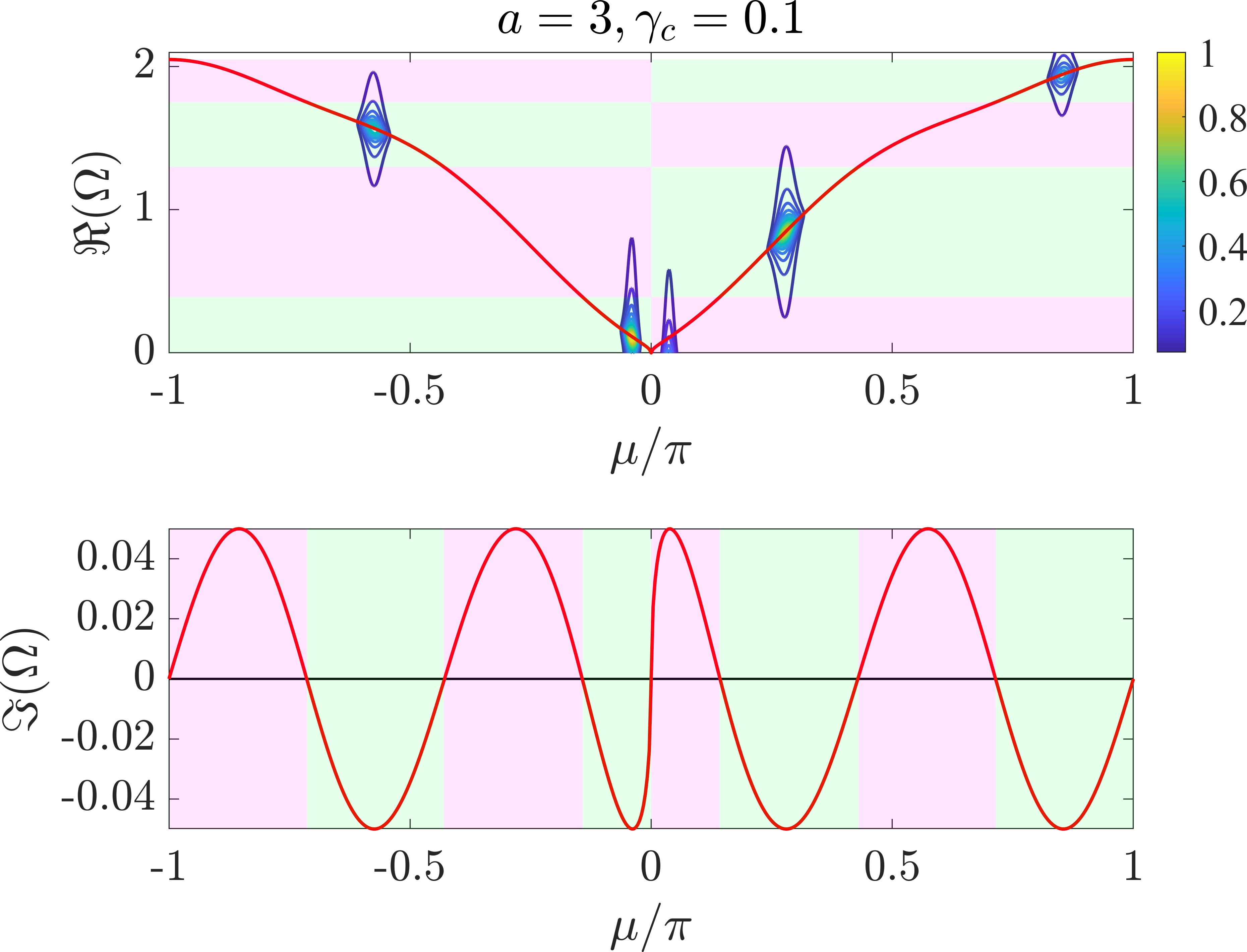}\label{Fig3c}}
		\subfigure[]{
			\includegraphics[height=0.35\textwidth]{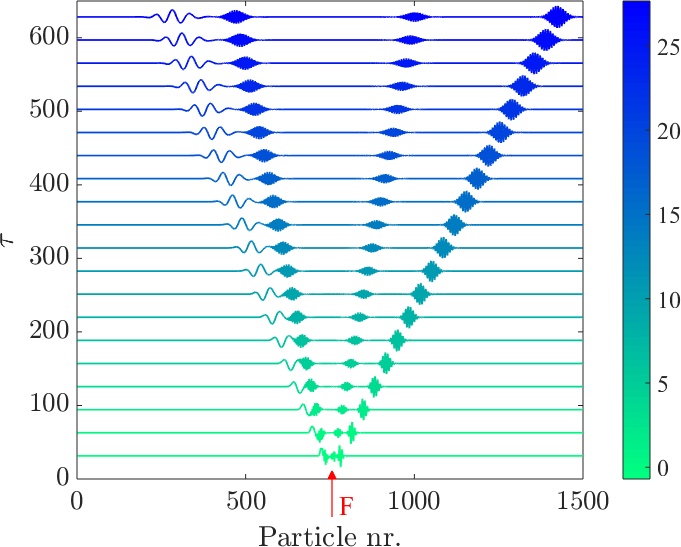}\label{Fig3d}}
			\subfigure[]{
			\includegraphics[width=0.45\textwidth]{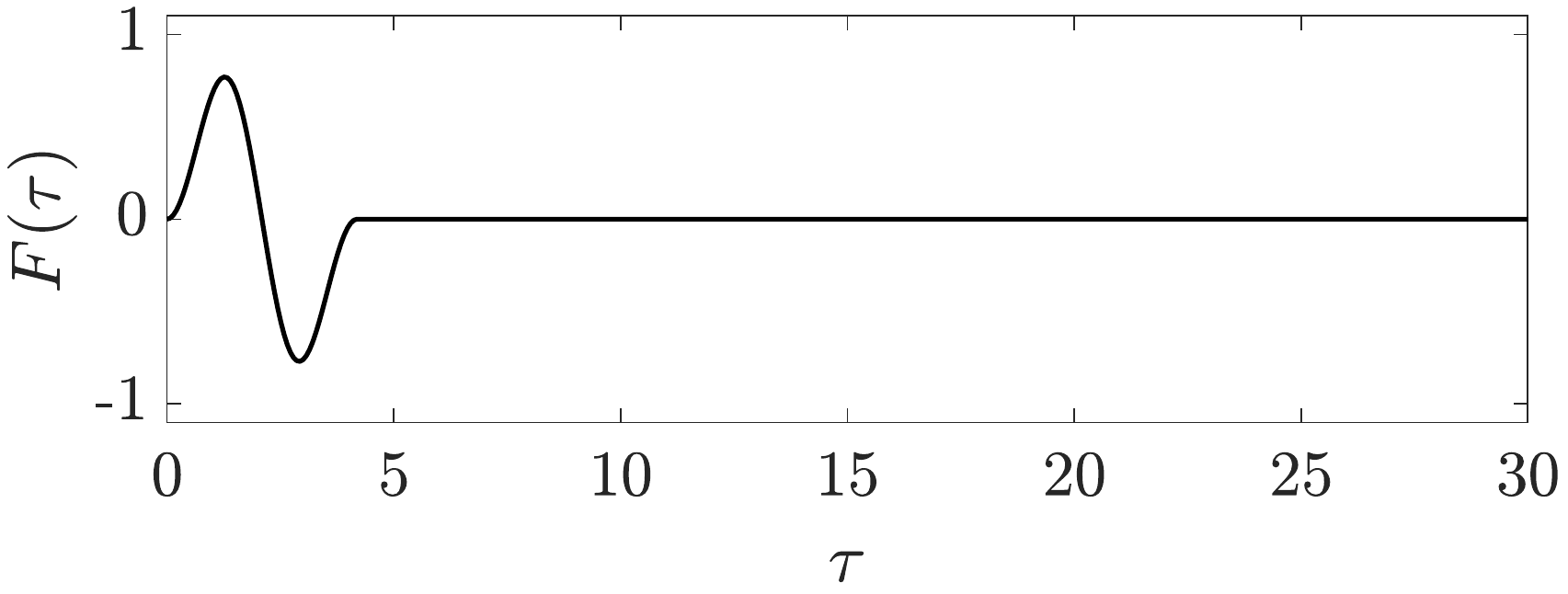}\label{Fig3e}}
		\subfigure[]{
			\includegraphics[width=0.45\textwidth]{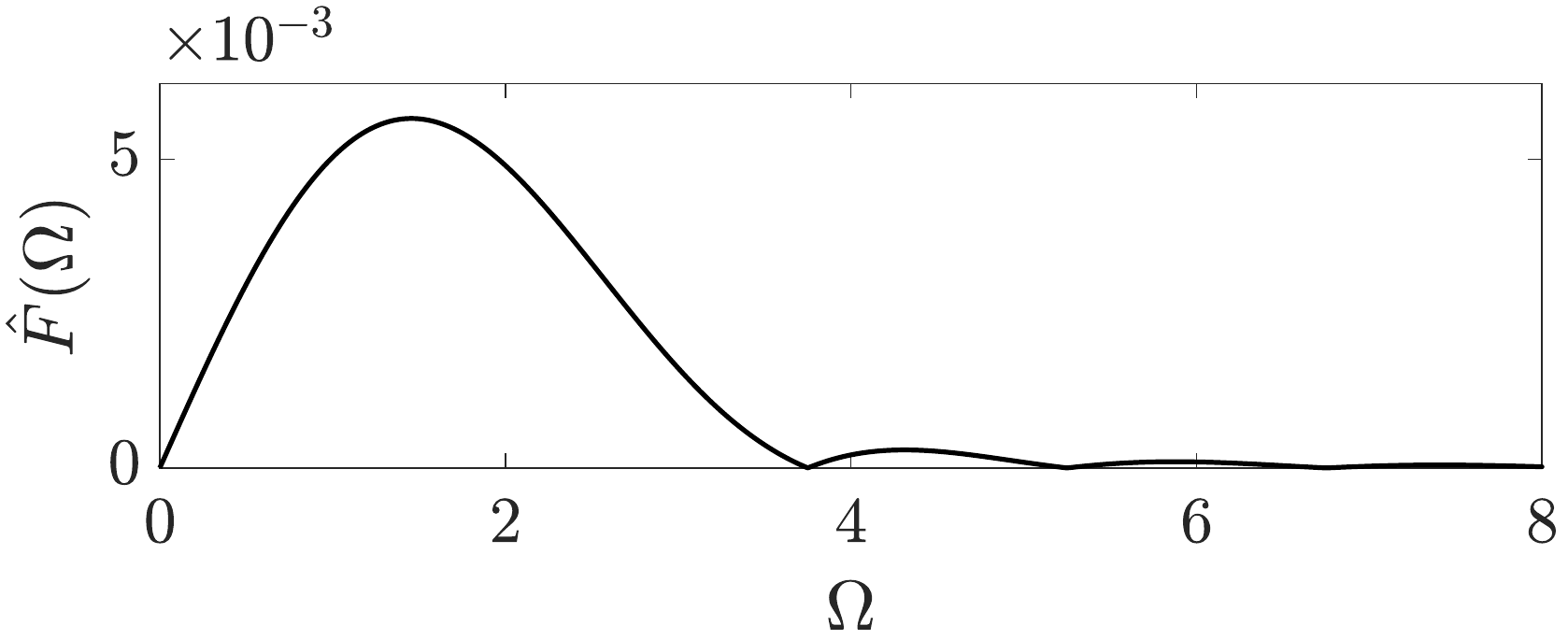}\label{Fig3f}}
	\caption{Non-reciprocal amplification and attenuation of waves in lattices with non-local feedback interactions ($a>0$). The dispersion $\Omega(\mu)$ for $a=1$ and $a=3$ are displayed in (a) and (c), respectively. The non-local feedback interactions result in multiple frequency bands with non-reciprocity in amplification and attenuation (shaded green and pink areas). Transient time domain simulations to a broad-band input force (e,f) illustrate the non-reciprocal behavior: one wave packet is amplified and propagates along each direction in (b), while two wave packets are amplified and propagate along each direction in (d). Their dispersion (contours in (a,c)) estimated through FT operations are in good agreement with the non-reciprocal bands. The broadband input force applied to mass $n=750$ is displayed in the time and in the frequency domains in (e) and (f), respectively.}
	\label{Fig3}
\end{figure}

Next, we investigate the role of non-local interactions defined by $a>0$ values. The dispersion for $a=1$ and $\gamma_c=0.1$ (Fig.~\ref{Fig3a}) features an imaginary frequency curve with two different regions of gain or loss for each propagation direction. The behavior is entirely non-reciprocal: positive and negative wavenumbers with the same absolute value correspond to attenuation along one direction, and amplification along the other, as highlighted by the shaded green and pink regions. In fact, one can verify in Eqn.~\eqref{disp1d} that $\Omega_r^2(\mu)=\Omega_r^2(-\mu)$ and $\Omega_i^2(\mu)=-\Omega_i^2(-\mu)$. By using basic properties related to the square roots of a complex number (not described here for brevity), one can confirm reciprocity for the real part of the dispersion ($\Omega_r(\mu)=\Omega_r(-\mu)$), and non-reciprocity for the imaginary part ($\Omega_i(\mu)=-\Omega_i(-\mu)$). Due to this property, the amplification and attenuation wavenumber ranges defined by the imaginary part of the dispersion can be translated to the real frequency dispersion curves by matching the corresponding wavenumber intervals (Fig.~\ref{Fig3a}). The procedure highlights two non-reciprocal frequency bands; the first amplifies waves traveling to the left, while the latter amplifies waves traveling to the right. In general, when considering higher $a$ values the number of non-reciprocal bands increases, usually being equal to $a+1$. For example, the dispersion for $a=3, \gamma_c=0.1$ displayed in Fig.~\ref{Fig3c} exhibits a total of four non-reciprocal frequency bands, as highlighted by shaded green and pink regions.

The non-reciprocal behavior of lattices with non-local feedback interactions is confirmed by transient time domain simulations, whose results are displayed in Figs.~\ref{Fig3}(b,d). To observe the behavior in the entire band, a broad-band input signal (Figs.~\ref{Fig3}(e,f)) is applied to the center mass of the lattice. The results show the de-multiplexing of the input signal, resulting from the amplification and propagation of one wave packet along each direction for $a=1$ (Fig.~\ref{Fig3b}), and of two wave packets along each direction for $a=3$ (Fig.~\ref{Fig3d}). The corresponding 2D-FTs are superimposed to the dispersion curves in Figs.~\ref{Fig3}(a,c), confirming the predicted amplification bands. We also note that the amplification of the wave packets is intensified around wavenumbers associated with local minima of $\Omega_i$, corresponding to the largest time amplification exponents. 


\subsection{Bulk topology and non-Hermitian skin effect}

\begin{figure}[b!]
	\centering
		\subfigure[]{
			\includegraphics[height=0.31\textwidth]{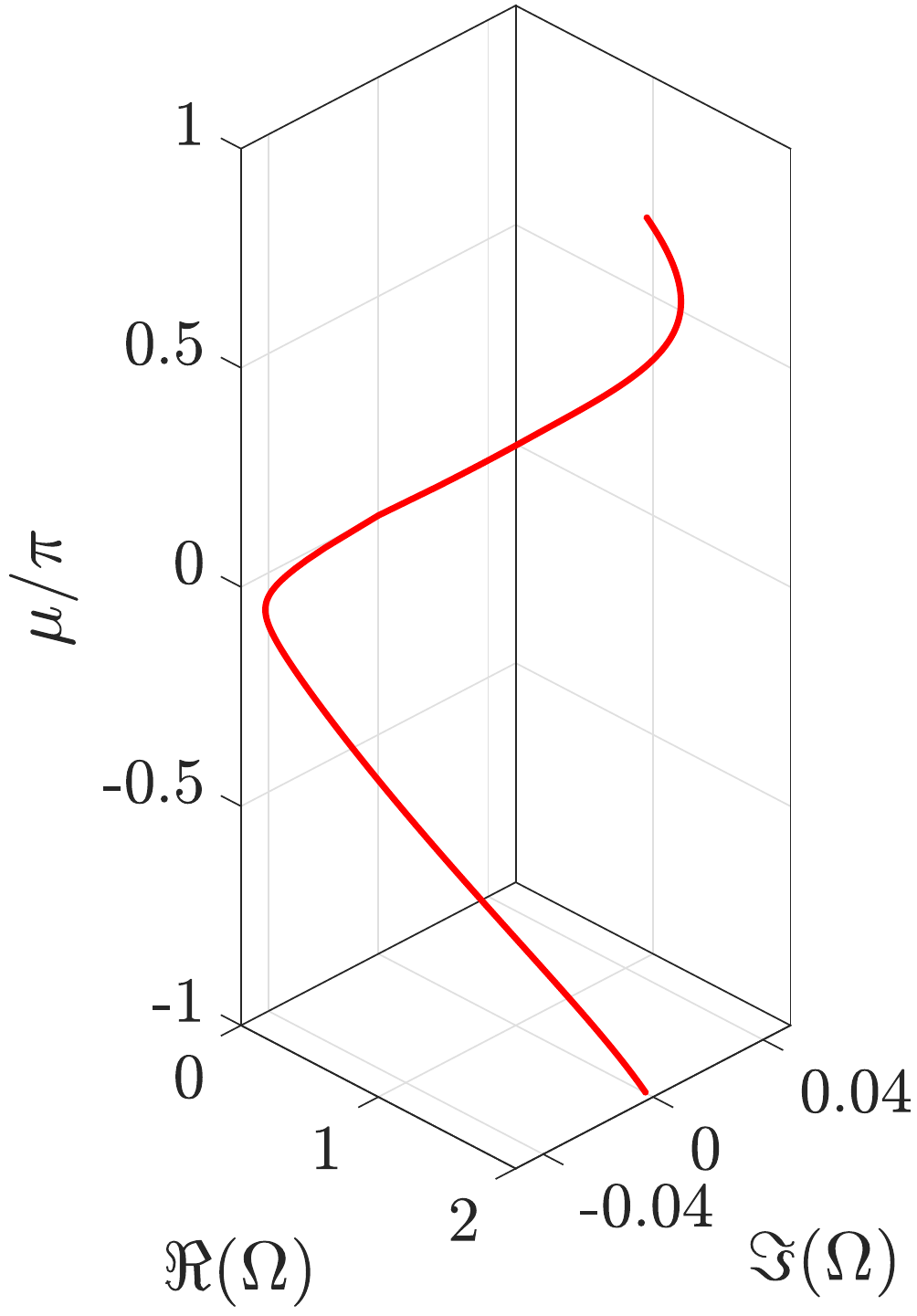}\label{Fig4a}}
		\subfigure[]{
			\includegraphics[height=0.31\textwidth]{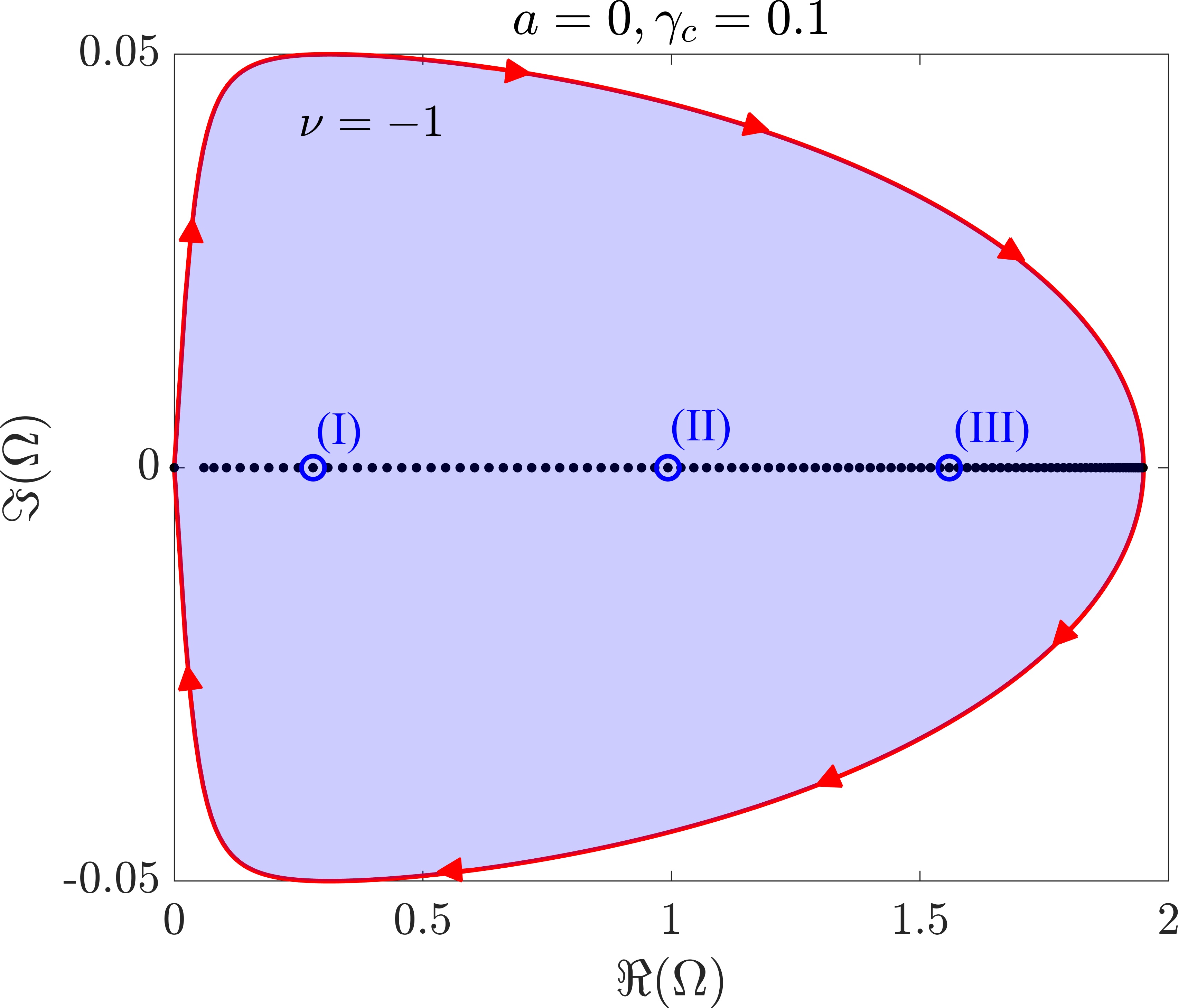}\label{Fig4b}}
		\subfigure[]{
			\includegraphics[height=0.31\textwidth]{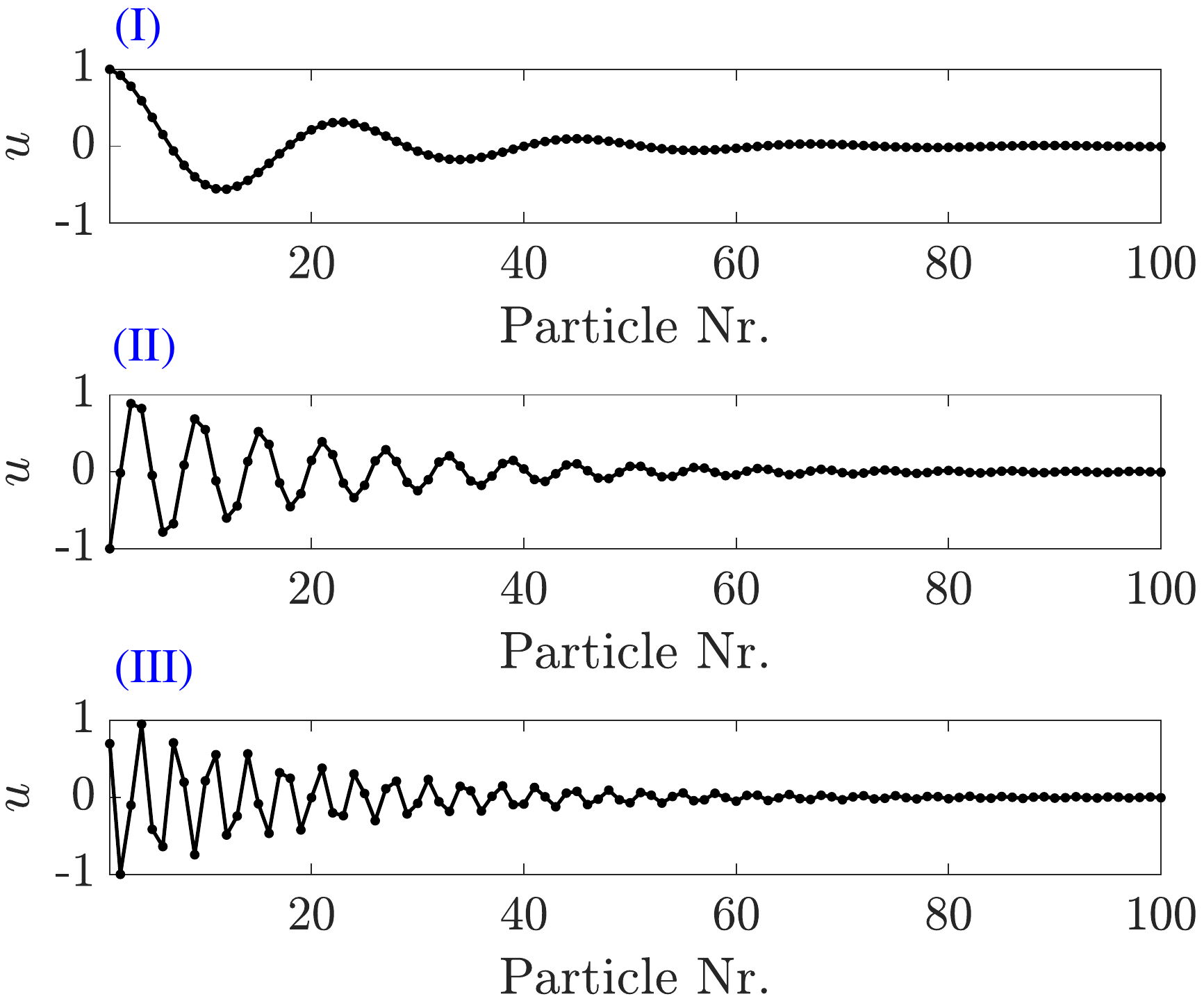}\label{Fig4c}}
			\subfigure[]{
			\includegraphics[height=0.31\textwidth]{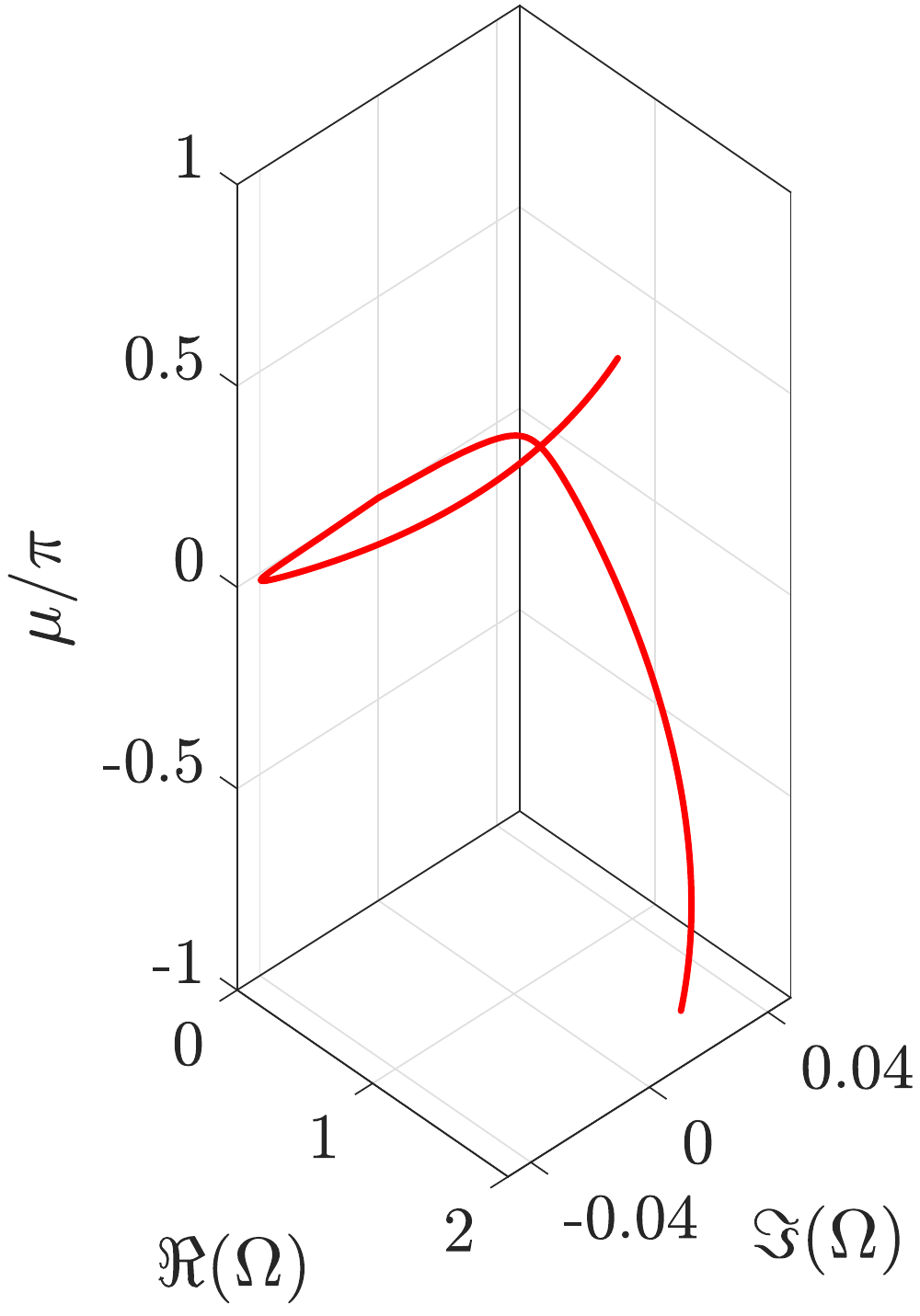}\label{Fig4d}}
		\subfigure[]{
			\includegraphics[height=0.31\textwidth]{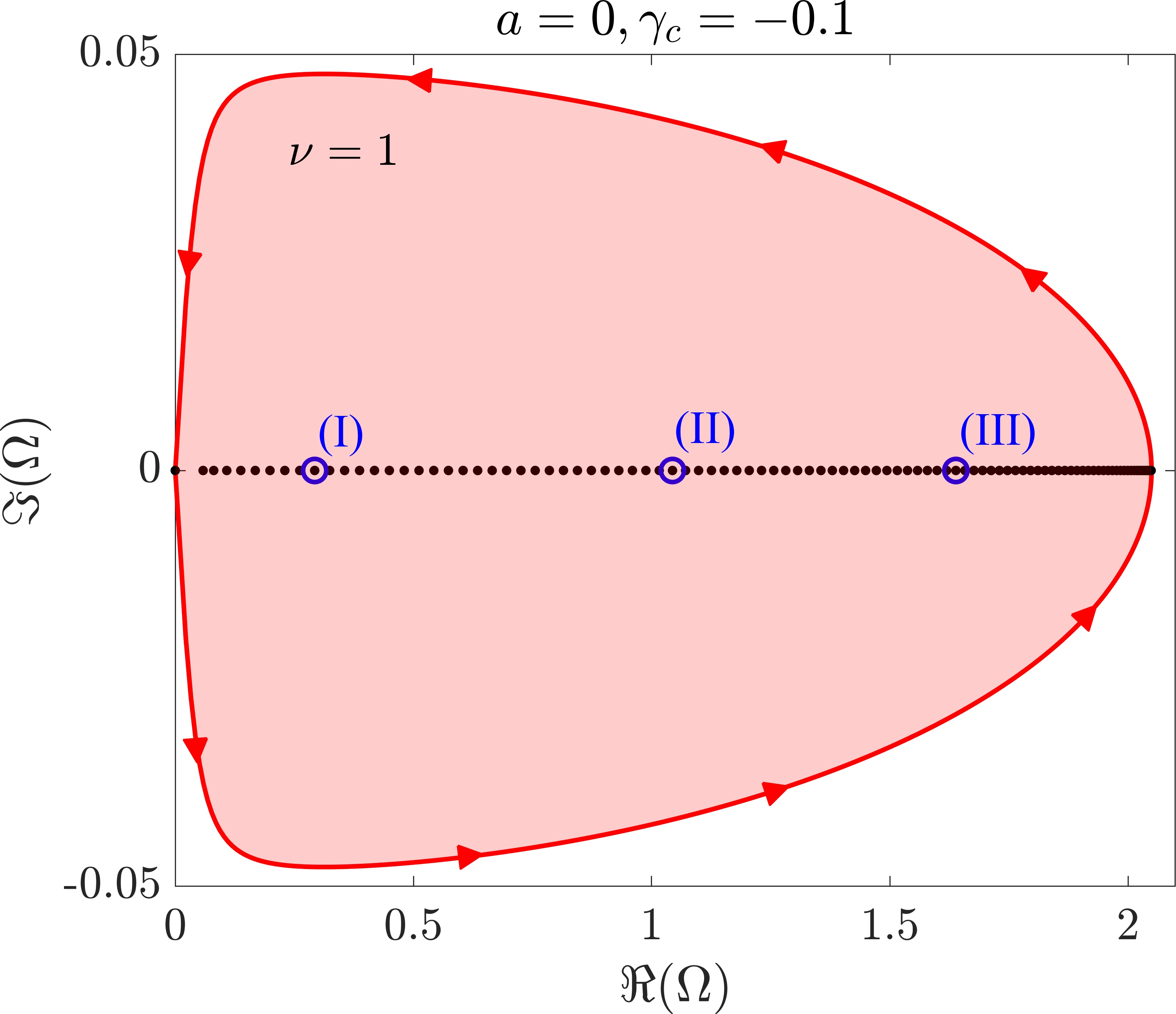}\label{Fig4e}}
		\subfigure[]{
			\includegraphics[height=0.31\textwidth]{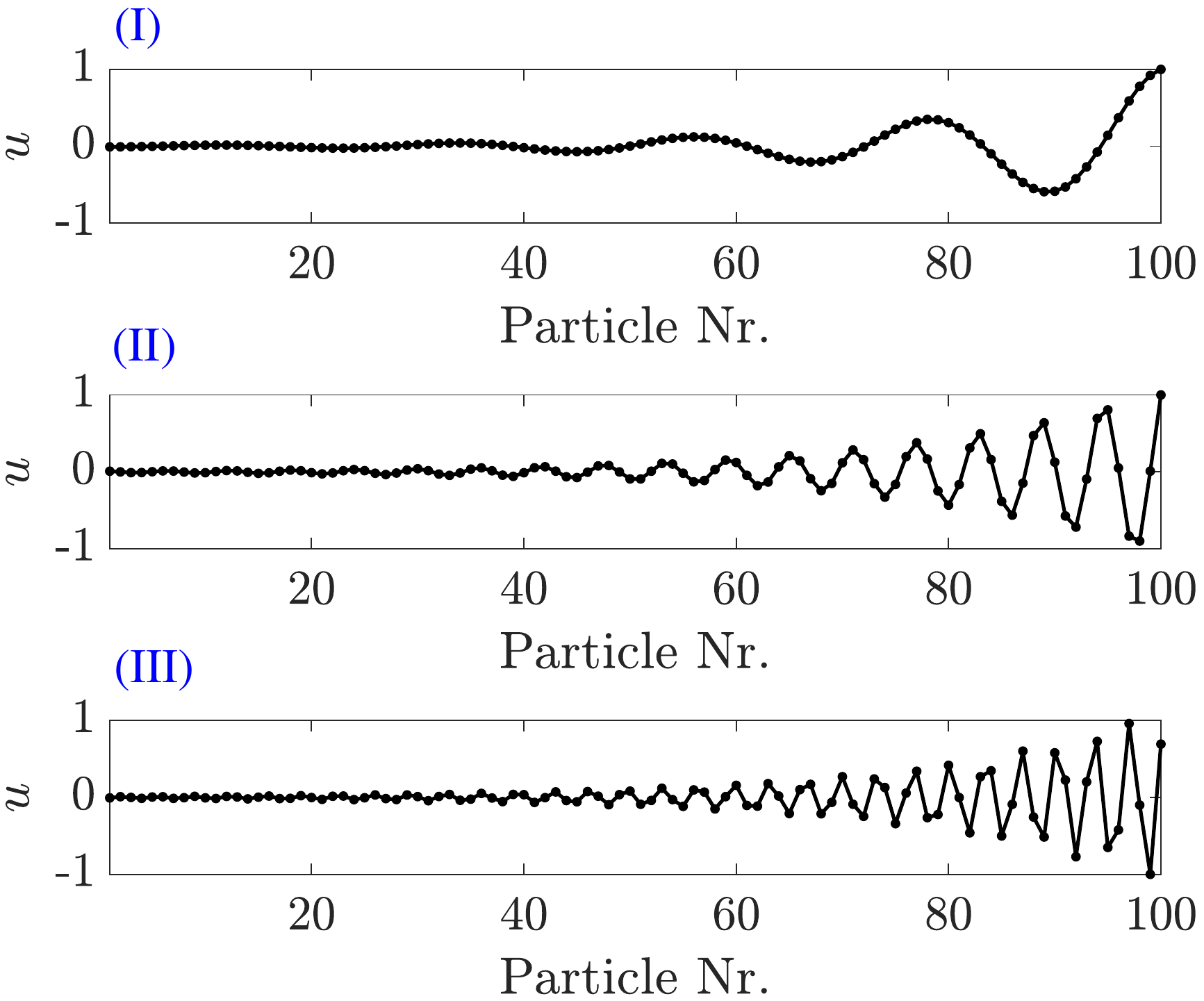}\label{Fig4f}}
	\caption{Dispersion topology and non-Hermitian skin effect in lattices with local feedback interactions ($a=0$). The complex dispersion bands $\Omega(\mu)$ for lattices with $\gamma_c=0.1$ and $\gamma_c=-0.1$ are displayed in (a) and (d), and their projection on the complex plane define closed loops as displayed in (b) and (e). Shaded blue and red areas represent regions with winding number $\nu=-1$ and $\nu=1$, respectively. Bulk modes of a finite lattice with $N=100$ masses whose eigenfrequencies (black dots) lie inside regions with $\nu=-1$ or $\nu=1$ are respectively localized at the left (c) or right (f) boundaries.}
	\label{Fig4}
\end{figure}

We now discuss the topological properties of the non-Hermitian lattices associated with the complex dispersion bands and how they are related to bulk modes localized at the boundaries of finite lattices. Starting with local feedback interactions ($a=0$), Figs.~\ref{Fig4}(a,d) display the complex representation of the dispersion for $\gamma_c=0.1$ and $\gamma_c=-0.1$, respectively, where both real and imaginary frequency components are plotted against the wavenumber $\mu$. The projections of the dispersion bands on the complex plane in Figs.~\ref{Fig4}(b,e) reveal closed loops (red lines) parameterized by $\mu$, with arrows denoting the direction of increasing $\mu$. As recently demonstrated in~\cite{gong2018topological}, the winding number of the loops define a topological invariant associated with the localization of bulk modes for finite lattices. The winding number of a dispersion band $\Omega(\mu)$ is given by~\cite{ahlfers1966complex}
\begin{equation}\label{eqwind}
\nu = \frac{1}{2\pi i}\int_{-\pi}^{\pi} \dfrac{\Omega'}{\Omega-\Omega_b} d\mu,
\end{equation}
where $\Omega'=\partial\Omega/\partial\mu$, and the base frequency $\Omega_b$ is an arbitrary point in the complex plane not belonging to the dispersion band~\cite{gong2018topological}, \textit{i.e.} $\Omega_b \neq \Omega(\mu)$. Geometrically, the winding number counts the number of times the dispersion loop encircles the base frequency, being positive for counterclockwise rotations. In the dispersion of Figs.~\ref{Fig4}(b,e), shaded blue and red areas denote regions for which any point has a winding number of $\nu=-1$ or $\nu=1$, respectively. In addition to simple observation of the winding numbers via geometrical interpretation, their values are confirmed by numerical integration of Eqn.~\eqref{eqwind} for a given point inside the loop, and by using the property that points inside a simply connected region have the same winding number~\cite{ahlfers1966complex}, which clarifies the arbitrary nature of the base frequency $\Omega_b$. Points outside the dispersion loops are trivially associated with a zero winding number $\nu=0$.

Hence, the feedback control interactions define distinct phases characterized by winding numbers which exhibit opposite behaviors for lattices with $\gamma_c=0.1$ and $\gamma_c=-0.1$. These behaviors manifest as localized bulk eigenmodes in finite lattices, a \textit{phenomenon} known as the non-Hermitian skin-effect~\cite{yao2018edge,alvarez2018non,lee2019anatomy,longhi2019probing}. As an illustration, the eigenfrequencies of a finite lattice with $N=100$ masses under free-free boundary conditions are displayed as black dots in Figs.~\ref{Fig4}(b,e), while representative eigenmodes marked by the blue circles are displayed in Figs.~\ref{Fig4}(c,f). Aligned with recent findings in quantum lattices~\cite{gong2018topological}, our results show that eigenfrequencies belonging to regions with $\nu<0$ define bulk modes localized at the left boundary (Fig.~\ref{Fig4c}), while $\nu>0$ values produce localization at the right boundary (Fig.~\ref{Fig4f}). This behavior is also in agreement with the non-reciprocal wave properties reported in Fig.~\ref{Fig2}: the phase with $\nu=-1$ is related to waves amplified to the left and attenuated to the right, hence the modes of a finite lattice are localized at the left boundary, while the opposite holds true for $\nu=1$.

\begin{figure}[b!]
	\centering
		\subfigure[]{
			\includegraphics[height=0.31\textwidth]{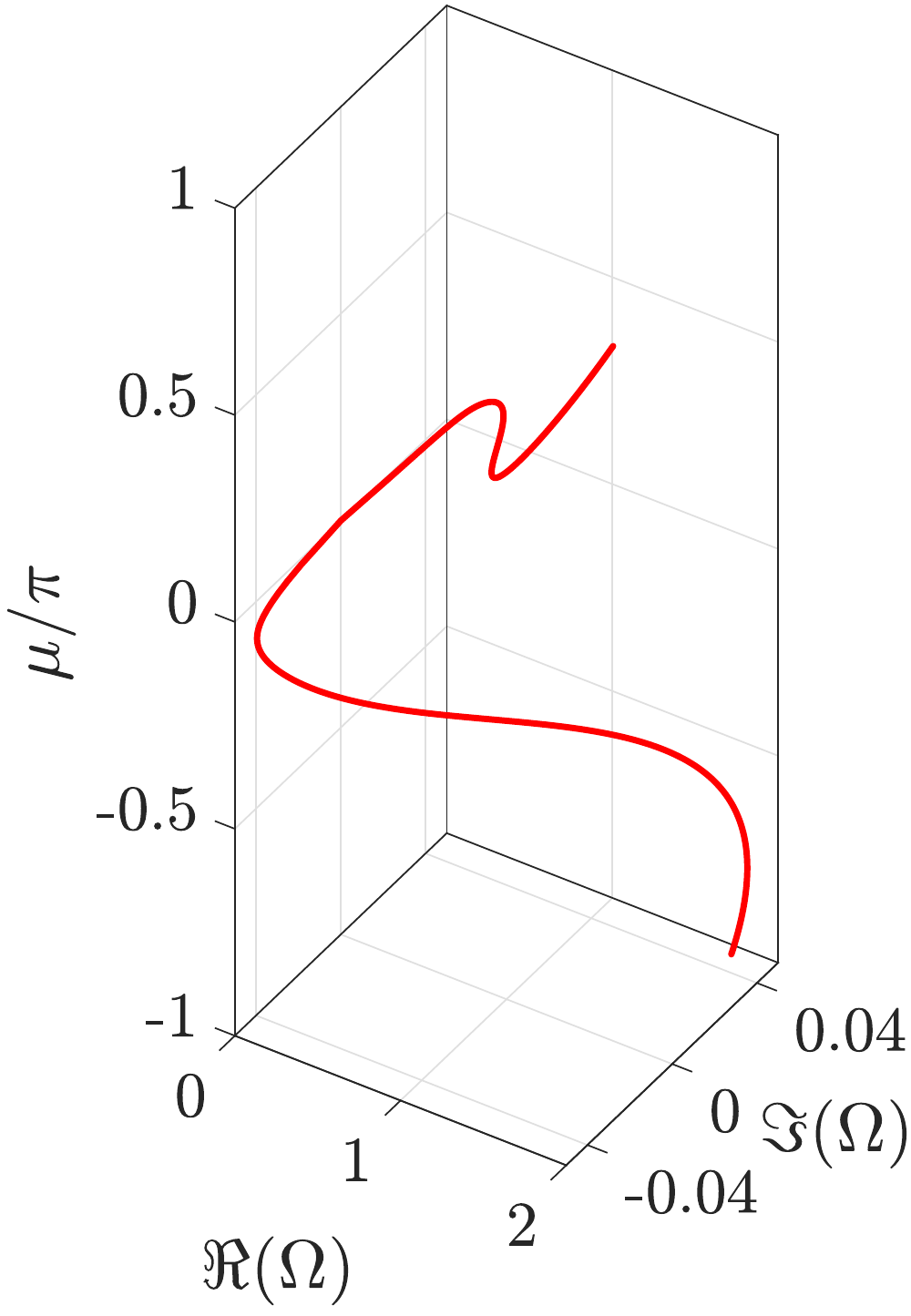}\label{Fig5a}}
		\subfigure[]{
			\includegraphics[height=0.31\textwidth]{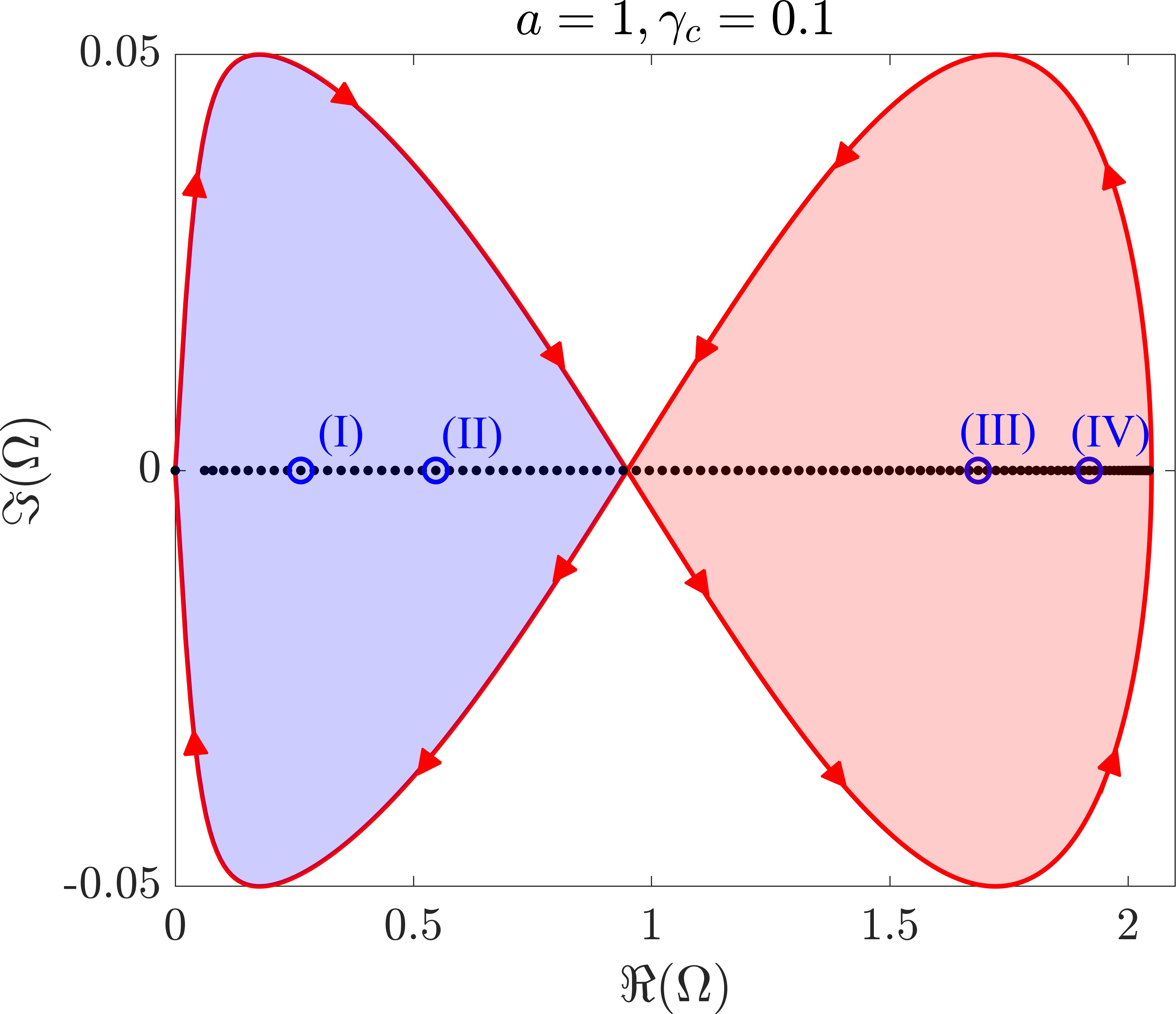}\label{Fig5b}}
		\subfigure[]{
			\includegraphics[height=0.31\textwidth]{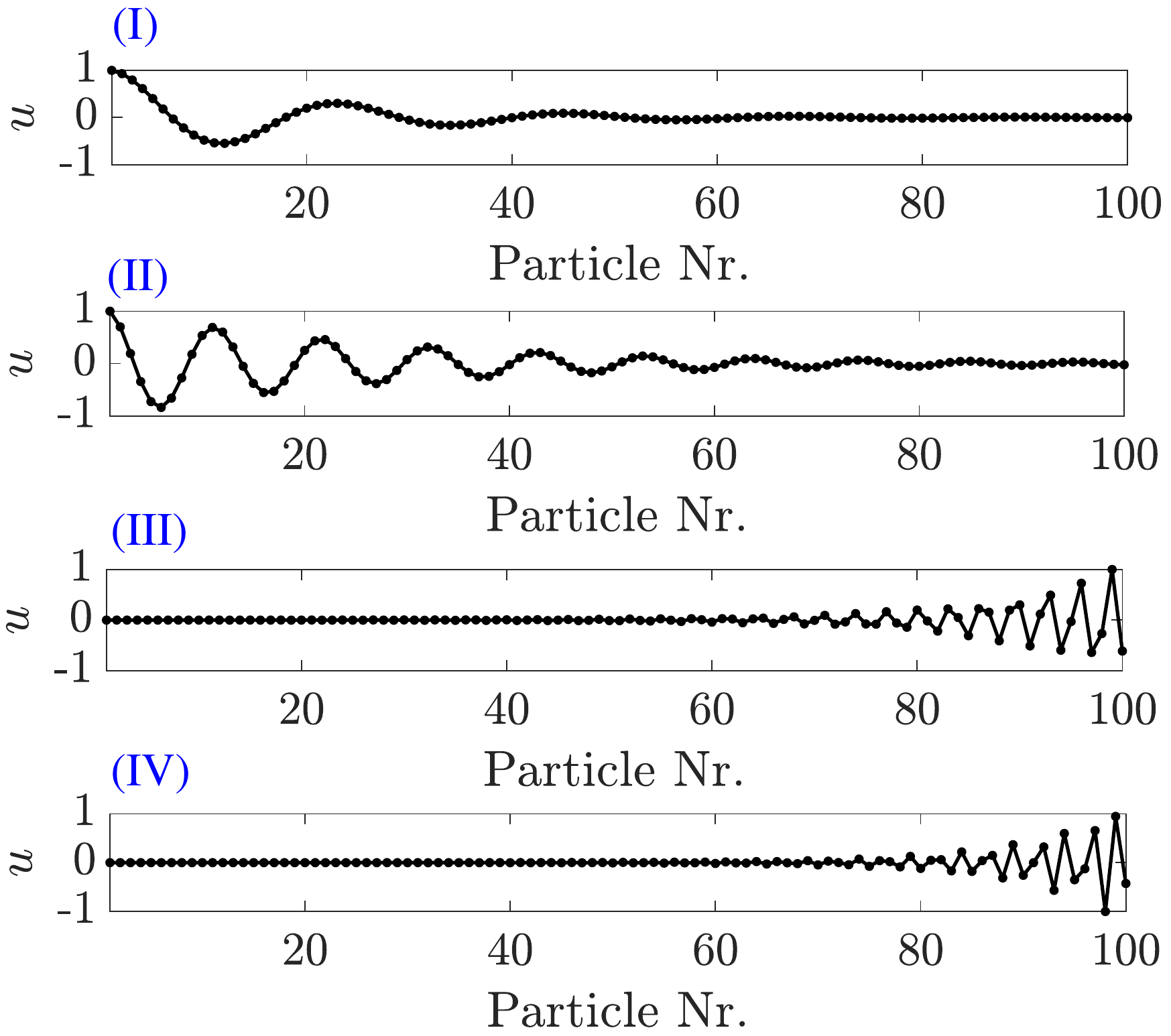}\label{Fig5c}}
			\subfigure[]{
			\includegraphics[height=0.31\textwidth]{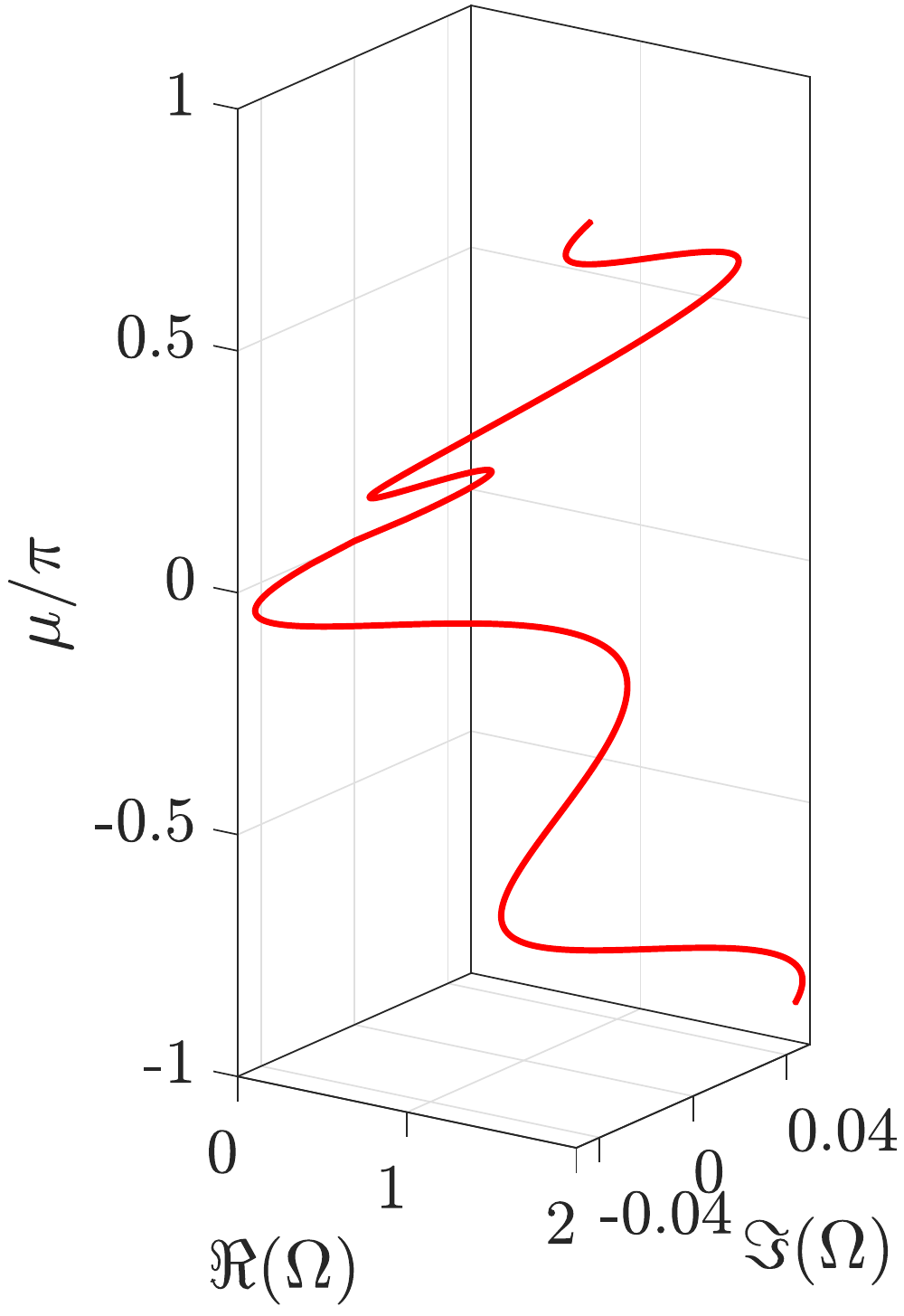}\label{Fig5d}}
		\subfigure[]{
			\includegraphics[height=0.31\textwidth]{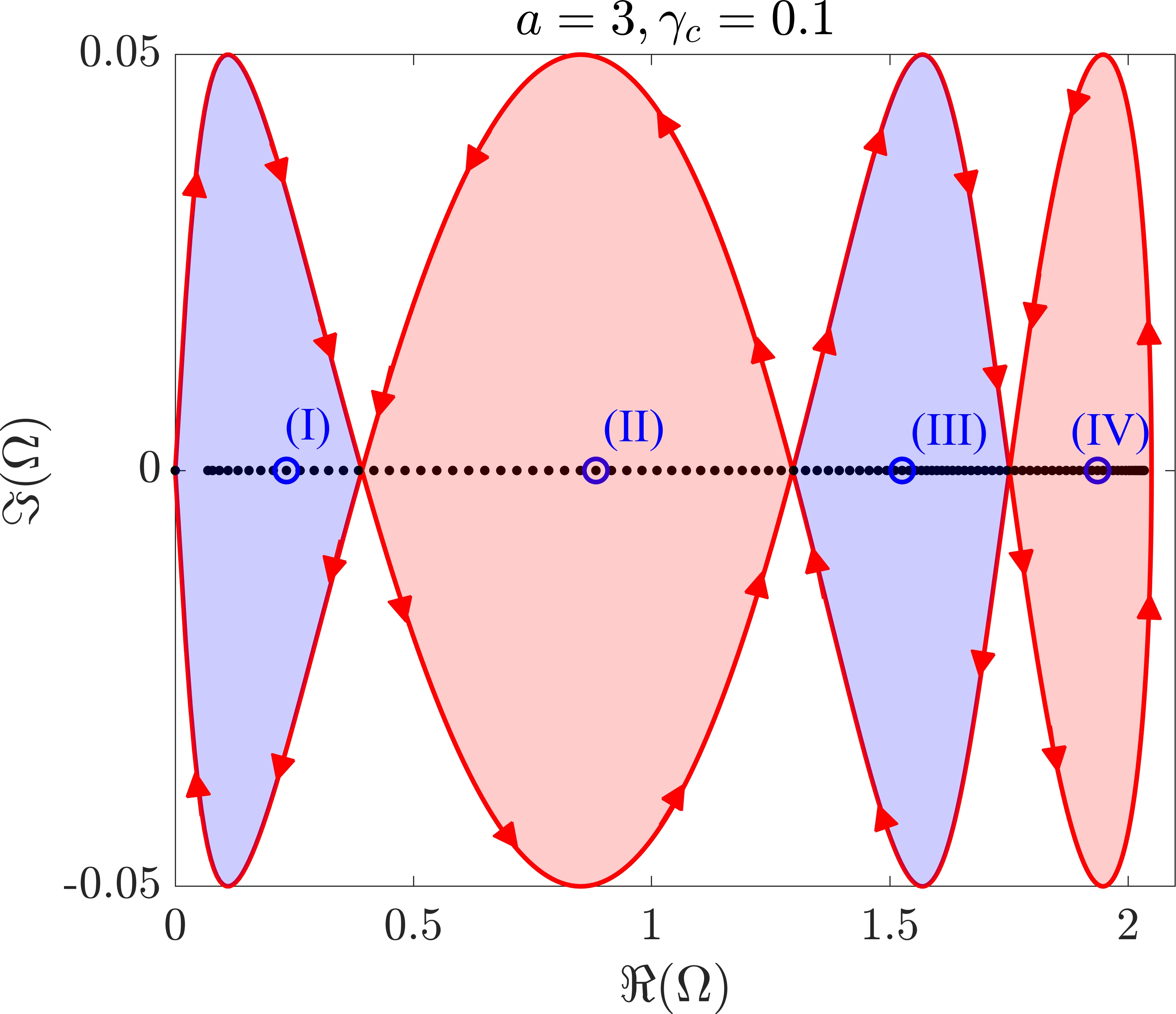}\label{Fig5e}}
		\subfigure[]{
			\includegraphics[height=0.31\textwidth]{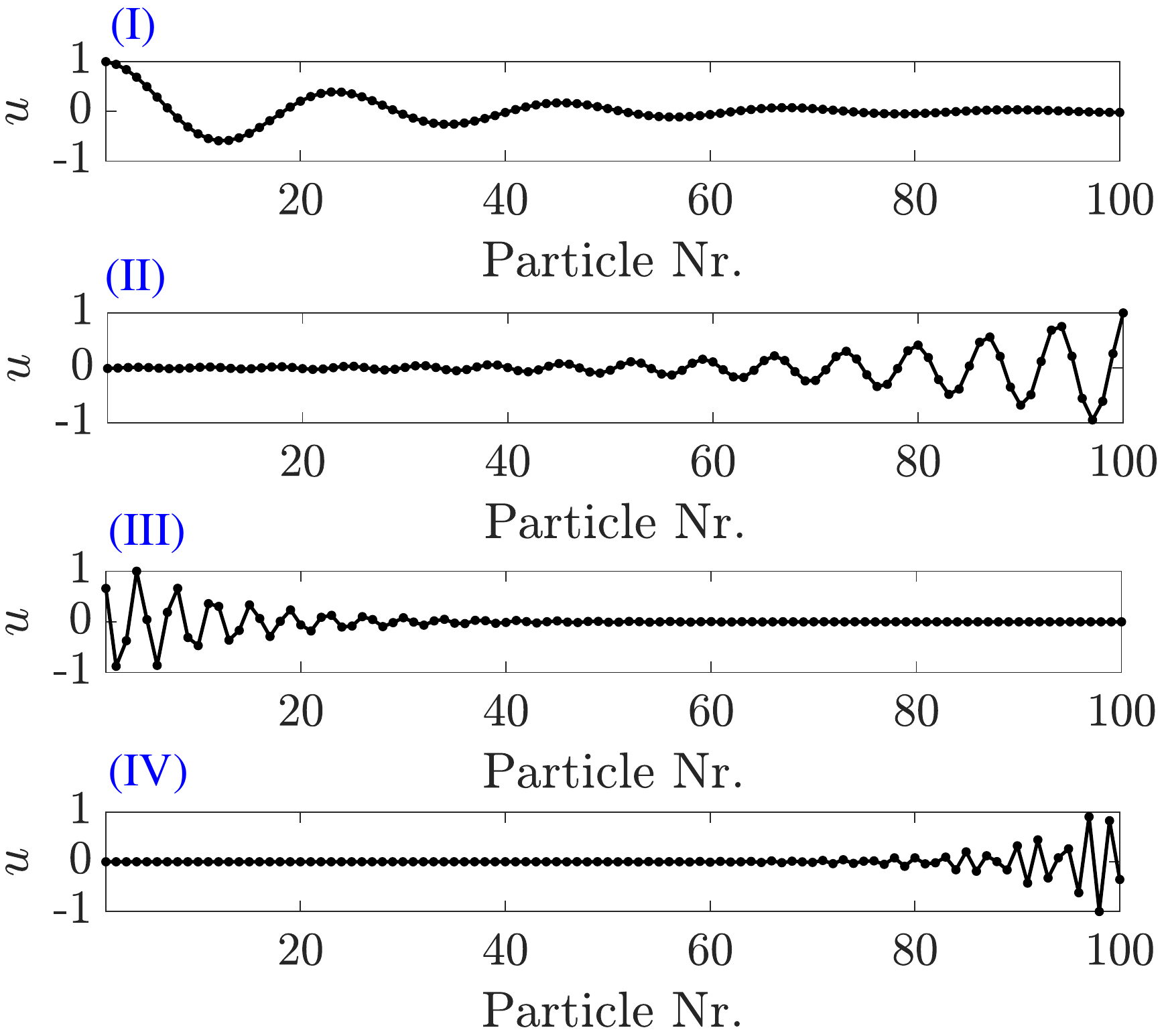}\label{Fig5f}}
	\caption{Dispersion topology and non-Hermitian skin effect in lattices with non-local feedback interactions ($a>0$). The complex dispersion bands $\Omega(\mu)$ for lattices with $a=1$ and $a=3$ are displayed in (a) and (d). Their projection on the complex plane define the closed loops shown in (b) and (e). The non-locality of the feedback interactions result in multiple phases defined within a single band, as highlighted by shaded blue and red areas representing regions with winding number $\nu=-1$ and $\nu=1$, respectively. Bulk modes of a finite lattice with $N=100$ masses whose eigenfrequencies (black dots) lie inside regions with $\nu=-1$ or $\nu=1$ are respectively localized at the left or right boundaries, as confirmed by the representative examples in (c) and (f).  }
	\label{Fig5}
\end{figure}

When non-local feedback interactions are considered ($a>0$), the dispersion topology is characterized by multiple phases defined within a single band. Figs.~\ref{Fig5}(a,b) displays the dispersion $\Omega(\mu)$ and its projection on the complex plane for a lattice with $\gamma_c=0.1$ and $a=1$, while results for $a=3$ are reported in Figs.~\ref{Fig5}(d,e). The dispersion loops feature multiple regions with interchanging winding numbers: the lattice with $a=1$ is characterized by two phases, while the lattice with $a=3$ is characterized by four phases, as highlighted by shaded blue and red areas denoting regions with $\nu=1$ and $\nu=-1$. We observe that the bulk modes of finite lattices (black dots) are localized at the left or the right boundary (Figs.~\ref{Fig5}(c,f)) when corresponding eigenfrequencies lie inside regions with $\nu=-1$ and $\nu=1$, respectively.

\begin{figure}[t!]
	\centering
		\subfigure[]{
			\includegraphics[width=0.49\textwidth]{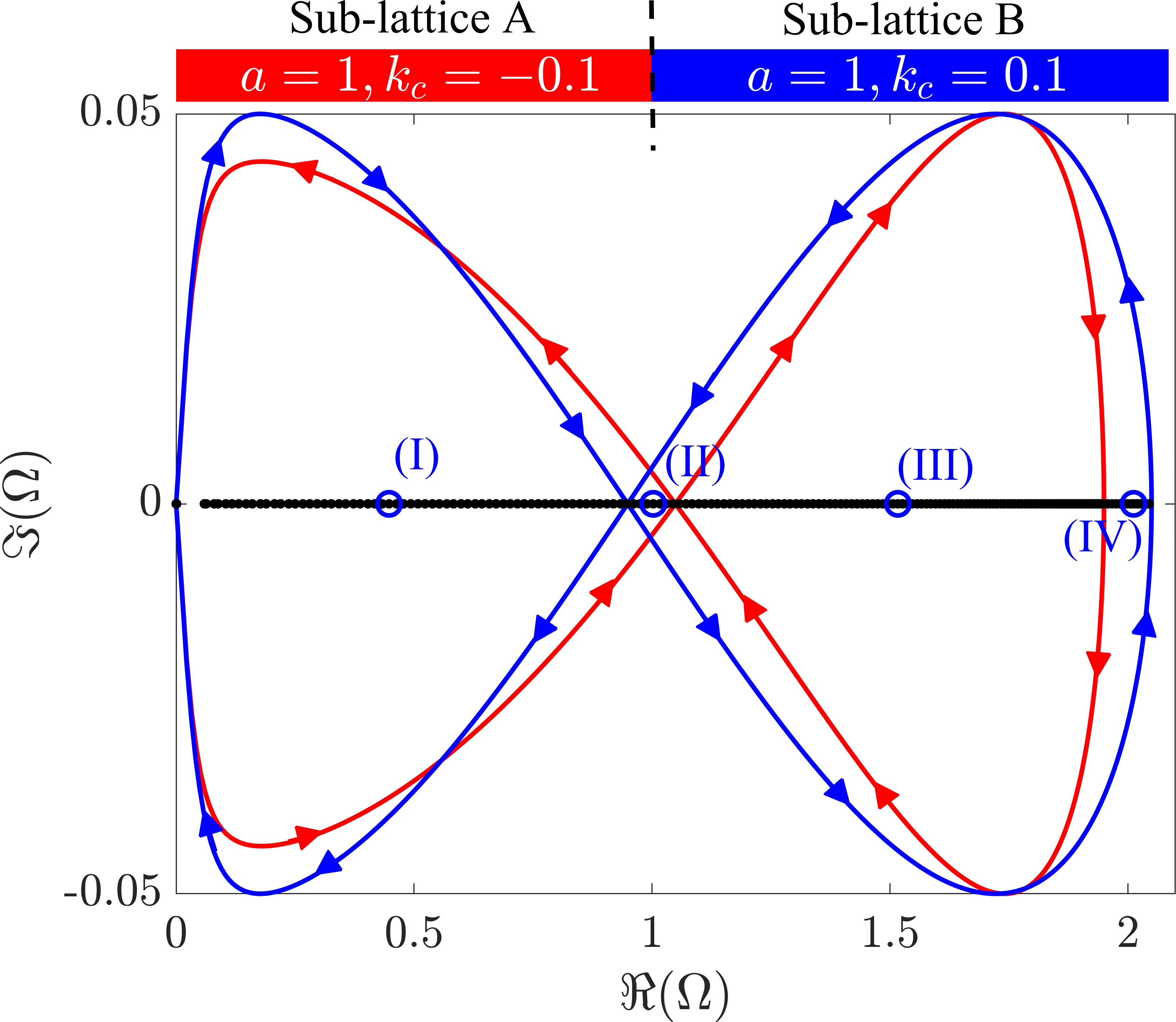}\label{Fig6a}}
		\subfigure[]{
			\includegraphics[width=0.49\textwidth]{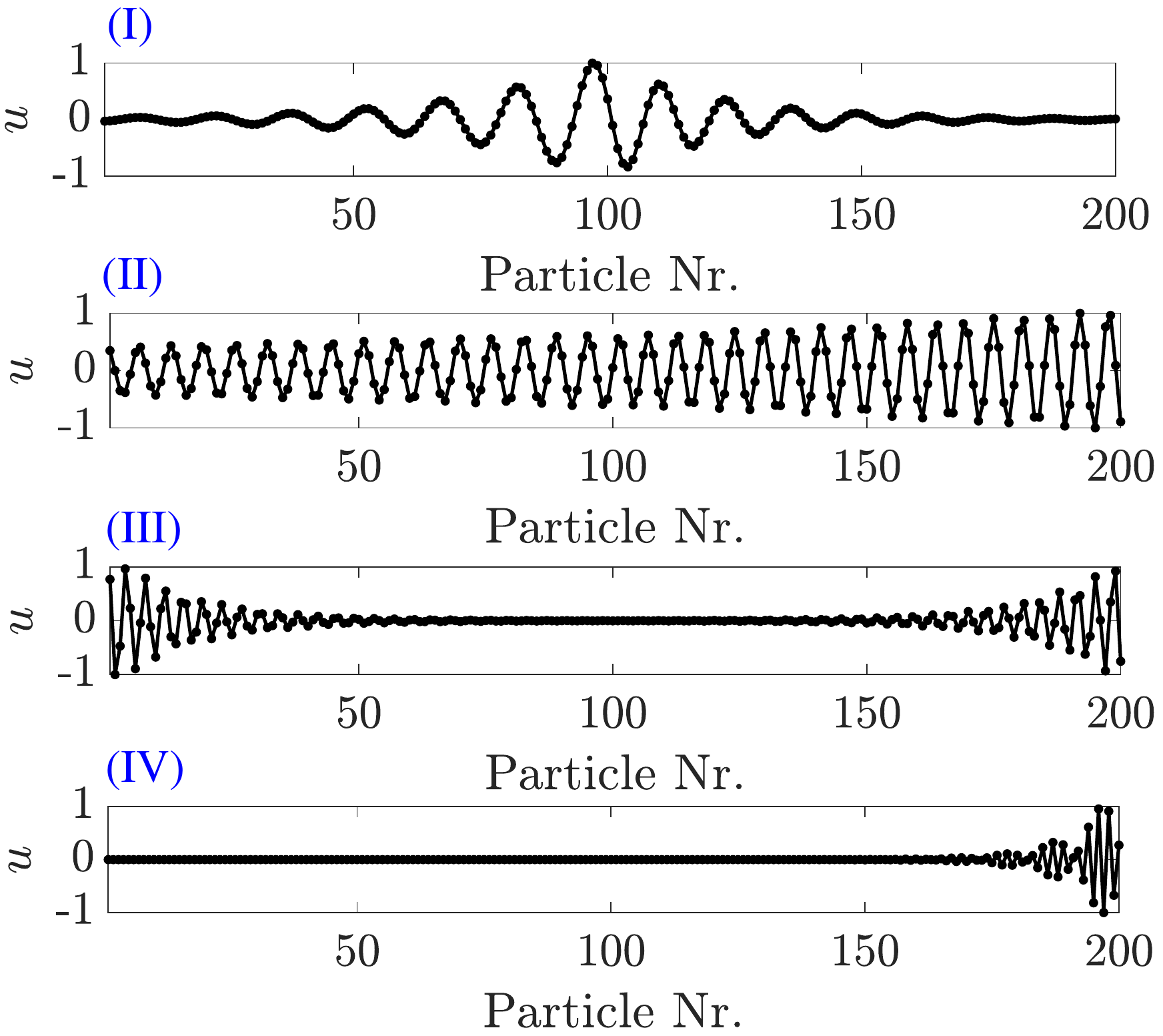}\label{Fig6b}}
	\caption{Bulk properties of finite lattice in a domain-wall configuration. Red and blue loops in (a) represent the dispersion of sub-lattices A and B, respectively, while black dots correspond to the eigenfrequencies of the finite lattice with $N=200$ masses. The localization properties of the representative modes displayed in (b) are interpreted based on where their eigenfrequencies lie in the complex plane. In the first region the effects of $\nu=1$ for sub-lattice A and $\nu=-1$ for sub-lattice B lead to bulk modes localized at the interface (mode I). In the second large region, modes are localized at both edges (mode III) since $\nu=-1$ for sub-lattice A and $\nu=1$ for sub-lattice B. Mode II lies in a region with $\nu=1$ for both sub-lattices and exhibits a slight tendency of amplification towards the right boundary, while mode IV lies outside the dispersion loop of sub-lattice A and inside a region with $\nu=1$ for sub-lattice B, resulting in localization at the right boundary.}
	\label{Fig6}
\end{figure}

The characterization of bulk properties through winding numbers can also be applied to systems coupled by a domain wall, which leads to the existence of bulk interface modes and of "double skin modes" (modes localized at both boundaries). We illustrate this by considering a finite lattice of $N=200$ masses, where the first $100$ masses are characterized by $a=1, \gamma_c=-0.1$ (sub-lattice A), and the masses on the second half by $a=1, \gamma_c=0.1$ (sub-lattice B). Figure~\ref{Fig6a} displays the spectral properties of the coupled system, where the red loop represents the dispersion of sub-lattice A, and the blue loop represents the dispersion of sub-lattice B. Also, black dots represent the eigenfrequencies of the finite lattice, while a few selected modes marked by blue circles have their mode shapes displayed in Fig.~\ref{Fig6b}. The localization properties of the bulk modes are related to the regions in the complex plane where their eigenfrequencies lie. The modes inside the first region (represented by mode I in Fig.~\ref{Fig6b}) are localized at the interface, since in that region $\nu=1$ for sub-lattice A implies a tendency of localization towards its right, while $\nu-1$ for sub-lattice B implies a tendency for localization towards its left. Modes inside a second large region (represented by mode III in Fig.~\ref{Fig6b}) exhibit an opposite behavior: $\nu=-1$ is associated with sub-lattice A, which implies a tendency for localization to its left, while $\nu=1$ is associated with sub-lattice B, which implies a tendency for localization to its right. The modes inside this region are therefore "double skin modes" simultaneously localized at both boundaries. In a small region between the two larger regions the modes are characterized by $\nu=1$ for both sublattices, and a slight tendendy of amplification towards the right boundary is observed (mode II). A final set of modes represented by mode IV lie in a region outside the dispersion loop for sub-lattice A, and in a region with $\nu=1$ for sub-lattice B, which results in localization to the right.


\section{Two-dimensional elastic lattices with feedback interactions}\label{sec3}

\begin{figure}[t!]
	\centering
			\includegraphics[height=0.35\textwidth]{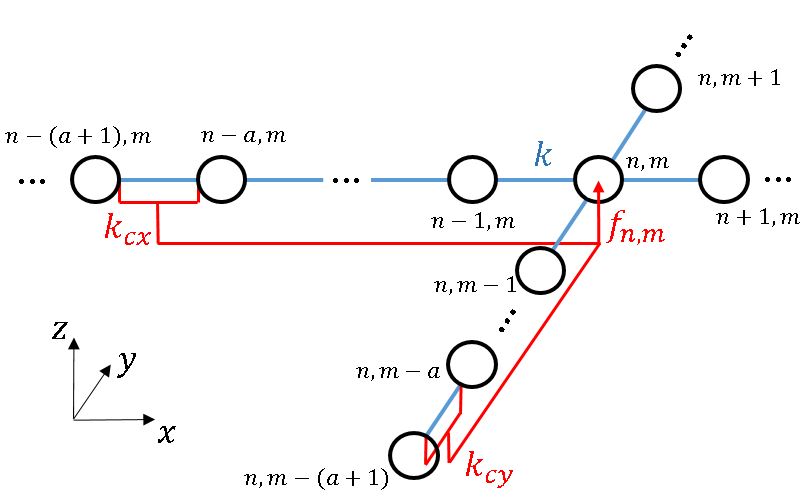}
	\caption{Two-dimensional lattice of equal masses $m$ connected by springs of stiffness $k$ with feedback control interactions. A force $F_{n,m}=k_{c_x}(u_{n-a,m}-u_{n-(a+1),m}) + k_{c_y}(u_{n,m-a} - u_{n,m-(a+1)})$ is applied to each mass in the lattice, corresponding to a reaction proportional to the strain of springs $a$ units behind in the $x$ direction, and $a$ units below in the $y$ direction.}
	\label{Fig7}
\end{figure}

We now extend the study to 2D lattices consisting of equal masses $m$ connected by springs $k$, separated by a unit distance in both $x$ and $y$ directions. Each mass moves along the perpendicular $z$ direction (Fig.~\ref{Fig7}), so that the springs react with a force proportional to the relative vertical motion of neighboring masses. Feedback interactions are defined by an additional force applied to mass $n,m$ proportional to the elongation of a spring $a$ units behind in the $x$ and $y$ directions, which is expressed as $f_{n,m}=k_{c_x}(u_{n-a,m}-u_{n-(a+1),m}) + k_{c_y}(u_{n,m-a} - u_{n,m-(a+1)})$, where $k_{c_x}$ and $k_{c_y}$ are the proportionality constants for elongations of springs aligned with the $x$ and $y$ directions, respectively. The governing equation of motion in the absence of external forces is
\begin{equation}\label{eq2D}
\begin{split}
(-\omega^2 m +4 k) u_{n,m}-k (u_{n-1,m}+u_{n+1,m}+u_{n,m-1}+u_{n,m+1})\\
-k_{c_x}(u_{n-a,m}-u_{n-(a+1),m}) - k_{c_y}(u_{n,m-a} - u_{n,m-(a+1)})=0.
\end{split}
\end{equation}


\subsection{Dispersion relations, non-reciprocity and and directionality}

We impose Bloch wave solutions in Eqn.~(\ref{eq2D})  of the form $u_{n,m}=Ue^{i(\omega t- \mu_x n -\mu_y m)}$, where $\mu_x$ and $\mu_y$ are the wave vector components along $x$ and $y$, respectively. This gives: 
\begin{equation}\label{eqdisp2D}
\Omega^2 = 2(2-\cos\mu_x-\cos\mu_y) - \gamma_x e^{i\mu_x a} (1-e^{i\mu_x})  -\gamma_y e^{i\mu_y a}(1-e^{i\mu_y}),
\end{equation}
where again $\Omega=\omega/\omega_0$, with $\omega_0=\sqrt{k/m}$, while $\gamma_x=k_{c_x}/k$ and $\gamma_y=k_{c_y}/k$. Similar to the one-dimensional case, we consider the solution with $\Omega_r>0$ to represent the dispersion, such that $\Omega_i<0$ is associated with wave amplification, while $\Omega_i>0$ with attenuation. 

\begin{figure}[t!]
	\centering
		\subfigure[]{
			\includegraphics[height=0.28\textwidth]{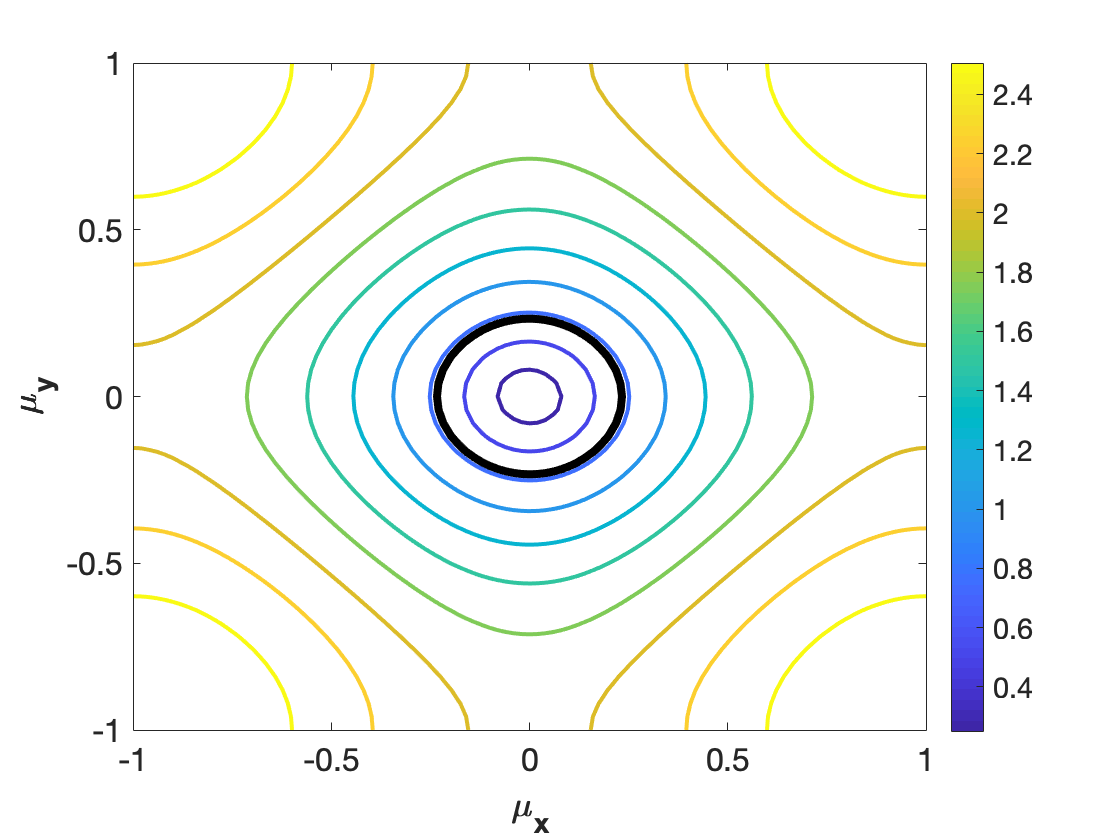}\label{Fig8a}}
		\subfigure[]{
			\includegraphics[height=0.28\textwidth]{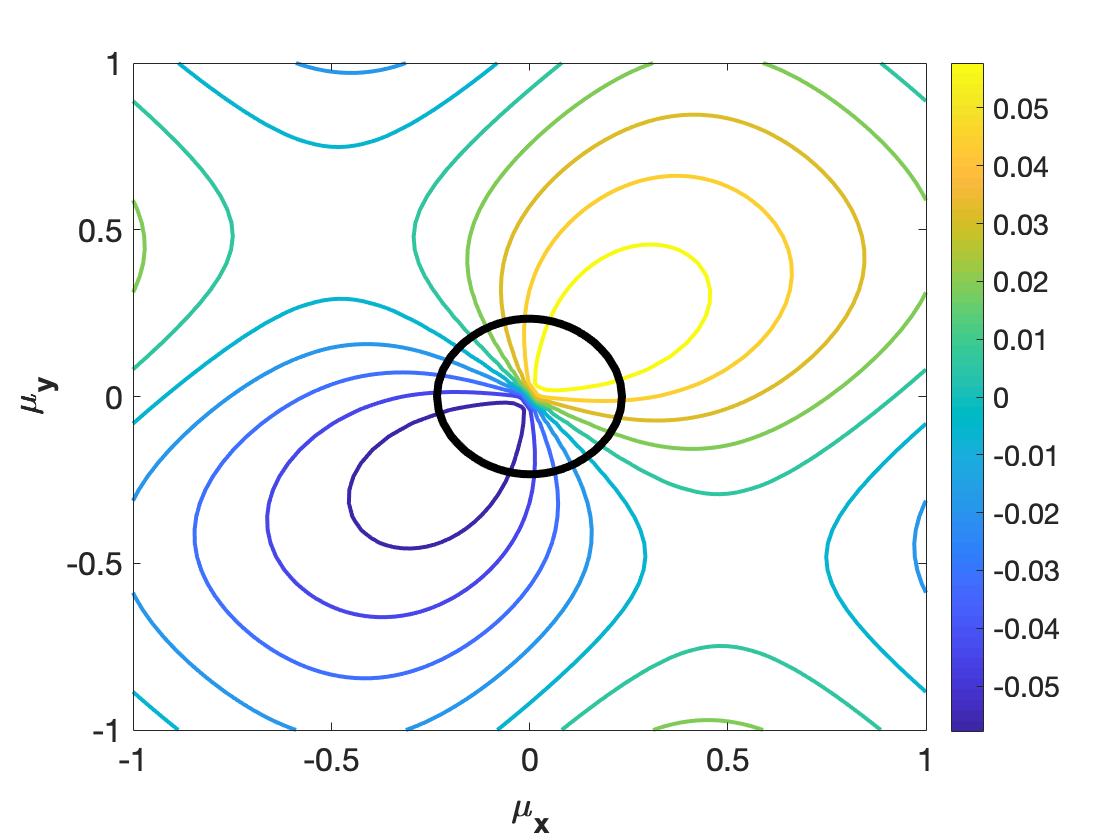}\label{Fig8b}}
		\subfigure[]{
			\includegraphics[height=0.27\textwidth]{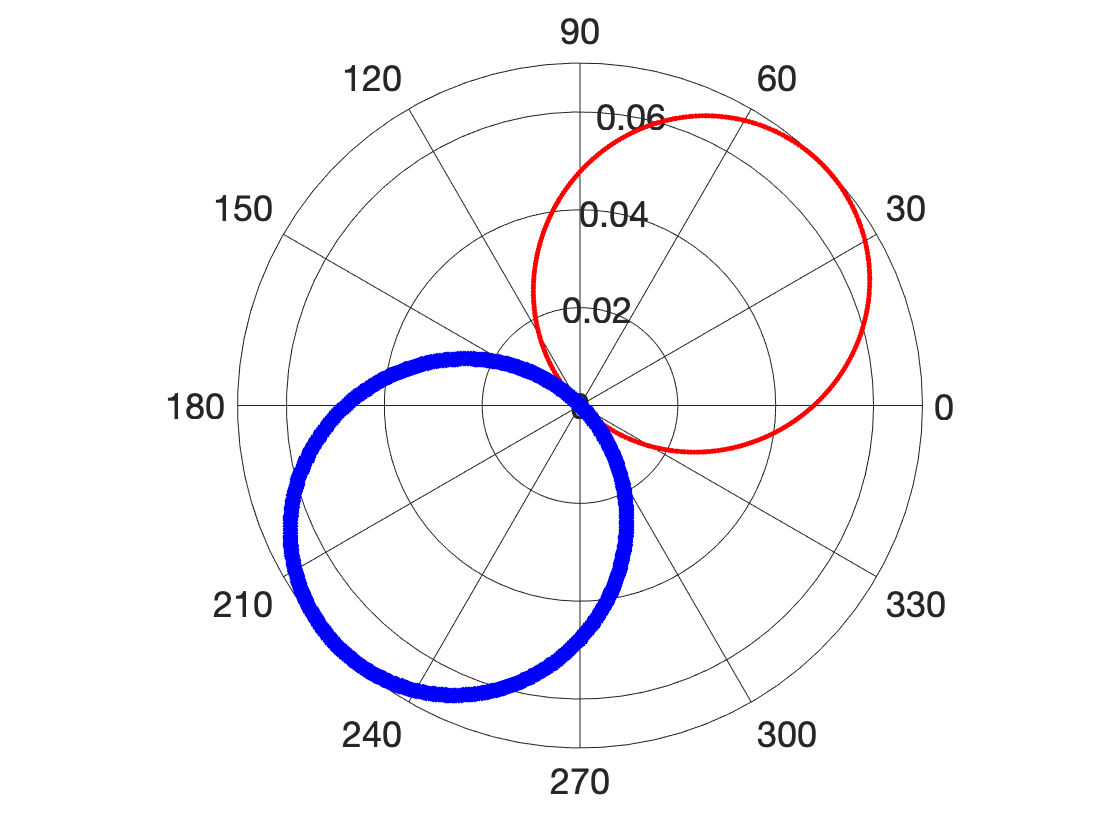}\label{Fig8d}}\\
		\subfigure[]{
			\includegraphics[height=0.28\textwidth]{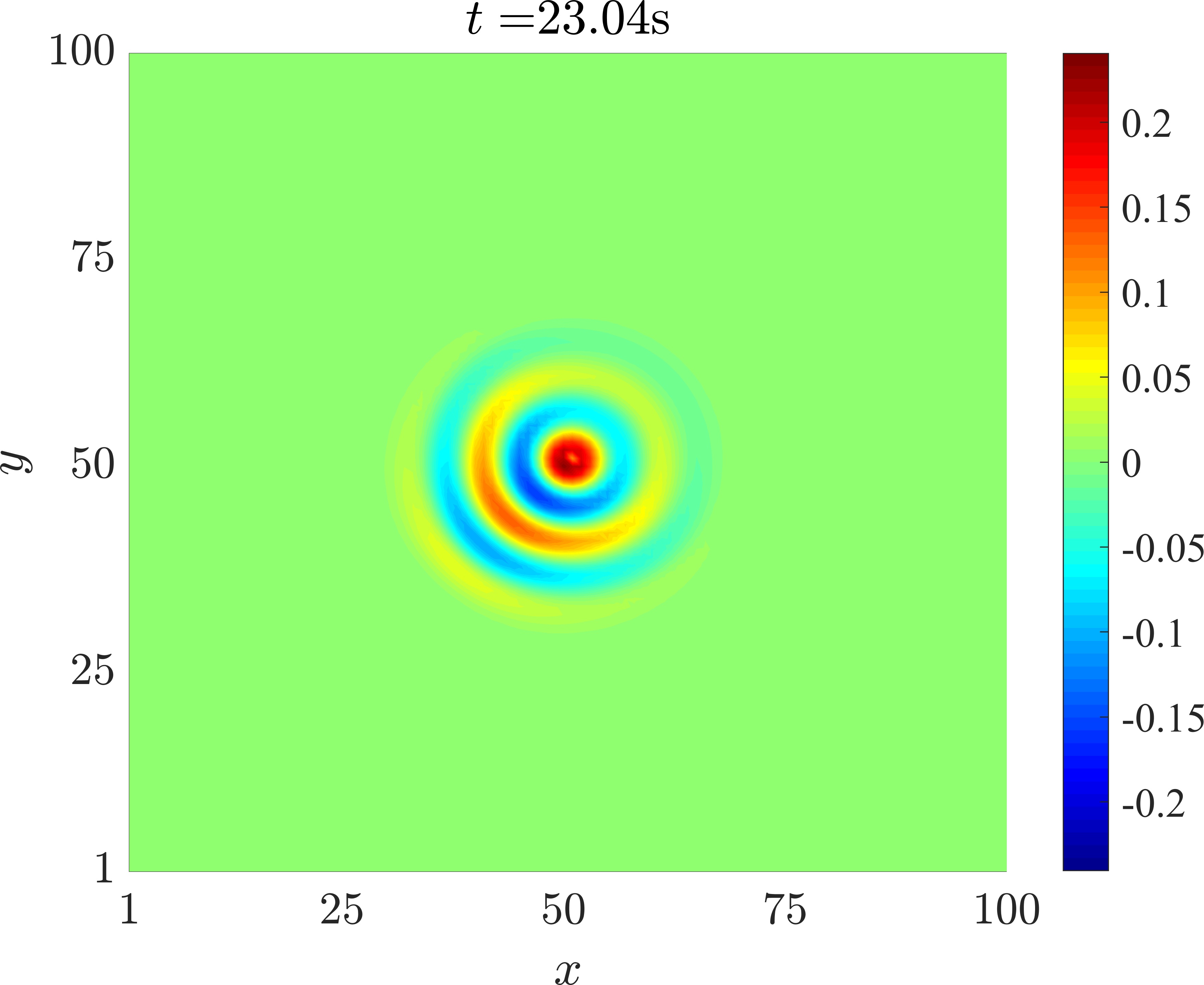}\label{Fig8c}}
		\subfigure[]{
			\includegraphics[height=0.27\textwidth]{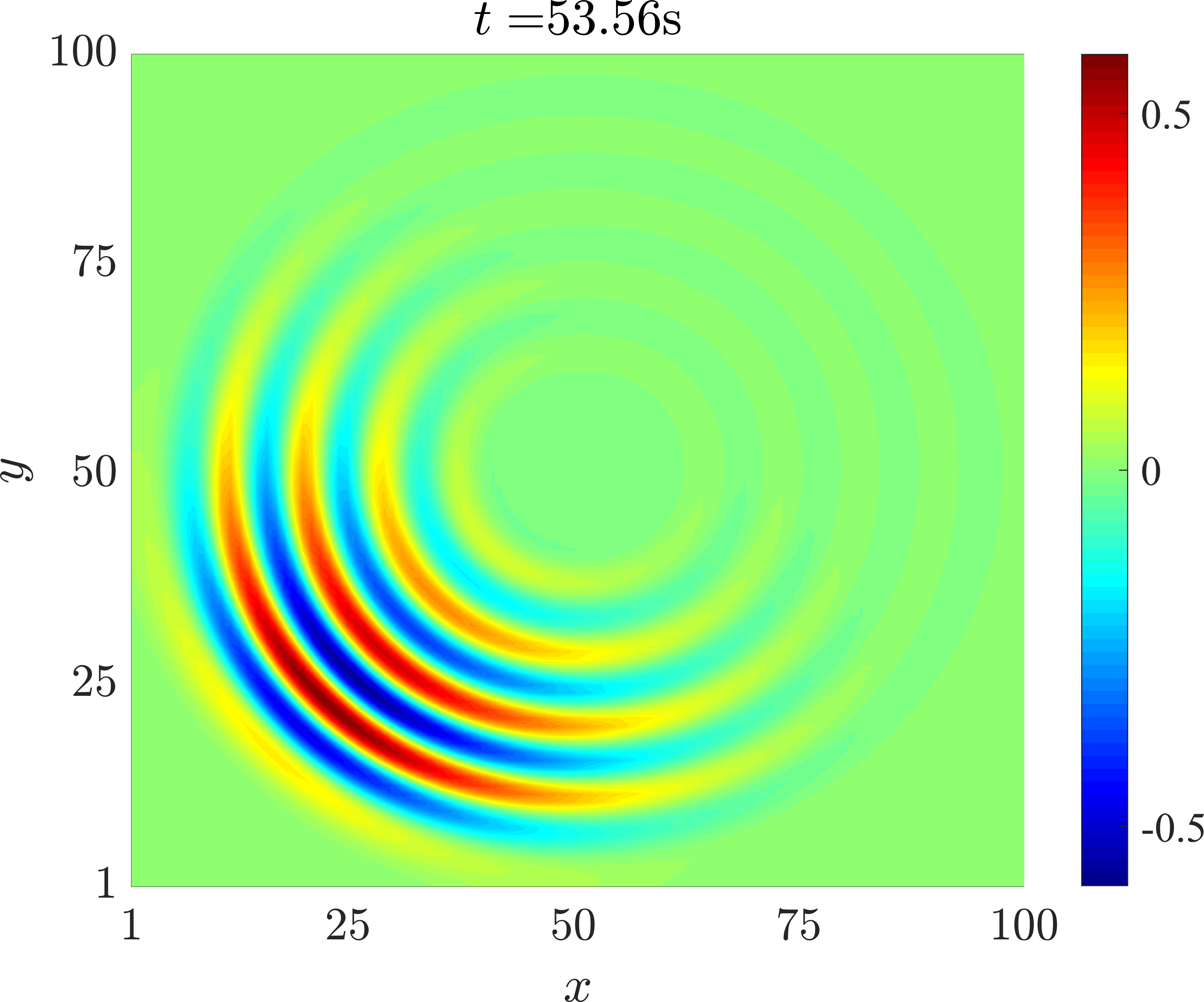}\label{Fig8f}}
	\caption{Non-reciprocal amplification and attenuation of waves in 2D lattice with feedback parameters $a=0$, $\gamma_x=\gamma_y=0.1$. Iso-frequency contours corresponding to the real part $\Omega_r$ (a), and imaginary part $\Omega_i$ (b) of the dispersion $\Omega(\mu_x,\mu_y)$. The thick black line outlines the contour for $\Omega_r=0.7$, and defines the wave vector components pairs $\mu_x,\mu_y$ governing propagation at the considered frequency. Angular variation of $\Omega_i$ highlighting the angular range of amplification (thick blue line) and attenuation (thin red line) (c). Snapshots of transient response to a tone-burst excitation at $\Omega=0.7$ applied to the center mass of the lattice illustrating that waves traveling towards the bottom left corner are preferentially amplified (d,e).}
	\label{Fig8}
\end{figure}

For the local control case ($a=0$), Figs.~\ref{Fig8}(a,b) display the real and imaginary iso-frequency contours of the dispersion surfaces of a lattice with $\gamma_x=\gamma_y=0.1$. While the real part (Fig.~\ref{Fig8a}) closely resembles that of a passive 2D lattice~\cite{hussein2014dynamics}, the imaginary part of the frequency contours (Fig.~\ref{Fig8b}) exhibits directional dependent attenuation and amplification zones. In particular, a region for which $\Omega_i<0$ is identified in the third quadrant of the $\mu_x,\mu_y$ plane (Fig.~\ref{Fig8b}), revealing a range of directions of wave amplification. This is further illustrated by considering a frequency of $\Omega=0.7$, whose corresponding contour in the $\Omega_r$ map, highlighted by the thick black line in Fig.~\ref{Fig8a}, is approximately circular, possibly suggesting isotropic propagation. However, the wave vector components at this frequency (also highlighted by the thick black circle in Fig.~\ref{Fig8b}), cross regions of positive and negative imaginary frequency. The angular dependence of $\Omega_i$ is shown in Fig.~\ref{Fig8c}, where it is plotted in polar form versus the propagation angle $\theta=\tan^{-1}(\mu_y/\mu_x)$. In the figure, the thick blue lobe denotes amplification corresponding to $\Omega_i<0$, while the thin red line defines the angular range associated with attenuation. The plot shows that maximum amplification is found for $\theta\approx225^o$, which corresponds to waves traveling towards the left bottom corner of a square lattice. We illustrate this by conducting a transient time domain simulation on a lattice with $100 \times 100$ masses, with a force consisting of a 5-cycle sinusoidal burst of frequency $\Omega=0.7$ (similar to that of Figs.~\ref{Fig2}(e,f)) applied to the center mass of the lattice. 
Snapshots of the lattice motion at two subsequent time instants displayed in Figs.~\ref{Fig8}(d,e) confirm that waves are preferentially amplified as they travel towards the bottom left corner of the lattice.

\begin{figure}[t!]
	\centering
		\subfigure[]{
			\includegraphics[height=0.225\textwidth]{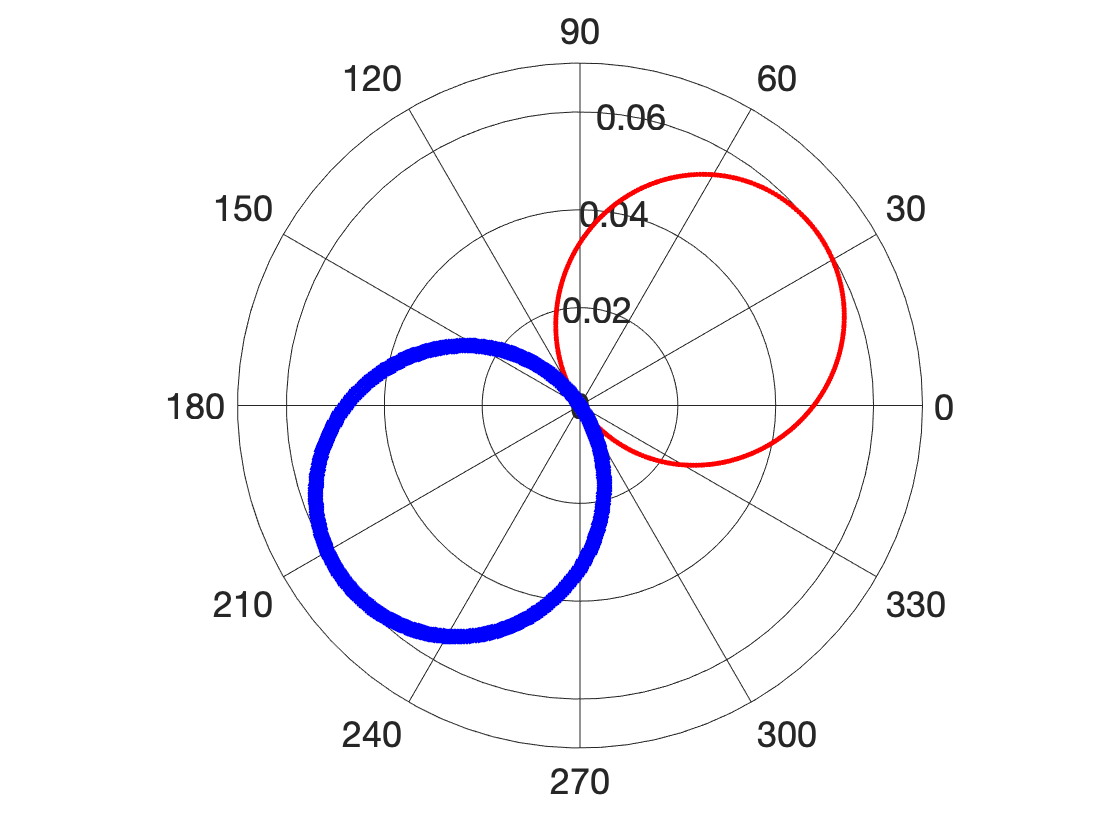}\label{Fig9a}}
		\subfigure[]{
			\includegraphics[height=0.225\textwidth]{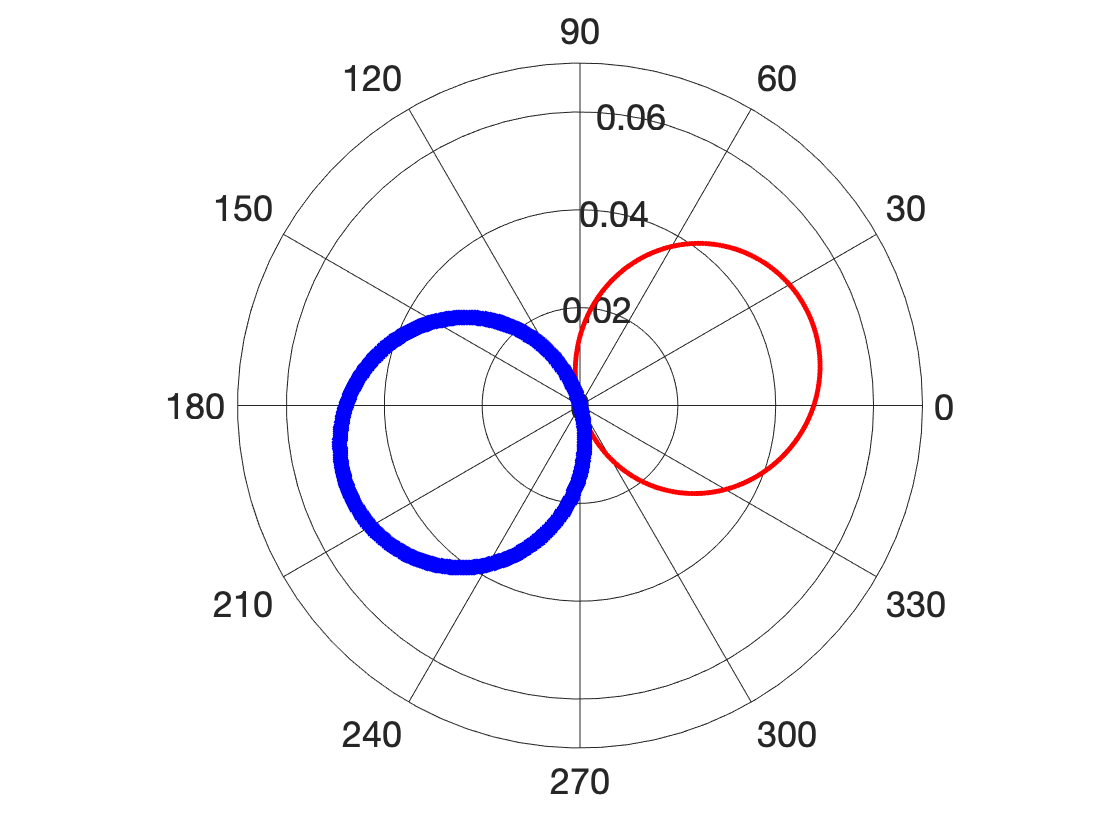}\label{Fig9b}}
		\subfigure[]{
			\includegraphics[height=0.225\textwidth]{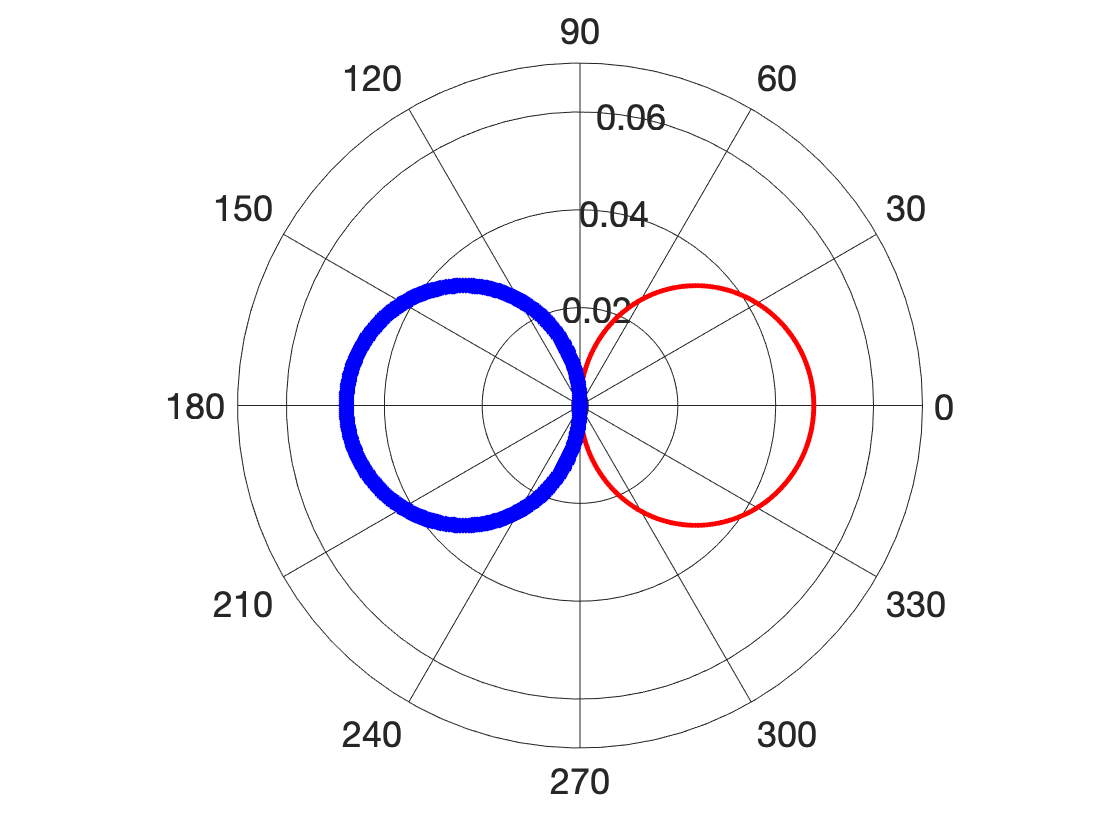}\label{Fig9c}}
		\subfigure[]{
			\includegraphics[height=0.26\textwidth]{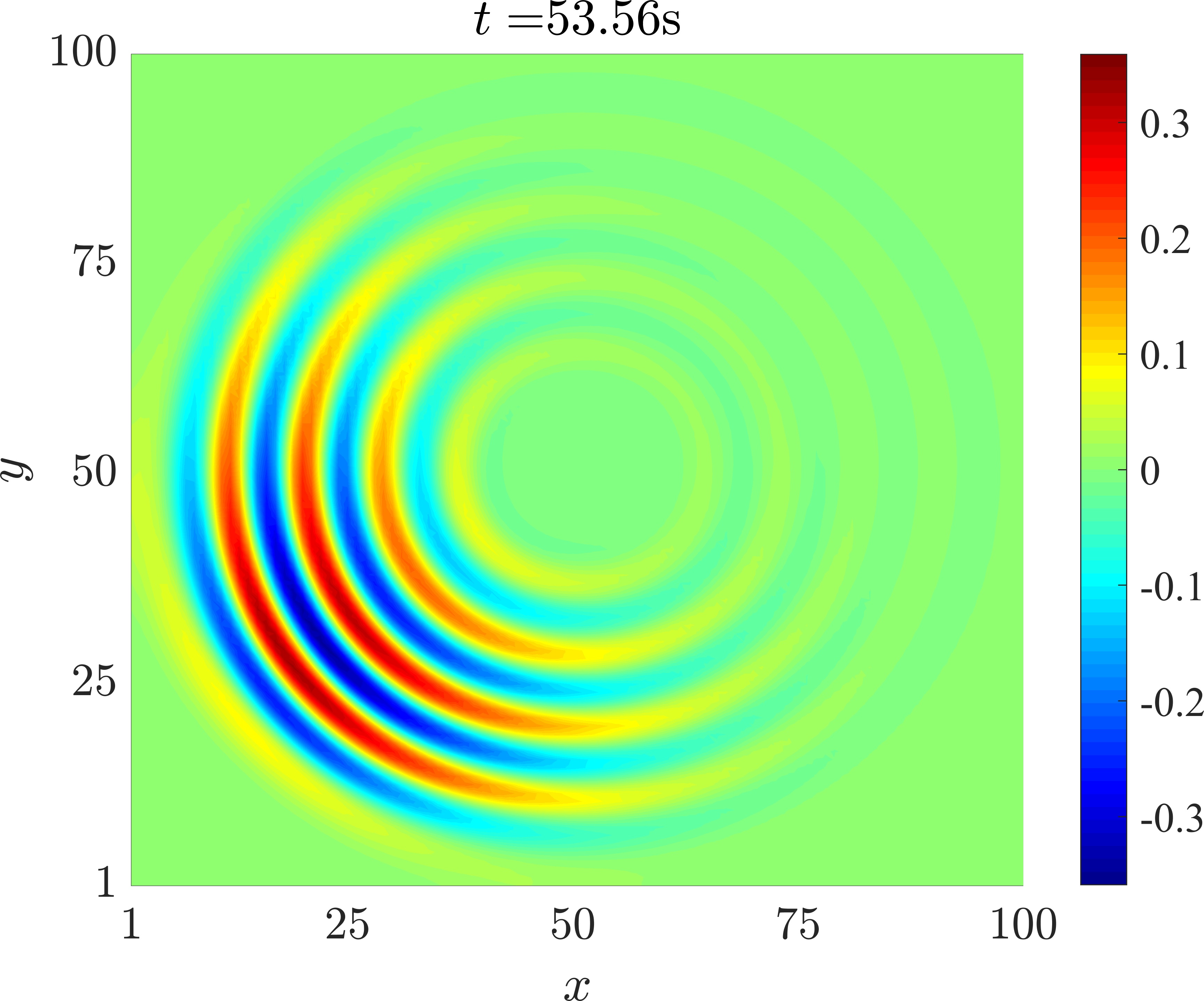}\label{Fig9d}}
		\subfigure[]{
			\includegraphics[height=0.26\textwidth]{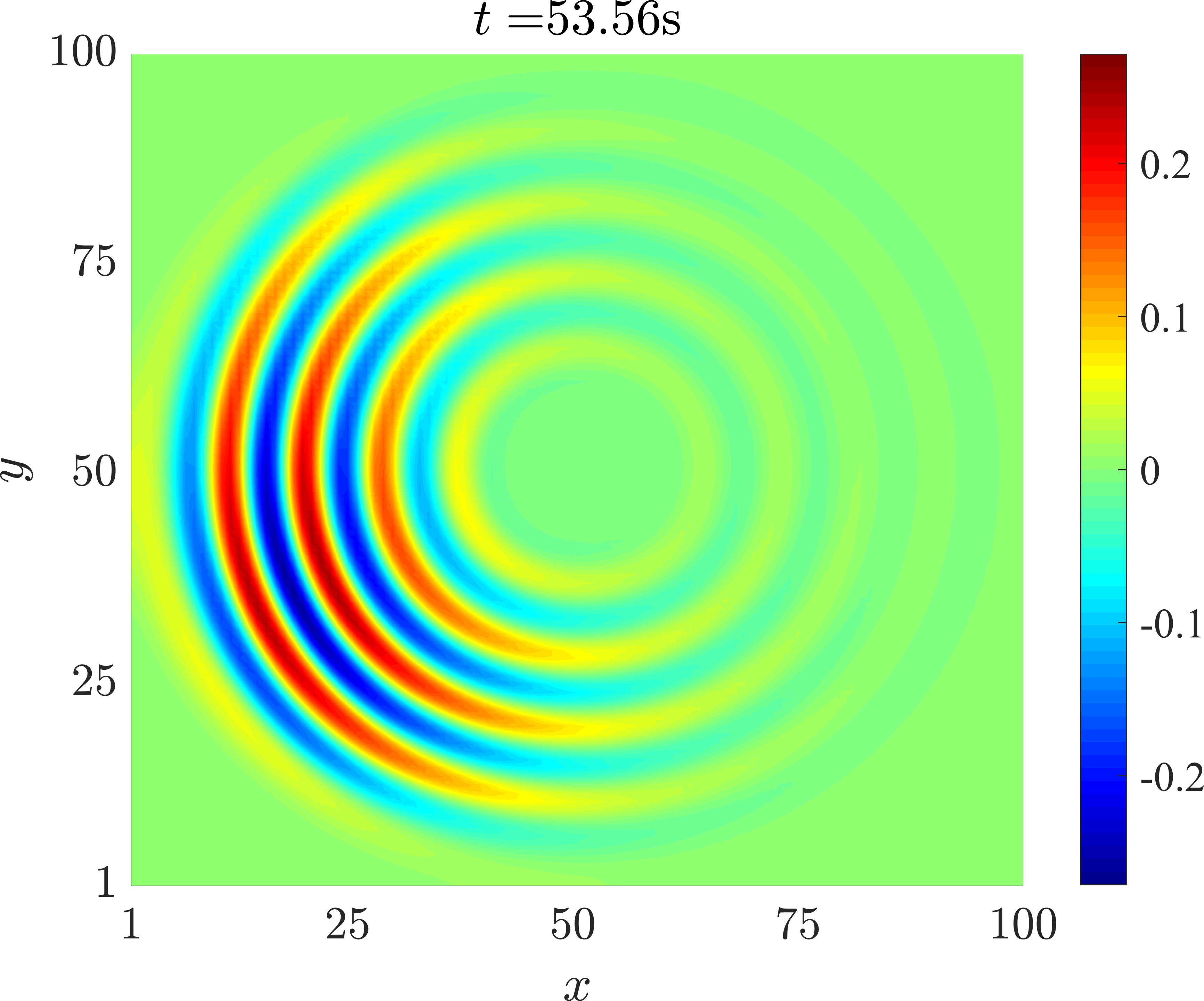}\label{Fig9e}}
		\subfigure[]{
			\includegraphics[height=0.26\textwidth]{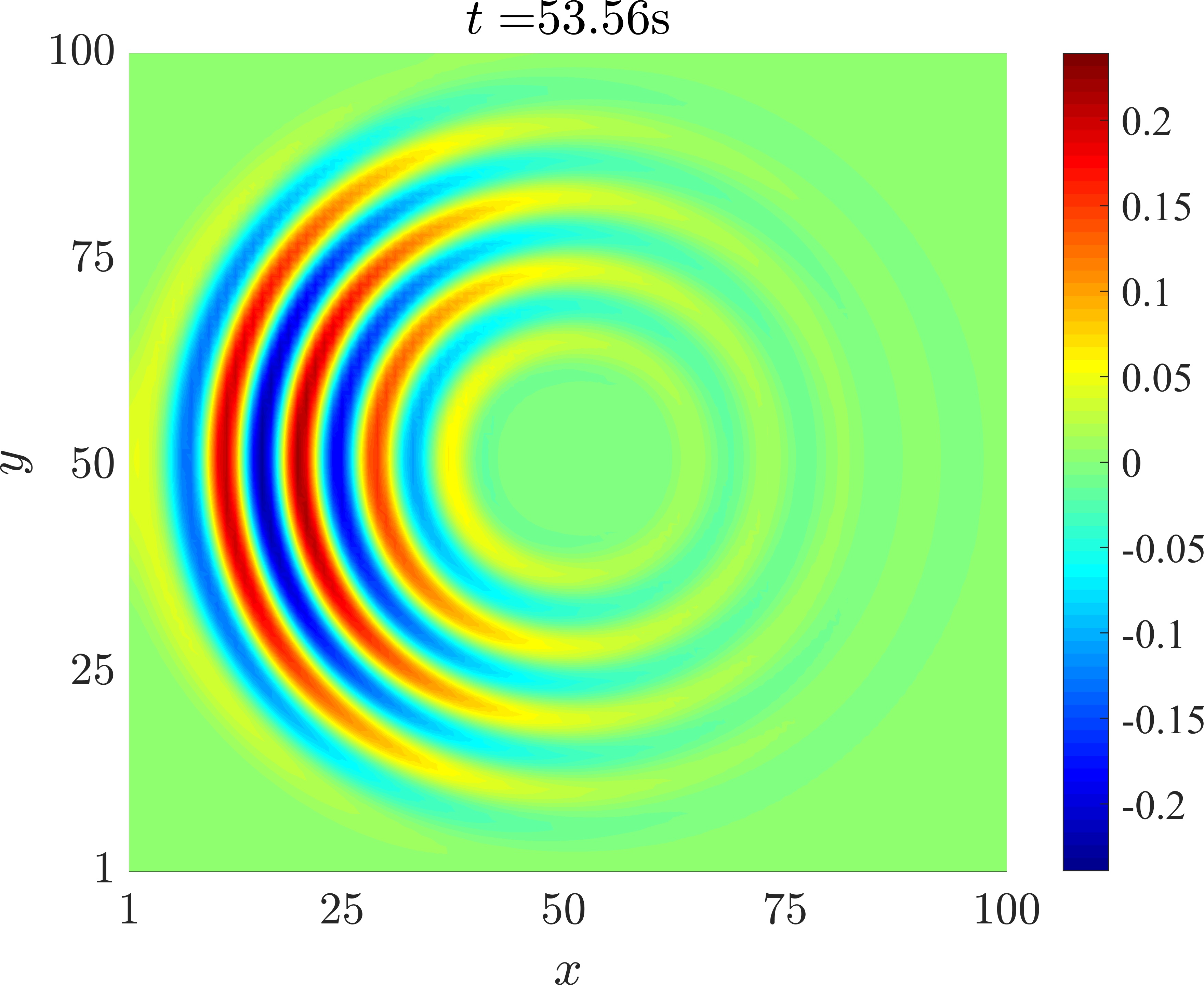}\label{Fig9f}}
	\caption{Tunability of non-reciprocal wave amplification in 2D lattices with feedback interactions ($a=0$, $\gamma_x=0.1$) for $\Omega_r=0.7$. Polar plots of the imaginary component of the dispersion showing angular ranges of amplification (thick blue line) and attenuation (thin red line) displayed for $\gamma_y=0.07$ (a), $\gamma_y=0.03$ (b), and $\gamma_y=0$ (c) demonstrate a transition of the preferential wave amplification direction based on feedback control. Snapshots of transient response to a sine-burst excitation of center frequency $\Omega=0.7$ are displayed in (d-e), illustrating the change in the directions of amplification for each case.}
	\label{Fig9}
\end{figure}

The direction of preferential amplification can be tuned based on the feedback parameters $\gamma_x,\gamma_y$, as illustrated in Fig.~\ref{Fig9}, again for $\Omega=0.7$. Letting $\gamma_x=0.1$ and varying $\gamma_y$ changes the direction of preferential amplification, as illustrated for 3 representative $\gamma_y$ values (0.07, 0.03 and 0) in the imaginary dispersion components polar plots of Figs.~\ref{Fig9}(a-c). Snapshots of the transient response in Figs.~\ref{Fig9}(d-e) confirm that waves are preferentially amplified according to the predicted directions. Other combinations of $\gamma_x,\gamma_y$ can tune the direction of amplification, which suggests that anisotropy in the control laws ($\gamma_x \neq \gamma_y$) can be employed for non-reciprocal directional amplification, which may significantly expand the functionality of reciprocal directionality encountered in passive 2D lattices with anisotropy in spring constants~\cite{hussein2014dynamics}.

\begin{figure}[t!]
	\centering
	\subfigure[]{
		\includegraphics[height=0.26\textwidth]{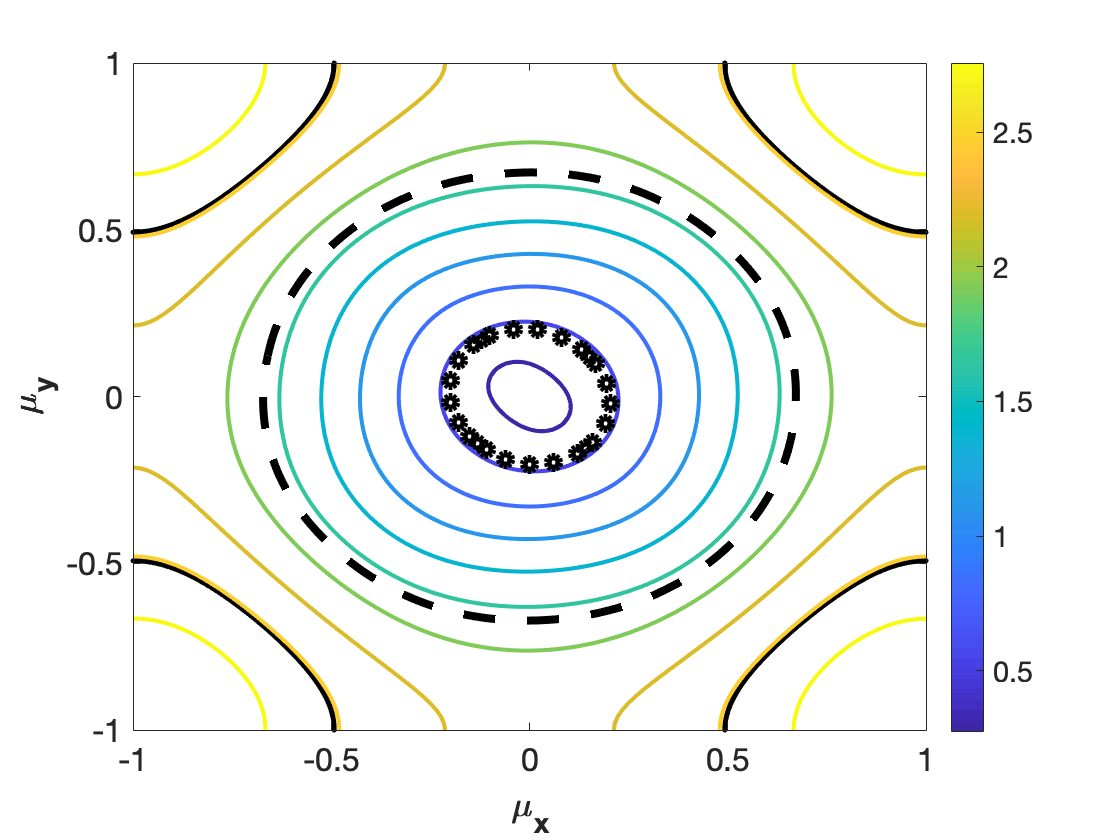}\label{Fig10a}}
	\subfigure[]{
		\includegraphics[height=0.26\textwidth]{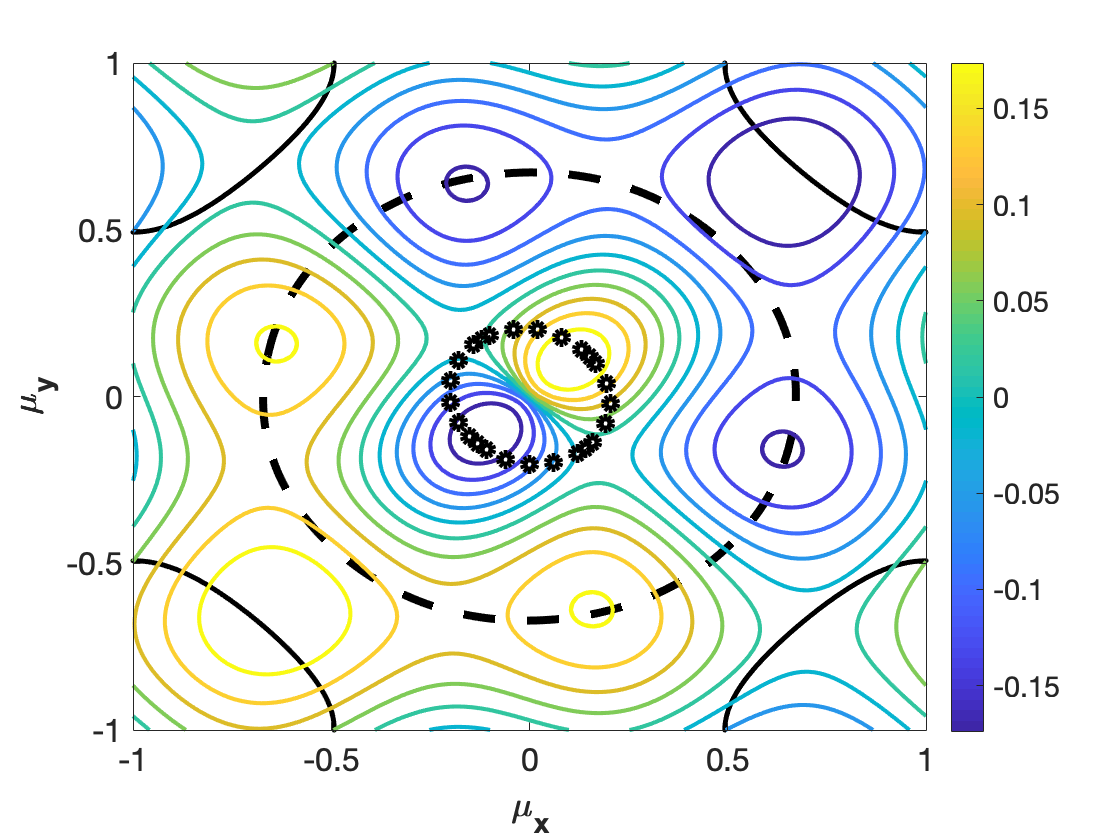}\label{Fig10b}}
	\subfigure[]{
		\includegraphics[height=0.225\textwidth]{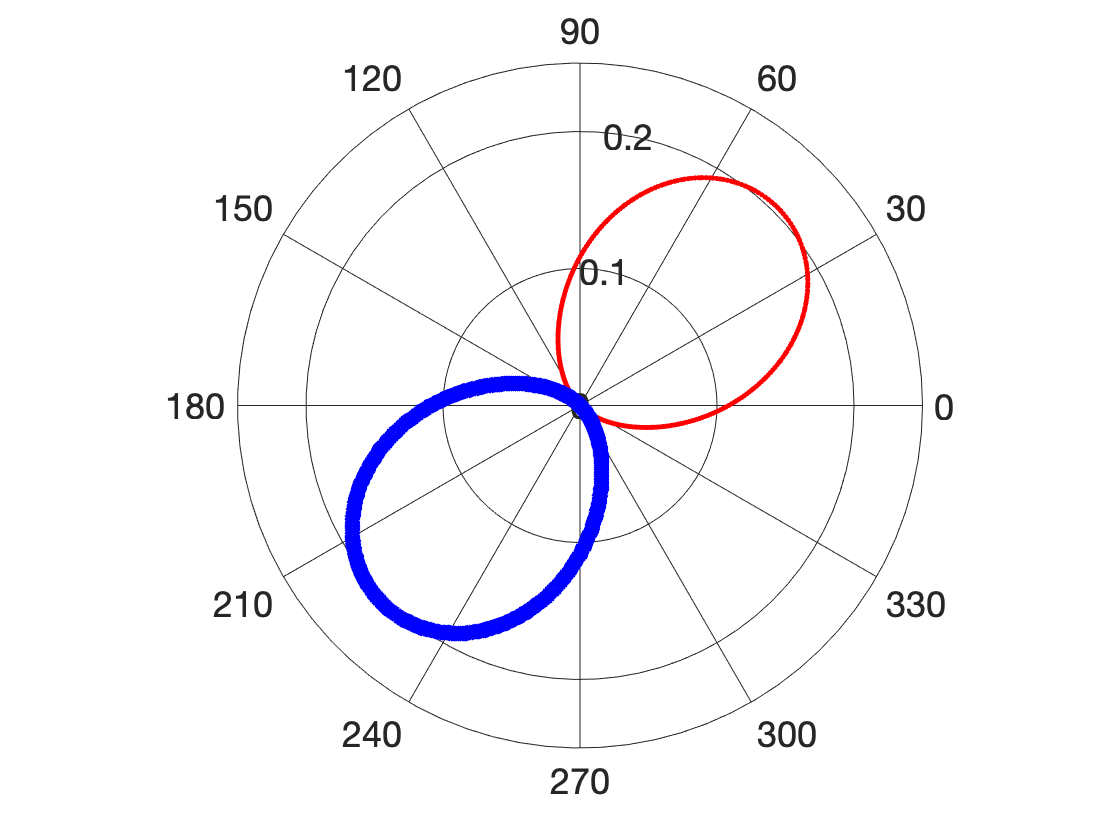}\label{Fig10c}}
	\subfigure[]{
		\includegraphics[height=0.225\textwidth]{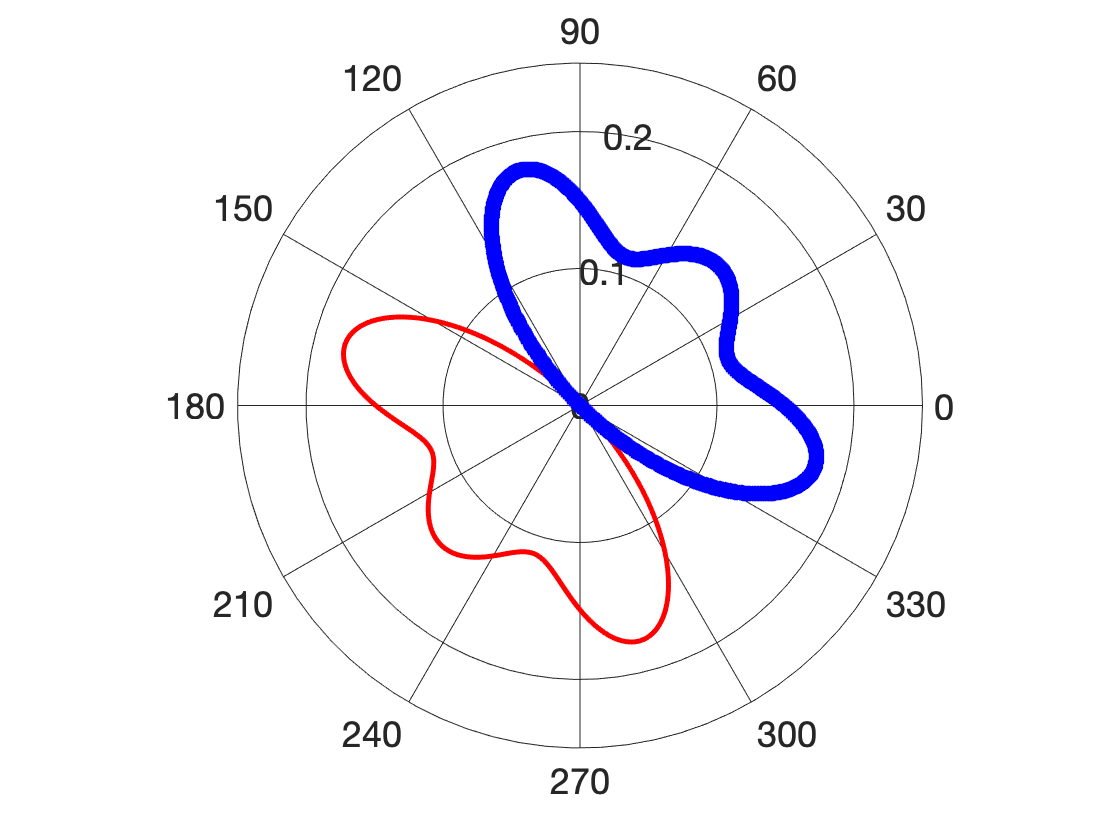}\label{Fig10d}}
	\subfigure[]{
	\includegraphics[height=0.225\textwidth]{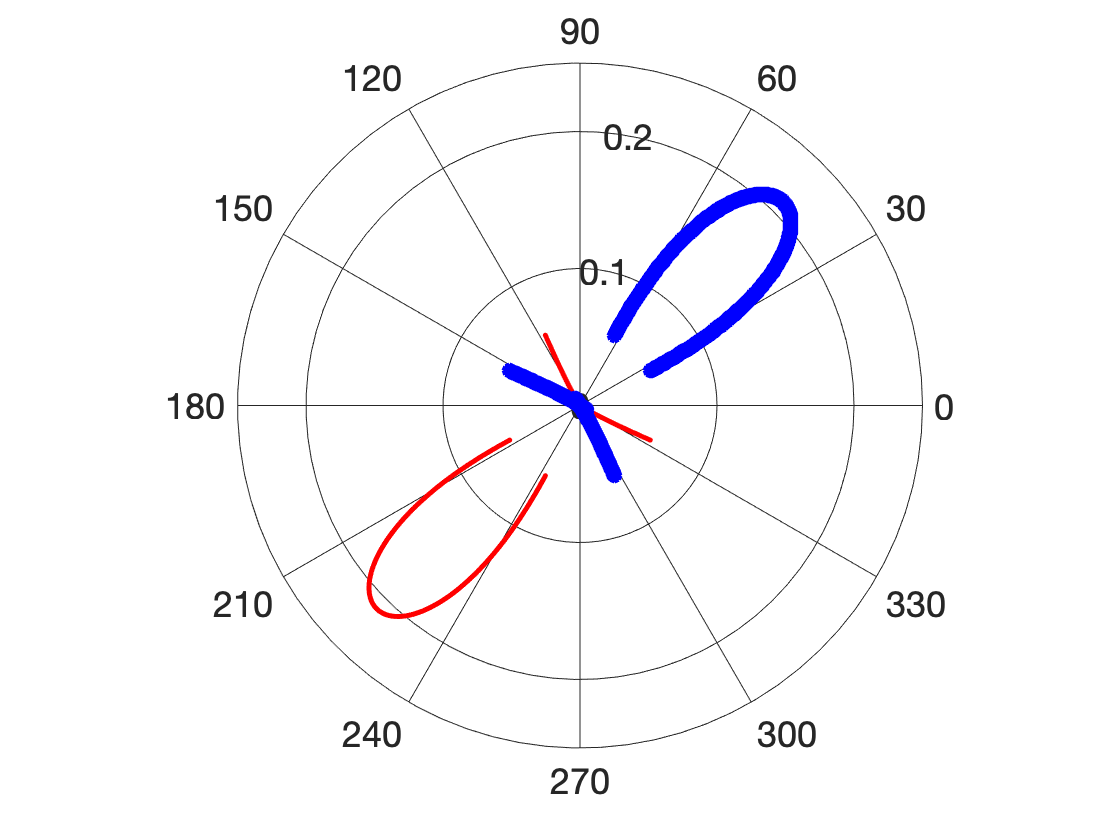}\label{Fig10e}}\\
	\subfigure[]{
		\includegraphics[height=0.26\textwidth]{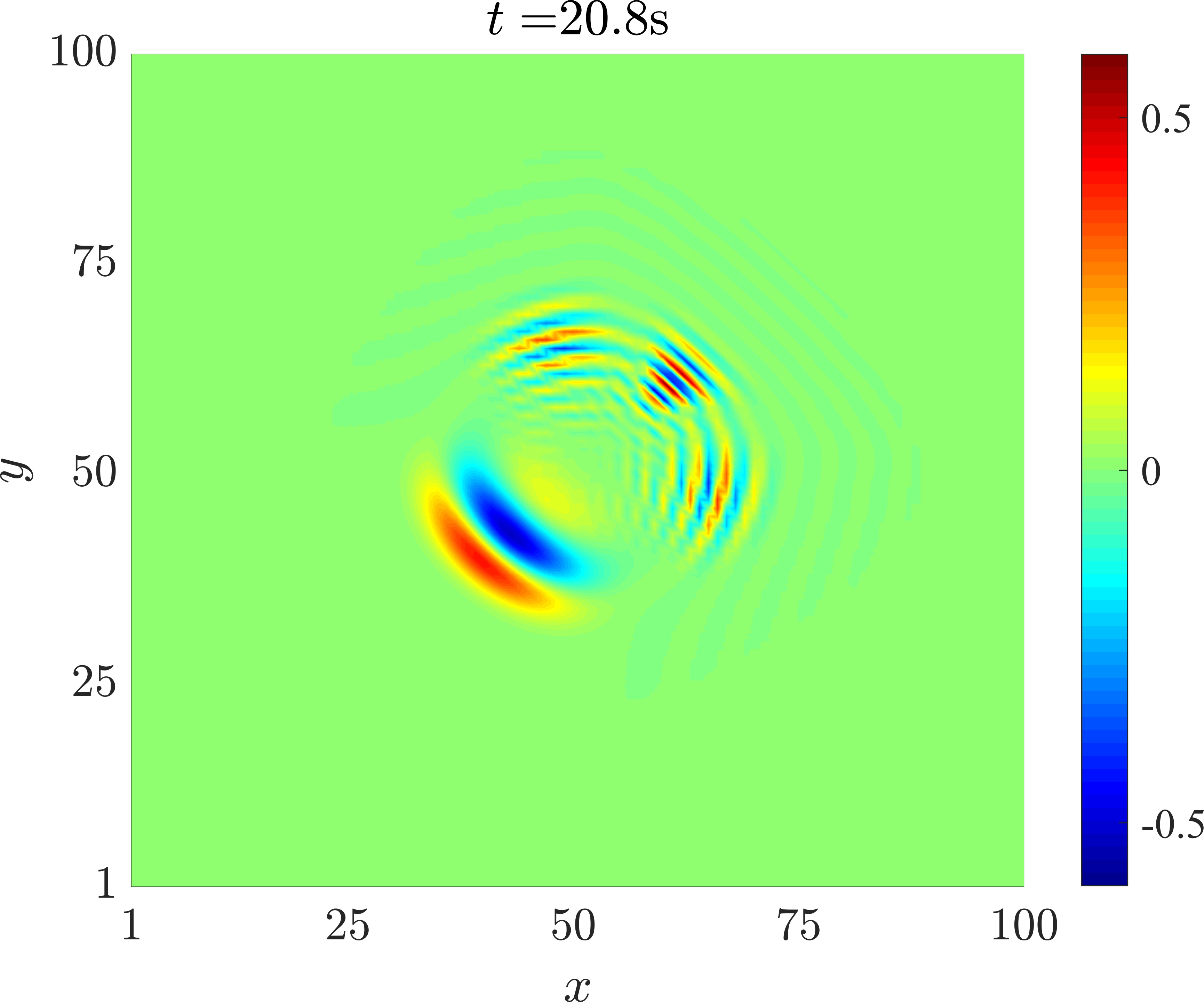}\label{Fig10f}}
	\subfigure[]{
		\includegraphics[height=0.26\textwidth]{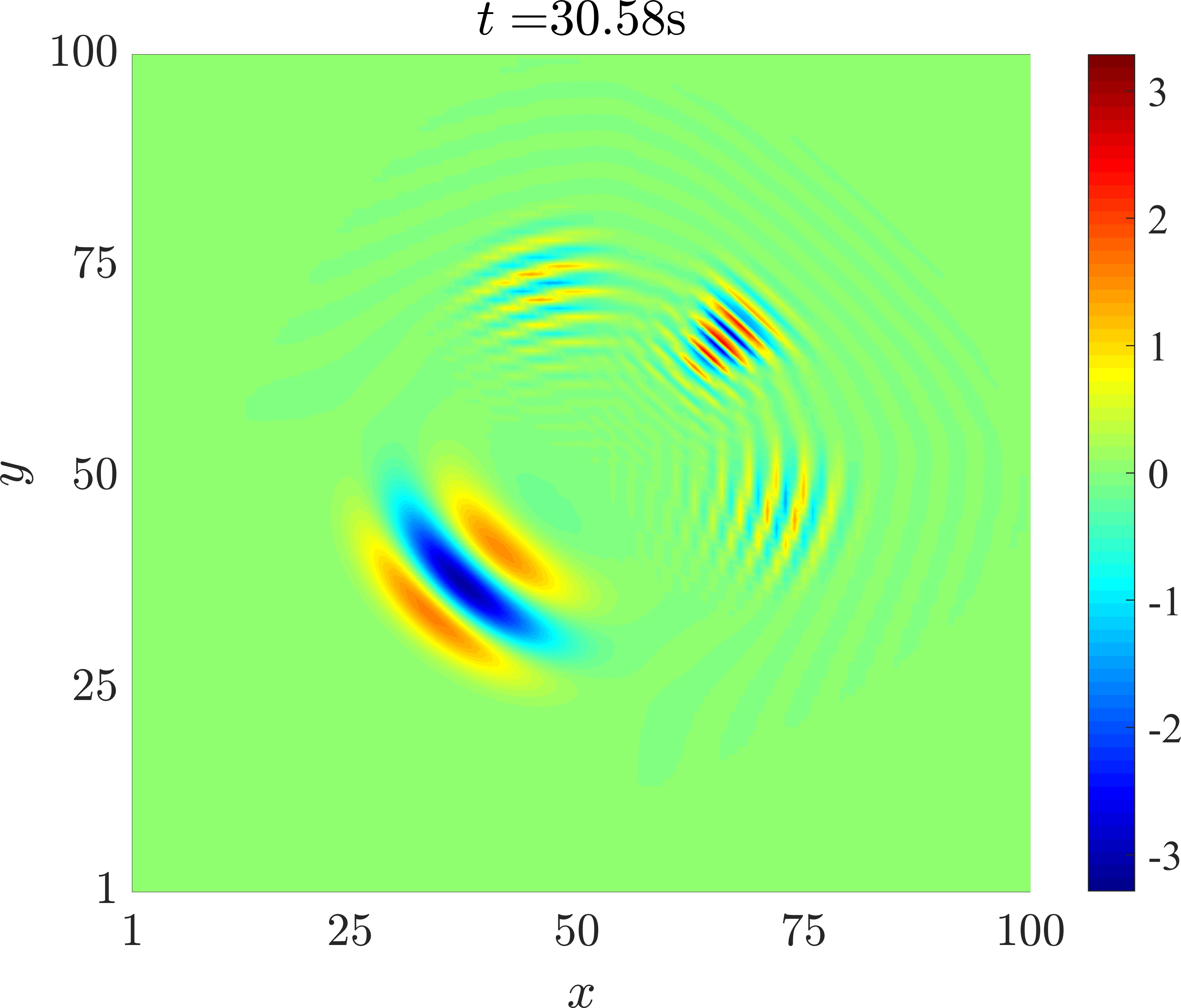}\label{Fig10g}}
	\caption{Non-reciprocal wave amplification for 2D lattices with non-local feedback interactions ($a=1$, $\gamma_x=\gamma_y=0.3$). Real (a) and imaginary (b) components of the dispersion are displayed along with contours associated with $\Omega_r=0.5$ - black circles, $\Omega_r=1.5$ - black dashed line, and $\Omega_r=2.5$ black solid line. Corresponding polar plots of the imaginary frequency components showing angular ranges of angular amplification (thick blue line), and attenuation (red thin line) (c-e): the non-locality of the feedback interactions result in multiple frequency/wavenumber bands where waves are amplified towards different directions. The behavior predicted by the dispersion analysis is illustrated by snapshots representing the lattice response to a broadband input (Fig.~\ref{Fig3}(e,f)) at subsequent time instants (f,g). }
	\label{Fig10}
\end{figure}

Similar to the 1D case, non-local feedback interactions in 2D lattices result in multiple non-reciprocal bands. This is illustrated for a lattice with $a=1$ and $\gamma_x=\gamma_y=0.3$ in Fig.~\ref{Fig10}. The real part of the dispersion displayed in Fig.~\ref{Fig10a} is similar to that of the local case (Fig.~\ref{Fig8a}). In contrast, the imaginary component of the dispersion (Fig.~\ref{Fig10b}) exhibits different regions of amplification and attenuation when compared to the local case (Fig.~\ref{Fig8b}). Contours at three different $\Omega_r$ values are highlighted in Fig.~\ref{Fig10a}: $\Omega_r=0.5$ - black circles, $\Omega_r=1.5$ - black dashed line, and $\Omega_r=2.5$ - black solid line. The corresponding wave vector component pairs are also shown in Fig.~\ref{Fig10b}, while angular variations of the amplification (thick blue line), and attenuation (red thin line) at these frequencies are shown in Figs.~\ref{Fig10}(c-e), illustrating how wave amplification occurs along different, and opposite directions within the range of frequencies defined by the dispersion relation of the lattice. Such behavior is confirmed by evaluating the transient response to a broad-band input (Figs.~\ref{Fig3}(e,f)) applied to the center mass of the lattice. Two subsequent snapshots of the lattice motion are displayed in Figs.~\ref{Fig10}(f,g), and illustrate how the broadband input is decomposed into approximately 4 wave packets that propagate along the distinct directions predicted by the imaginary component of dispersion. 

\subsection{Bulk topology, skin modes and corner modes}

\begin{figure}[t!]
	\centering
		\subfigure[]{
			\includegraphics[height=0.305\textwidth]{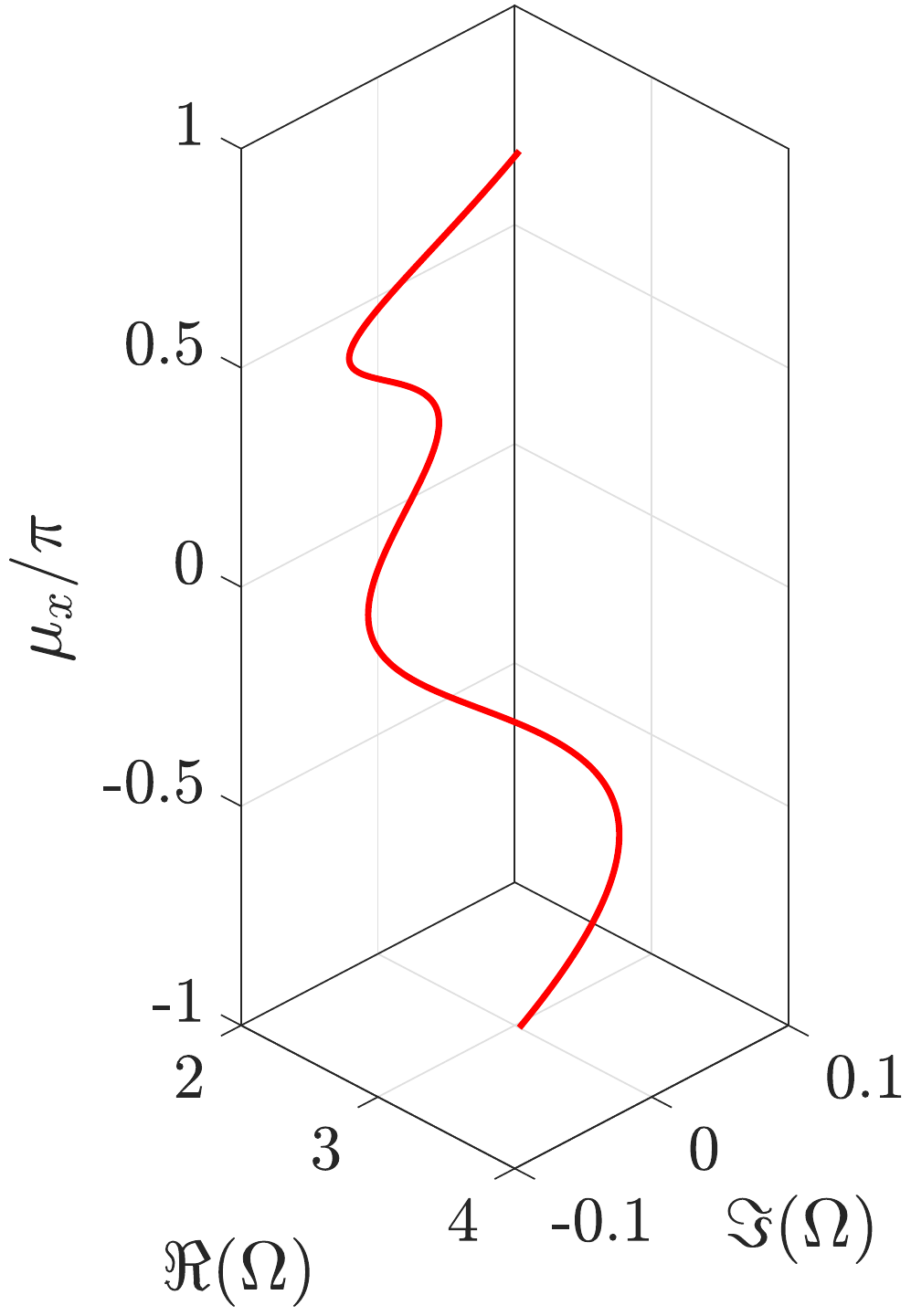}\label{Fig11a}}
		\subfigure[]{
			\includegraphics[height=0.305\textwidth]{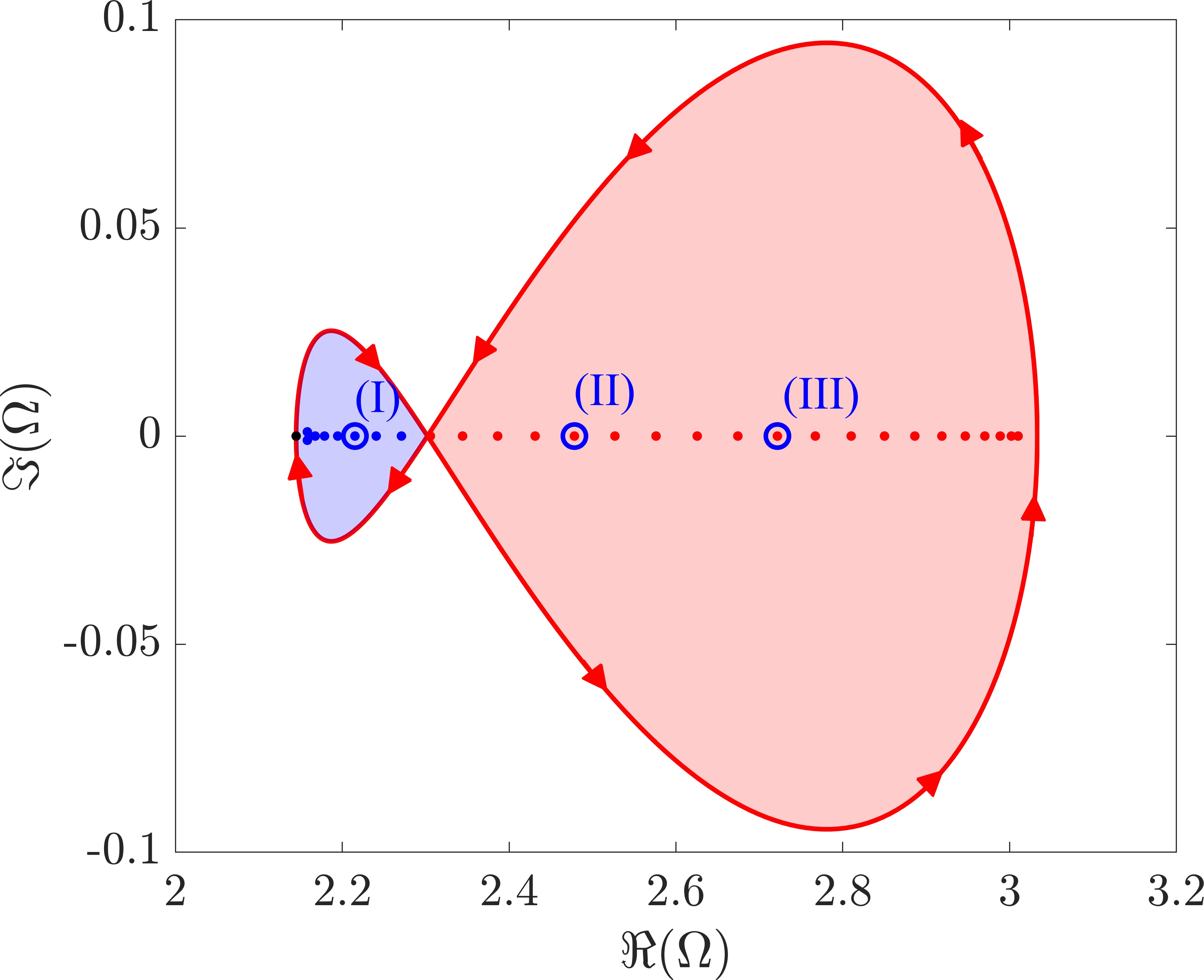}\label{Fig11b}}
		\subfigure[]{
			\includegraphics[height=0.305\textwidth]{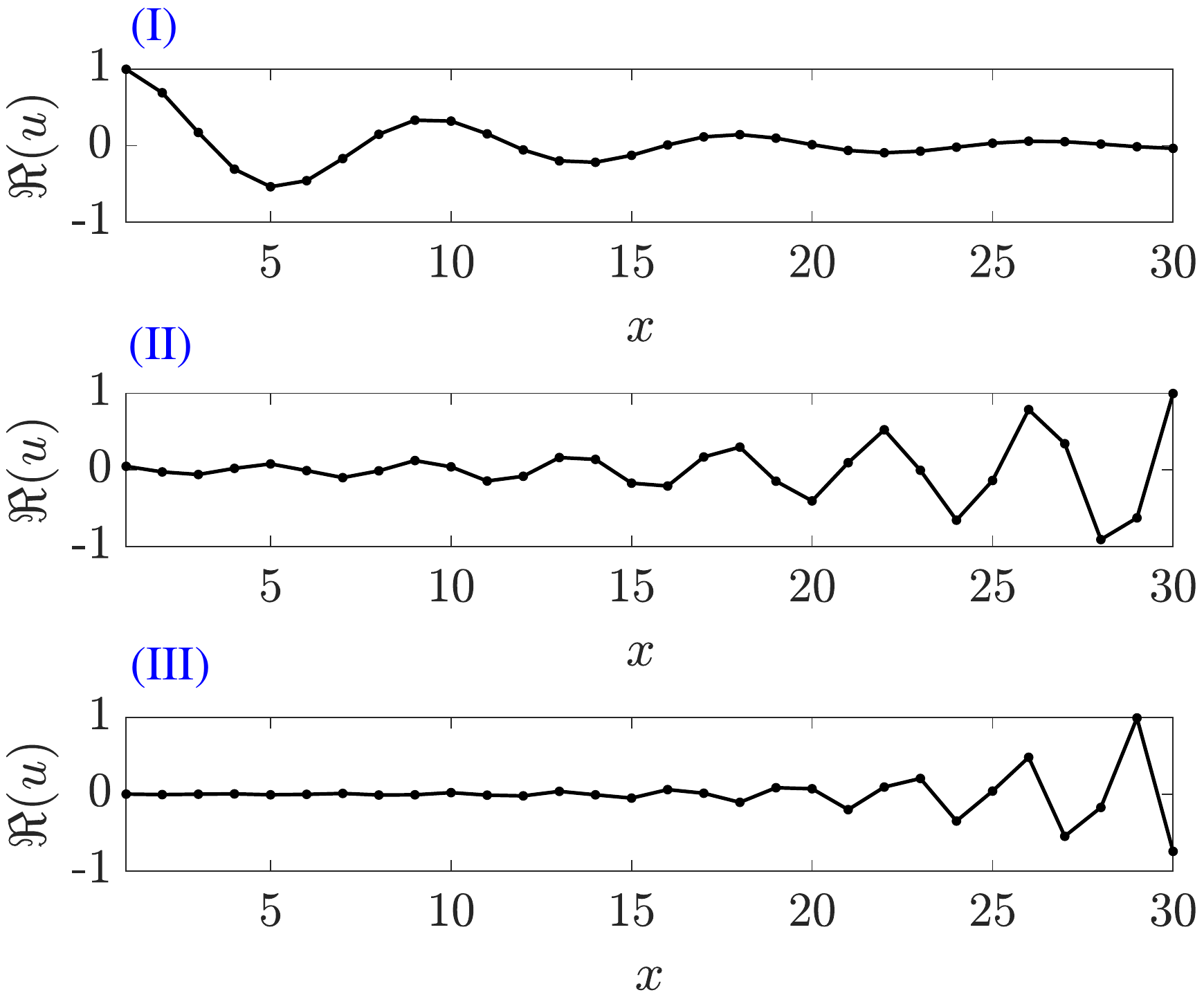}\label{Fig11c}}
		\begin{minipage}{0.59\textwidth}
		\subfigure[]{
		\includegraphics[height=0.78\textwidth]{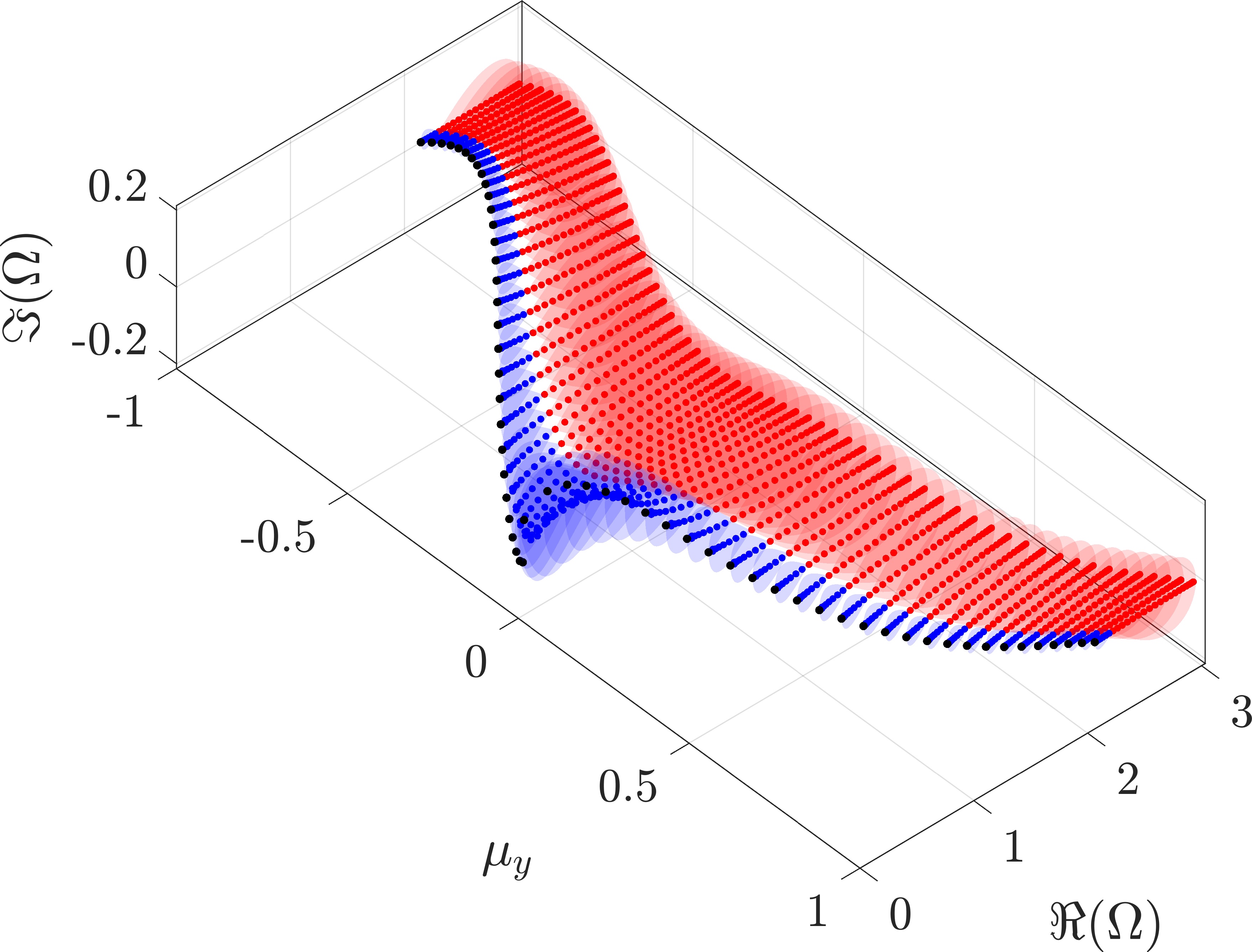}\label{Fig11d}}
		\end{minipage}
		\begin{minipage}{0.4\textwidth}
		\subfigure[]{
			\includegraphics[height=0.65\textwidth]{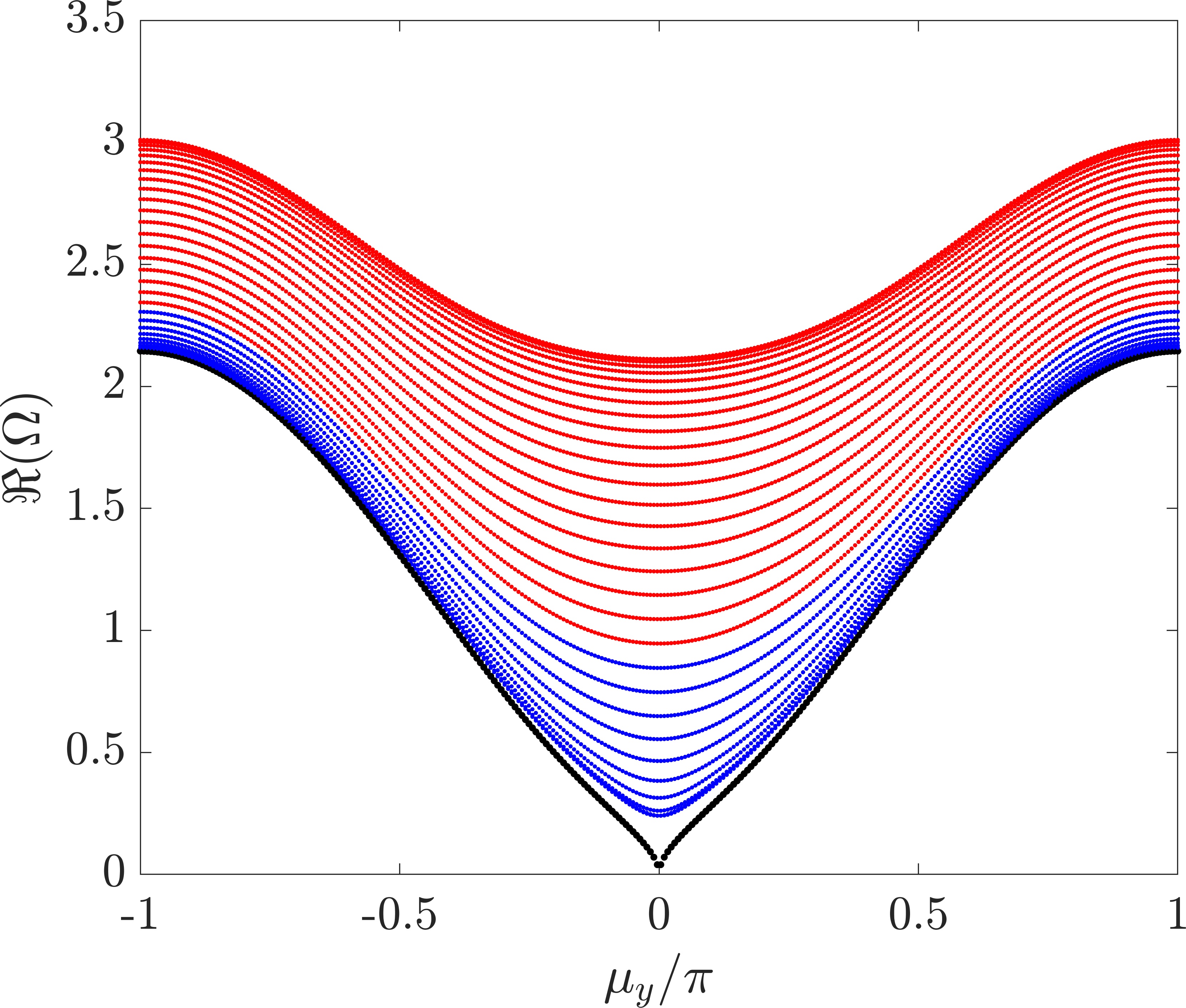}\label{Fig11e}}	
		\subfigure[]{
			\includegraphics[height=0.65\textwidth]{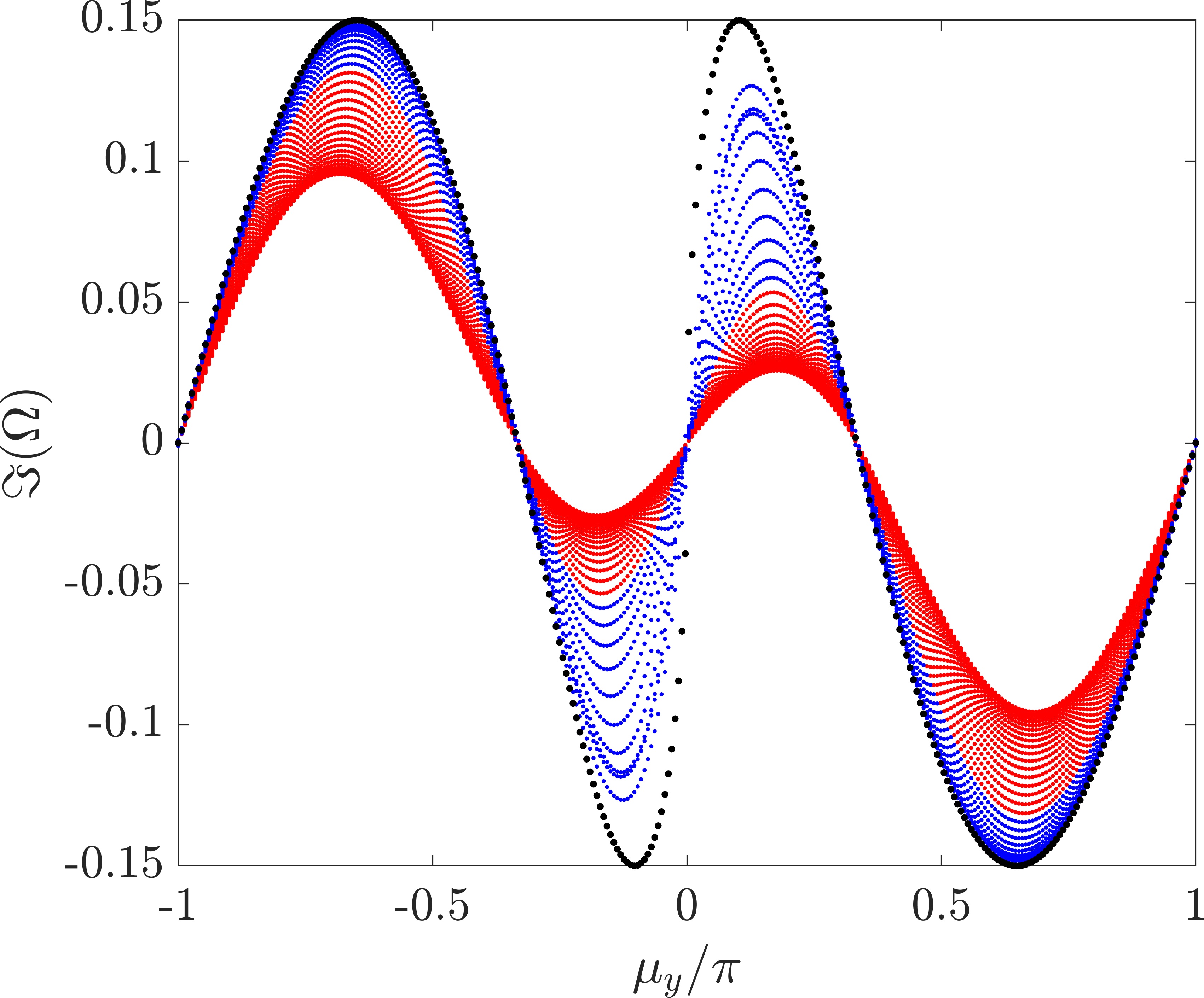}\label{Fig11f}}	
		\end{minipage}
	\caption{Bulk topology and skin modes of finite strip with $N=30$ masses along $x$ for lattice with feedback parameters $a=1$, $\gamma_x=\gamma_y=0.3$. The dispersion $\Omega(\mu_x,\mu_y=-\pi)$ (a) defines a closed loop on the complex plane (red lines in b), identifying two regions with winding numbers $\nu=-1$ (blue) and $\nu=1$ (red). Modes of the finite strip for $\mu_y=-\pi$ are localized at the left or right boundary (c) when their eigenfrequencies lie inside regions with $\nu=-1$ (blue dots) or $\nu=1$ (red dots), respectively. The procedure repeated for $\mu_y \in \{-\pi, \pi \}$ results in a representation of the bulk topology as red and blue regions representing different winding numbers, and modes of the finite strip spanning such regions are therefore localized at the left (blue dots) or right (red dots) boundary. The real and imaginary frequency components of the finite strip dispersion are displayed in (e,f) revealing non-reciprocity in amplification/attenuation of the localized modes propagating along the $y$ direction.}
	\label{Fig11}
\end{figure}

We extend the winding number analysis conducted for 1D lattices to describe the topological properties of 2D lattices and demonstrate the presence of skin edge and corner modes. We start by considering a finite lattice strip and show that its dispersion is associated with modes localized at one of the boundaries, that are either amplified or attenuated as they propagate along the other (infinite) direction. We then show that the combined effect of localization for finite strips in two directions ($x$ and $y$) produce modes that are localized at the corners of finite lattices.

As a representative case, we consider the set of parameters $a=1$, $\gamma_x=\gamma_y=0.3$, corresponding to the lattice whose wave properties are described in Fig.~\ref{Fig10}. We consider a finite lattice strip with $N=30$ masses along the $x$ direction, and infinite along the $y$ direction. To understand the topological properties and localization of the strip modes, we first consider a single wavenumber $\mu_y=-\pi$, for which the dispersion $\Omega(\mu_x,\mu_y=-\pi)$ is displayed in Fig.~\ref{Fig11a}, while its projection on the complex plane defines a loop represented by red lines in Fig.~\ref{Fig11b}. The eigenfrequencies of the finite strip for $\mu_y=-\pi$ are represented by dots in Fig.~\ref{Fig11b}, while a few representative modes marked by blue circles in Fig.~\ref{Fig11b} have their mode shapes displayed in Fig.~\ref{Fig11c}, revealing localization at the boundaries. Our analysis reveal that the localization of the strip modes for a given $\mu_y$ is related to the topology of the dispersion $\Omega(\mu_x)$ at that $\mu_y$ value. In Fig.~\ref{Fig11b}, blue and red zones again define regions for which $\nu=-1$ and $\nu=1$, and similar to the 1D lattices, modes of the finite strip whose eigenfrequencies lie inside such regions are respectively localized at the left (blue dots) or right (red dots) boundary (Fig.~\ref{Fig11c}). One particular mode marked by the black dot lies on top of the left end of the dispersion loop, and is characterized by displacements of all masses uniform along $x$ due to the free-free boundaries. Repeating this procedure for $\mu_y \in [-\pi, \pi]$ leads to the complete characterization of the finite strip dispersion (Fig.~\ref{Fig11d}): blue and red areas represent regions with bulk invariants $\nu=-1$ and $\nu=1$ (the dispersion loops are not shown for better visualization), hence modes of the finite strip spanning such regions are localized at the left (blue dots) or right (red dots) boundary. The real and imaginary components of the finite strip dispersion are separately displayed in Figs.~\ref{Fig11}(e,f), revealing that the dispersion of each mode exhibits non-reciprocity in amplification/attenuation associated with the imaginary frequency component, similarly to the behavior of the 1D lattice with $a=1$ (Fig.~\ref{Fig3a}). This opens the possibility of establishing non-reciprocal wave amplification as demonstrated in Fig.~\ref{Fig3} at the edges of a 2D lattice. 

\begin{figure}[t!]
	\centering
		\subfigure[]{
			\includegraphics[height=0.26\textwidth]{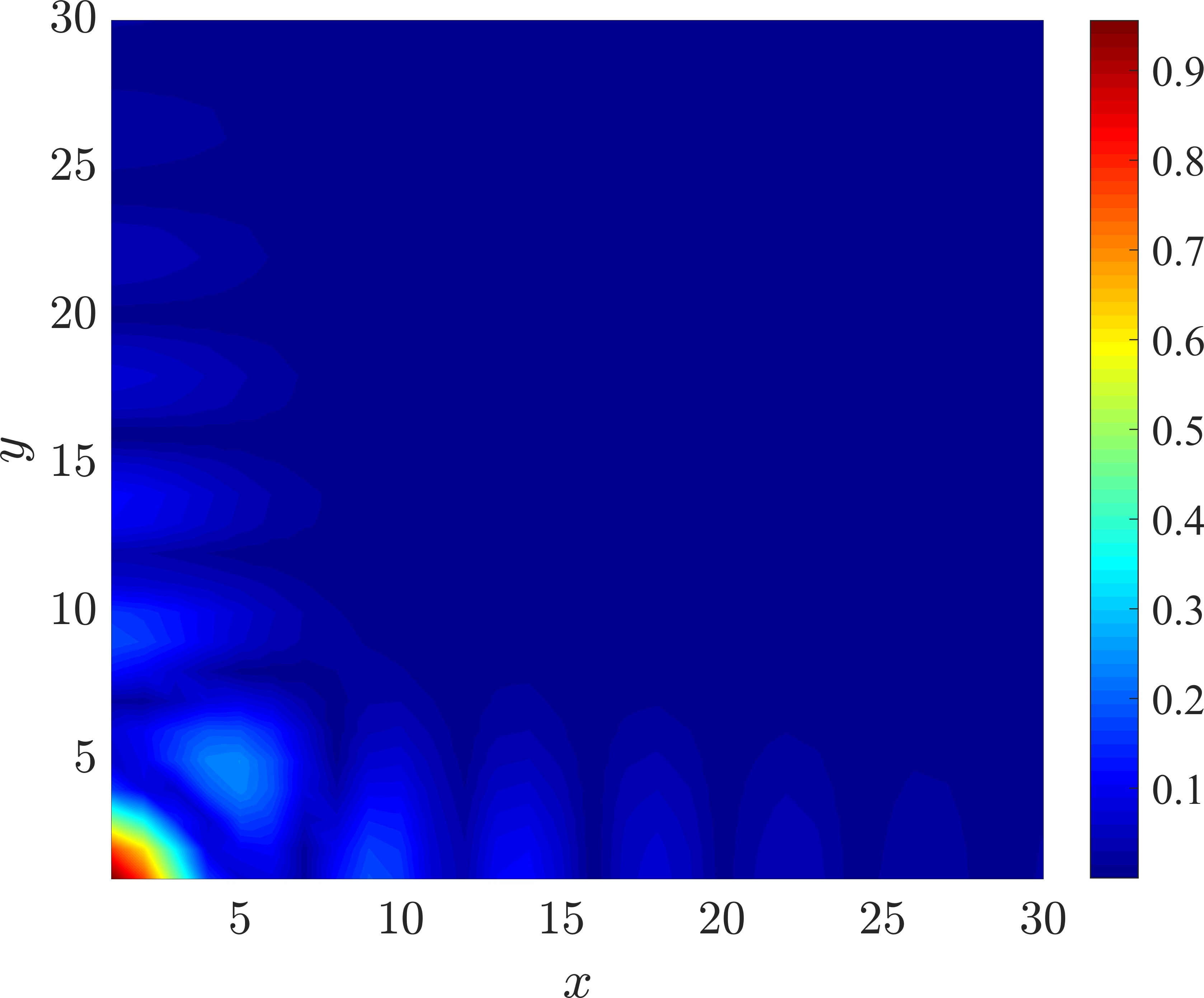}\label{Fig12a}}
		\subfigure[]{
			\includegraphics[height=0.26\textwidth]{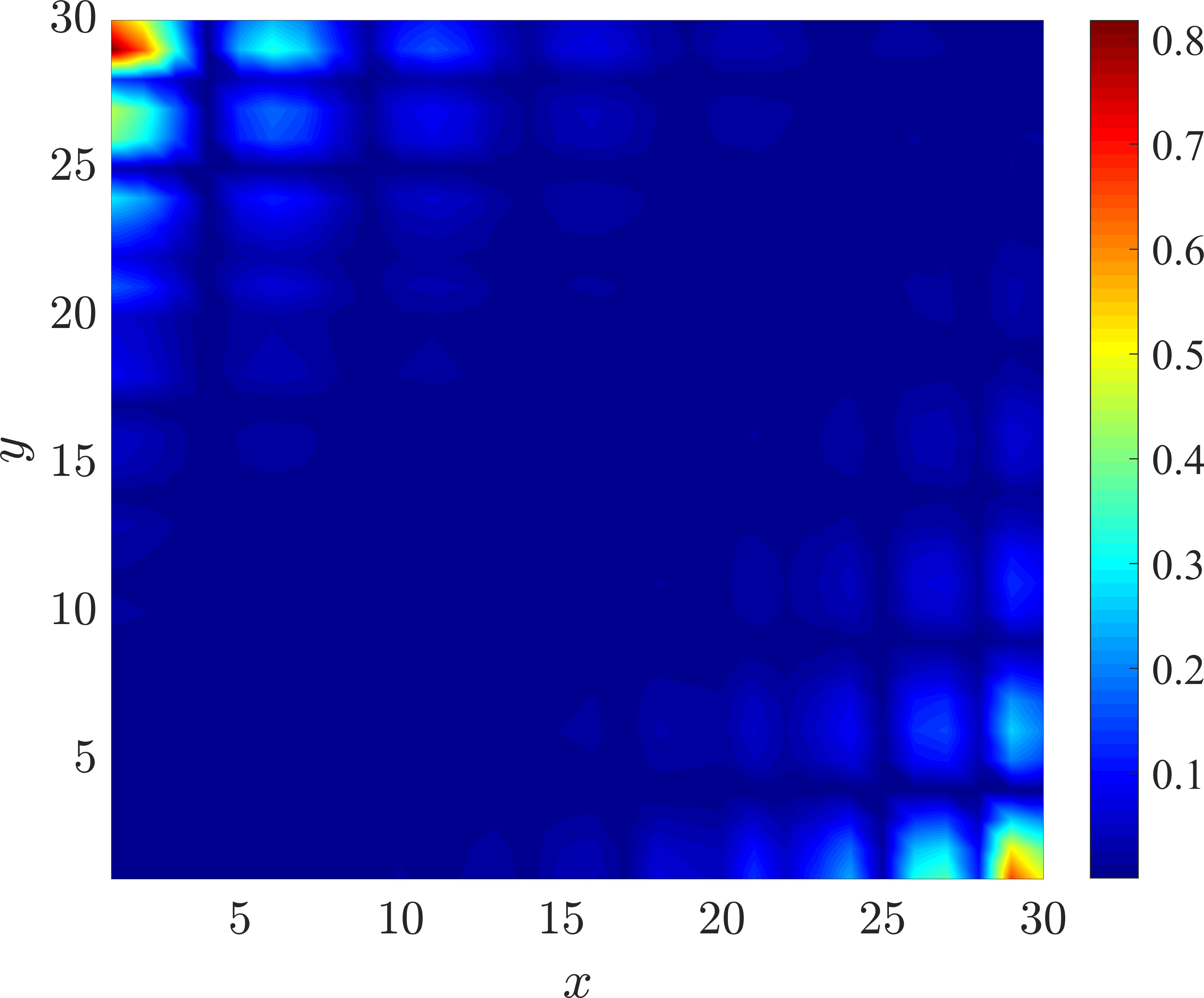}\label{Fig12b}}
		\subfigure[]{
			\includegraphics[height=0.26\textwidth]{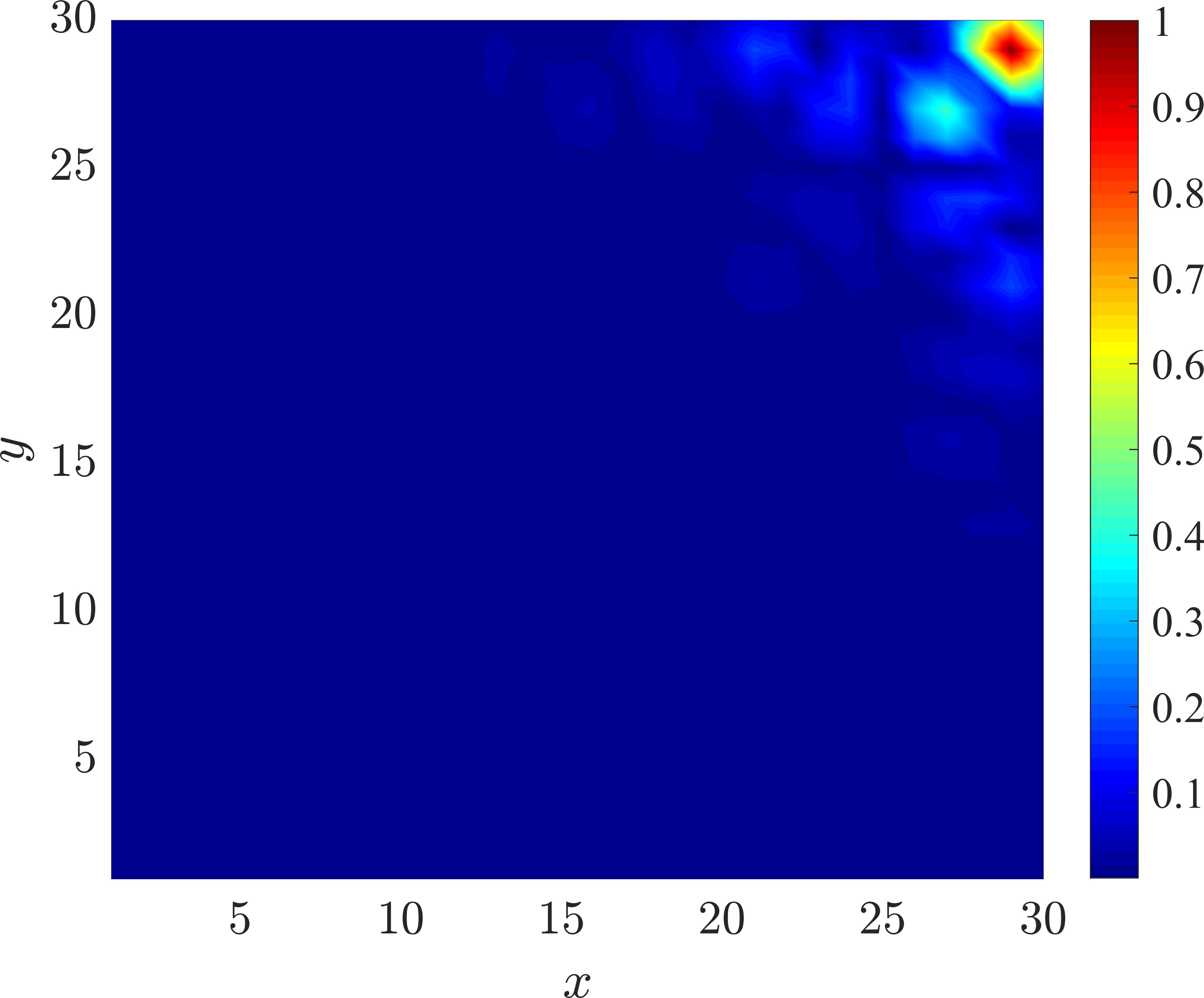}\label{Fig12c}}
	\caption{Representative bulk modes localized at corners of finite 2D lattice with $30 \times 30$ masses and feedback parameters $a=1$, $\gamma_x=\gamma_y=0.3$. The localization occurs due to the combined effect of localized skin modes for both the $x$ and $y$ directions.  }
	\label{Fig12}
\end{figure}

The same procedure applied to the finite strip along $x$ can be repeated for a finite strip with $N=30$ masses along $y$ instead, and infinite along the $x$ direction. Analogous results are obtained and modes localized at the bottom or upper boundary of the lattice strip are identified, which are not shown here for brevity. In line with recent work in quantum lattices~\cite{lee2019hybrid}, we find that the combined effect of the localized strip modes for the $x$ and $y$ directions lead to bulk modes of finite 2D lattices localized at one or more corners. Examples for a lattice with $30 \times 30$ masses are displayed in Fig.~\ref{Fig12}, where representative modes localized at the bottom left corner, upper right corner, or simultaneously localized at both upper left corner and bottom right corner are displayed. The localization at these corners is also in line with the non-reciprocal wave amplification behavior reported in Fig.~\ref{Fig10}, where contours at different frequencies define wave amplification either towards the bottom left corner, upper right corner, or simultaneously towards the upper left corner and bottom right corner. Although the localization of bulk modes at the corners of finite lattices is observed, to the authors knowledge, a prediction of their localization region based solely on where their frequency lie on the complex plane is still missing, and can be the subject of future investigations.

\section{Conclusions}\label{sec4}
In this paper, we investigate a family of elastic lattices where non-local feedback interactions lead to a series of unconventional phenomena associated with the physics of non-Hermitian systems. Among the key results, we demonstrate non-reciprocity associated with attenuation and amplification for waves propagating in different directions in 1D and 2D lattices, along with their topological properties associated with winding number of the complex dispersion bands, and localization of bulk modes at edges and corners. The presented results open new possibilities for the design of active metamaterials with novel functionalities such as those related to selective wave filtering, splitting, amplification and localization, both in one and two dimensions. Our results also corroborate recent observations~\cite{ghatak2019observation,brandenbourger2019non} that feedback control may be a fruitful strategy to investigate the physics and topology of non-Hermitian systems. While this work focuses on single-banded systems (already exhibiting a series of interesting properties), multiple possibilities are open for future work, such as exploring lattices with different geometries, modulations of control parameters and/or modification of control laws (\textit{e.g.} derivative and integral controls), as well as the introduction of non-linearities.

\begin{acknowledgments}
The authors gratefully acknowledge the support from the National Science Foundation (NSF) through the EFRI 1741685 grant and from the Army Research office through grant W911NF-18-1-0036.
\end{acknowledgments}

\bibliographystyle{unsrt}
\bibliography{References}

\end{document}